\documentclass[a4paper,fleqn,usenatbib]{mnras}
\pdfoutput=1

\usepackage[T1]{fontenc}
\usepackage{ae,aecompl}

\usepackage{graphicx}	% Including figure files
\usepackage{amsmath}	% Advanced maths commands
\usepackage{amssymb}	% Extra maths symbols
\usepackage{pdflscape}

\title[Young Star Clusters]{Young Star Clusters In Nearby Molecular Clouds}

\author[K. V. Getman et al.]{
K. V. Getman,$^{1}$\thanks{E-mail: kug1@psu.edu (KVG)}
M. A. Kuhn,$^{2,3}$
E. D. Feigelson,$^{1}$\newauthor
P. S. Broos,$^{1}$
M. R. Bate,$^{4}$
G. P. Garmire$^{5}$
\\
$^{1}$Department of Astronomy \& Astrophysics, 525 Davey Laboratory, Pennsylvania State University, University Park PA 16802\\
$^{2}$Instituto de Fisica y Astronomia, Universidad de Valparaiso, Gran Bretana 1111, Playa Ancha, Valparaiso, Chile\\
$^{3}$Millenium Institute of Astrophysics, Av. Vicuna Mackenna 4860, 782-0436 Macul, Santiago, Chile\\
$^{4}$Department of Physics and Astronomy, University of Exeter, Stocker Road, Exeter, Devon EX4 4QL, UK\\
$^{5}$Huntingdon Institute for X-ray Astronomy, LLC, 10677 Franks Road, Huntingdon, PA 16652, USA
}

\date{Accepted for publication in MNRAS, 2018 February 19}

\pubyear{2018}

\begin{document}
\label{firstpage}
\pagerange{\pageref{firstpage}--\pageref{lastpage}}
\maketitle

\begin{abstract}
The SFiNCs (Star Formation in Nearby Clouds) project is an X-ray/infrared study of the young stellar populations in 22 star forming regions with distances $\la1$~kpc designed to extend our earlier MYStIX survey of more distant clusters.  Our central goal is to give empirical constraints on cluster formation mechanisms.  Using parametric mixture models applied homogeneously to the catalog of SFiNCs young stars, we identify 52 SFiNCs clusters and 19 unclustered stellar structures. The procedure gives cluster properties including location, population, morphology, association to molecular clouds, absorption, age ($Age_{JX}$), and infrared spectral energy distribution (SED) slope. Absorption, SED slope, and $Age_{JX}$ are age indicators. SFiNCs clusters are examined individually, and collectively with MYStIX clusters, to give the following results. (1) SFiNCs is dominated by smaller, younger, and more heavily obscured clusters than MYStIX. (2) SFiNCs cloud-associated clusters have the high ellipticities aligned with their host molecular filaments indicating morphology inherited from their parental clouds. (3) The effect of cluster expansion is evident from the radius-age, radius-absorption, and radius-SED correlations.  Core radii increase dramatically from $\sim0.08$ to $\sim0.9$~pc over the age range $1-3.5$~Myr. Inferred gas removal timescales are longer than 1 Myr. (4) Rich, spatially distributed stellar populations are present in SFiNCs clouds representing early generations of star formation. An Appendix compares the performance of the mixture models and nonparametric Minimum Spanning Tree to identify clusters. This work is a foundation for future SFiNCs/MYStIX studies including disk longevity, age gradients, and dynamical modeling.
\end{abstract}


%%%%%%%%%%%%%%%%%%%%%%%%%%%%%%%%%%%%%%%%%%%%%%%%%%

%%%%%%%%%%%%%%%%% BODY OF PAPER %%%%%%%%%%%%%%%%%%
\section{Introduction} \label{intro_section}

Most stars in the Galaxy were formed in compact bound stellar clusters \citep{Lada2003} or distributed, unbound stellar associations \citep{Kruijssen2012}. Short-lived radioisotopic daughter nuclei in the Solar System meteorites indicate that our Sun formed in a modest-sized cluster on the edge of a massive OB-dominated molecular cloud complex \citep{Gounelle2012,Pfalzner2015}.  It is well accepted that most clusters quickly expand and disperse upon the removal of the residual molecular gas via stellar feedback so only a small fraction survive as bound open clusters \citep{Tutukov1978,Krumholz2014}.

But the formation mechanisms of rich clusters is still under study.  Dozens of sophisticated numerical (radiation)-(magneto)-hydrodynamic simulations of the collapse and fragmentation of turbulent molecular clouds followed by cluster formation have been published \citep{Krumholz2014,Dale2015}. The inclusion of stellar feedback is generally found to better reproduce important characteristics of star formation such as the stellar initial mass function (IMF)\citep{Bate2009,Offner2009,Krumholz2012}, gas depletion time, and star formation efficiency \citep{Krumholz2014,VazquezSemadeni2017}.

Constraints can be imposed on the simulations through quantitative comparison with the detailed properties of large cluster samples. For instance, realistic simulations of stellar clusters in the solar neighborhood, such as \citet{Bate2009,Bate2012}, are anticipated to obey the empirical cluster mass-size relationship seen in infrared (IR) cluster catalogs \citet{Carpenter2000, Lada2003}, X-ray/IR catalogs \citet{Kuhn2014,Kuhn2015a,Kuhn2015b}, and compilations of massive stellar clusters \citet{Portegies10, Pfalzner2013, Pfalzner2016}.

Our effort called MYStIX, Massive Young Star-Forming Complex Study in Infrared and X-ray, produces a large and homogeneous dataset valuable for analysis of clustered star formation \citep[][\url{http://astro.psu.edu/mystix}]{Feigelson2013,Feigelson2018}. MYStIX identified $>30,000$ young diskless (X-ray selected) and disk-bearing (IR excess selected) stars in 20 massive star forming regions (SFRs) at distances from 0.4 to 3.6 kpc.  Using quantitative statistical methods, \citet{Kuhn2014} identify over 140 MYStIX clusters with diverse morphologies, from simple ellipsoids to elongated, clumpy substructures. \citet{Getman2014a,Getman2014b} develop a new X-ray/IR age stellar chronometer and discover spatio-age gradients across MYStIX SFRs and within individual clusters. \citet{Kuhn2015a,Kuhn2015b} derive various cluster properties, discover wide ranges of the cluster surface stellar density distributions, and provide empirical signs of dynamical evolution and cluster expansion/merger.

More recently, the Star Formation in Nearby Clouds (SFiNCs) project \citep{Getman2017} extends the MYStIX effort to an archive study of 22 generally nearer and smaller SFRs where the stellar clusters are often dominated by a single massive star --- typically a late-O or early-B --- rather than by numerous O stars as in the MYStIX fields. Utilizing the MYStIX-based X-ray and IR data analyses methods, Getman et al. produce a catalog of nearly 8500 diskless and disk-bearing young stars in SFiNCs fields.  One of our objectives is to perform analyses similar to those of MYStIX in order to examine whether the behaviors of clustered star formation are similar --- or different --- in smaller (SFiNCs) and giant (MYStIX) molecular clouds.

In the current paper, the MYStIX-based parametric method for identifying clusters using finite mixture models \citep{Kuhn2014} is applied to the young stellar SFiNCs sample. Fifty two SFiNCs clusters and nineteen unclustered stellar structures are identified across the 22 SFiNCs SFRs. Various basic SFiNCs stellar structure properties are derived, tabulated, and compared to those of MYStIX.  In an Appendix, we compare our parametric method to the common non-parametric method for identifying stellar clusters based on the minimum spanning tree (MST) \citep{Gutermuth2009,Schmeja11}.

The method of finite mixture models for spatial point processes is briefly described in \S \ref{model_section}. The SFiNCs stellar sample is provided in \S \ref{sample_section}. The cluster surface density maps, model validation, error analysis, and membership are presented in \S\S \ref{surface_density_maps_subsection}-\ref{membership_subsection}. A multivariate analysis of SFiNCs$+$MYStIX clusters is given in \S\S \ref{ma_section}-\ref{cloud_subsection}. The main science results are discussed in \S \ref{discussion_sec}. The Appendices discuss each SFiNCs region and compare the performance of cluster identification methods.  

\section{Cluster Identification with Mixture Models} \label{model_section} 

Identification and morphological characterization of SFiNCs clusters uses the parametric statistical mixture model developed in MYStIX \citep{Kuhn2014}. The mixture model is the sum of multiple clusters fit to the sky distribution of young stellar objects (YSOs) using maximum likelihood estimation.  The number of clusters is one of the model parameters optimized in the procedure.

Briefly, this analysis investigates the distribution of a set of points $\mathbf{X}=\{\mathbf{x}_1,...,\mathbf{x}_N\}$ giving the (RA,Dec) coordinates of each YSO.  Each point belongs to one of $k$ clusters, or to an unclustered component of scattered stars characterized by complete spatial randomness. Each cluster is described by the probability density function $g(\mathbf{x};\mathbf{\theta}_j)$, where $\theta_j$ is the set of parameters for the component.  Mixture models of point processes commonly use a normal (Gaussian) function \citep{McLachlan2000,Fraley2002, KuhnFeigelson2017}. But, following \citet{Kuhn2014}, we choose $g$ to be a two-dimensional isothermal ellipsoid defined in their equation 4. This is the radial profile of a dynamically relaxed, isothermal, self-gravitating system where the stellar surface density having a roughly flat core and power-law halo.  The effectiveness of the isothermal ellipsoidal model is shown in \S\ref{model_validation} below for SFiNCs and by \citet{Kuhn2014, Kuhn2017} for other young clusters.  The relative contribution of each cluster is given by mixing coefficients $\{a_1,...,a_k\}$, where $0\le a_j\le1$ and $\sum_{j=1}^{k}a_j=1$. The distribution of the full sample $f(\mathbf{x})$ is then the sum of the $k$ cluster components, $\sum_{j=1}^{k+1}a_j g(\mathbf{x};\mathbf{\theta}_j)$, plus a constant representing the unclustered component.  

The mixture model is fit to the (RA,Dec) point distribution for each SFiNCs field from \citet{Getman2017} by maximum likelihood estimation.  Each cluster has six parameters, $\theta=\{x_0,y_0,R_c,\epsilon,\phi,\Sigma_c\}$: coordinates of cluster center ($x_0,y_0$), isothermal core radius $R_c$ defined as the harmonic mean of the semi-major and semi-minor axes, ellipticity $\epsilon$, position angle $\phi$, and central stellar surface density $\Sigma_c$.  The full model, including the unclustered surface density, has $6k+1$ parameters. The density in our chosen model never vanishes at large distances from the cluster center. However, the absence of a parameter for the outer truncation radius in our model does not strongly affect the results, because the spatial distribution of the MYStIX and SFiNCs stars is examined over a finite field of view. The core radius parameter can be better constrained than the truncation radius \citep[e.g.,][]{Kroupa2001}, and our previous work shows that inclusion of a truncation radius parameter is not necessary to get a good fit \citep[e.g.,][]{Kuhn2010}.

To reduce unnecessary computation on the entire parameter space, an initial superset of possible cluster components is obtained by visual examination of the YSO adaptively smoothed surface density maps.  This initial guess then iteratively refined using the Nelder-Mead simplex optimization algorithm implemented in the $R$ statistical software environment.   The optimal model set is chosen by minimizing the penalized likelihood Akaike Information Criterion (AIC).  Additional software for post-fit analysis and visualization is provided by $R$'s {\it spatstat} CRAN package, a comprehensive statistical tool for analyzing spatial point patterns \citep{Baddeley2015}.  See \citet{Kuhn2014} for further details; the $R$ code for cluster identification using parametric mixture models is presented in Appendix of \citet{Kuhn2014}.

Figure \ref{fig_fitting_example_be59} shows an example of the fitting process for SFiNCs field containing cluster Berkeley 59. Panel (a) shows an initial guess for a cluster that is purposefully misplaced.  Its location is corrected by the algorithm in panel (b) giving $AIC= -2340$. In the remaining panels, new components are iteratively added until $AIC$ reaches its minimum value of $AIC=-2395$ for two clusters (panel d).  The attempt to add a third cluster raises the AIC (panel f), so the $k=2$ model is considered to be optimal for this field.

%\clearpage
\begin{figure}
\centering
\includegraphics[angle=0.,width=80mm]{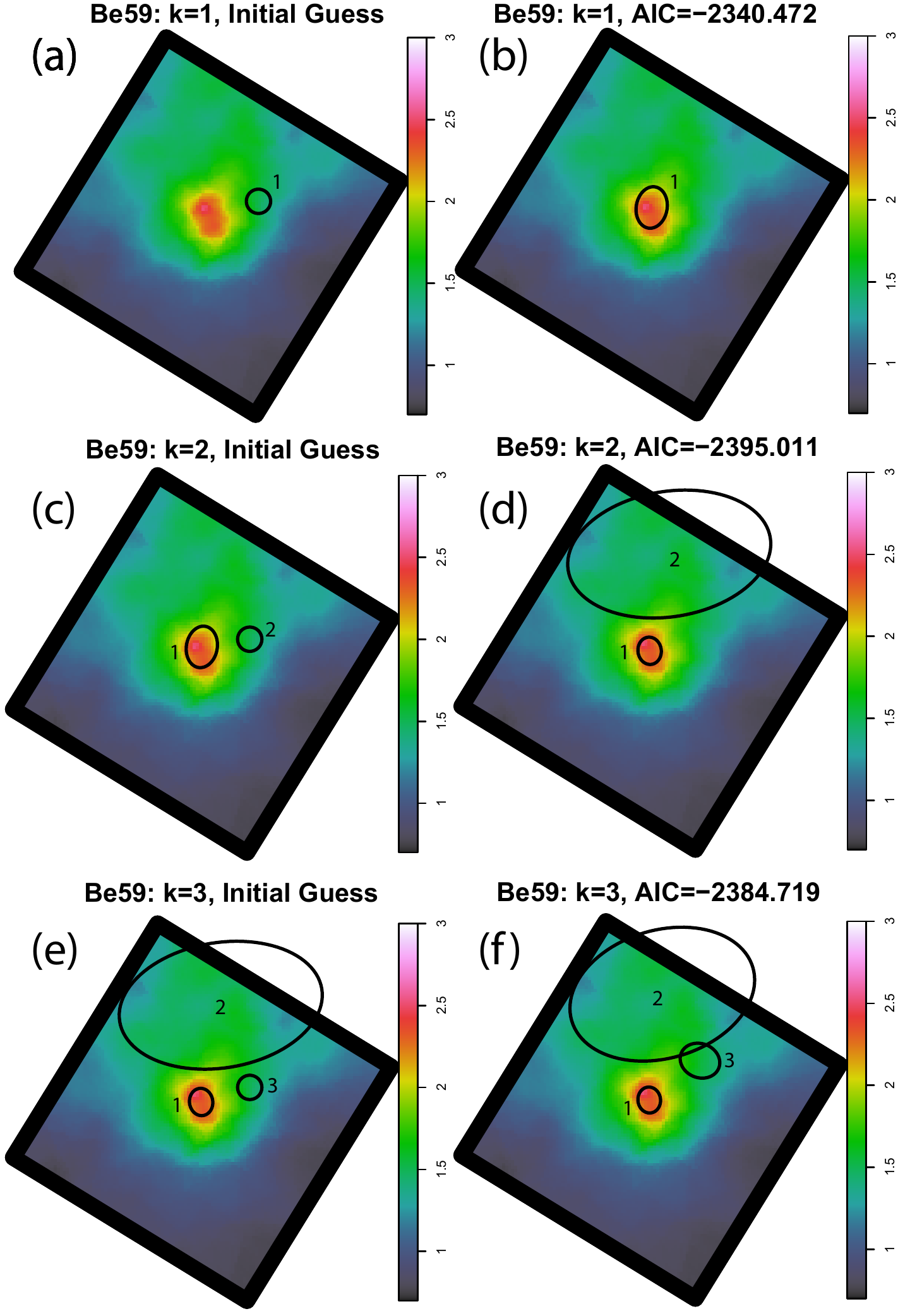}
\caption{An example of the mixture model fitting for the ``flattened'' star sample in Be 59 (see explanation for ``flattened'' sample in \S \ref{sample_section}). The fitting consists of a sequence of main operations, starting with panel (a) and ending with panel (f). The panels show the smoothed projected stellar surface density of the YSOs in Be 59, with a color-bar in units of observed stars per pc$^{-2}$ (on a logarithmic scale). The core radii of the considered clusters are marked by the black ellipses and labeled by numbers. Panels (a), (c), and (e) display the refinement of the initial guess for the number of clusters. Panels (b), (d), and (f) display the fitting results using the initial guesses presented in panels (a), (c), and (e), respectively. The panel titles give the number of considered clusters ($k$), and, for the fitting operations, the resulting values of the Akaike Information Criterion. \label{fig_fitting_example_be59}}
\end{figure} 
%\newpage

We are aware that the use of parametric mixture models for star cluster identification is unusual; most researchers use nonparametric methods.  One important nonparametric procedure is the pruned Minimal Spanning Tree (MST) of the stellar celestial locations, also known as the astronomers' friends-of-friends algorithm and the statisticians' single linkage hierarchical clustering algorithm.  In Appendix, we compare the performances of the mixture model and MST methods both for simulated situations and our SFiNCs datasets.

\section{Sample Selection} \label{sample_section}

Using MYStIX-based data analysis methods described by \citet{Feigelson2013}, \citet{Getman2017} perform a homogeneous reanalysis of the archived {\it Chandra}-ACIS X-ray, {\it Spitzer}-IRAC mid-IR data for the 22 nearby SFiNCs SFRs. Table \ref{tbl_cluster_sample} lists the target fields and their estimated distances.  This analysis resulted in $\simeq 15,300$ X-ray and $\sim 1,630,000$ mid-IR point sources. Further combining these X-ray and mid-IR source data with the archived 2MASS near-IR catalog and applying a decision tree classification method (based on the photometry and spatial distributions of the X-ray and IR point sources), Getman et al. identify 8,492 SFiNCs probable cluster members (SPCMs) across the 22 SFiNCs SFRs. 

The SPCMs are a union of {\it Chandra} X-ray selected diskless and disk-bearing YSOs, {\it Spitzer} IR excess (IRE) disk-bearing YSOs, and published OB stars.  SPCMs were considered as disk-bearing (diskless) when their infrared spectral energy distributions in 2MASS+IRAC IR bands  exhibited (did not exibit) an IRE when compared to the dereddened median SED templates of IC 348 PMS stellar photospheres \citep{Lada2006}. A fraction of SPCMs that lack IR SEDs were classified as ``PMB'' (possible member). Out of 8492 SPCMs, 66\%, 30\%, and 4\% were classified as disk-bearing, diskless, and ``PMB'', respectively. The total numbers of SPCMs for each field are listed in Column 3 of Table \ref{tbl_cluster_sample}.

As in MYStIX, substantial spatial variations in X-ray sensitivity are present in the SPCM dataset due to the off-axis mirror vignetting and degradation of the point-spread function, and further complicated by the disorganized mosaics of {\it Chandra} fields with different exposures. To mitigate these effects, weak X-ray SPCMs with their X-ray photometric flux below the X-ray completeness limit ($F_{X,lim}$) were excluded from the cluster identification analysis. $F_{X,lim}$ is calculated as in MYStIX \citep{Kuhn2014}.  The values of $F_{X,lim}$ flux and the number of the remaining X-ray SPCMs are given in Columns 4 and 6 of Table \ref{tbl_cluster_sample}, respectively.  

The SPCM dataset is further culled of the non-X-ray {\it Spitzer} selected disk-bearing YSOs that lie outside the {\it Chandra}-ACIS-I fields. The number of the remaining {\it Spitzer} disk-bearing YSOs is given in Column 7. Column 8 gives the number of the OB-type SPCMs that are either X-ray sources with $F_{X} < F_{X,lim}$ or non-X-ray disk-bearing YSOs lying within the {\it Chandra}-ACIS-I fields. The final number of SPCMs $N_\star$ left for the cluster identification analysis is 5,164 (Column 5) constituting 61\% of the original SPCM dataset. Referred to hereafter as the ``flattened'' sample, these stars are used for identifying SFiNCs clusters and deriving their morphological properties. However, the membership analysis (\S \ref{membership_subsection}) and derivation of cluster's absorption, SED slope, and age (\S \ref{sample_subsection}) are based on the original sample of 8492 SPCMs.  

\newpage
%\begin{table*}\scriptsize
\begin{table*}\small
%\begin{table*}\tiny
\centering
 \begin{minipage}{180mm}
 \caption{SFiNCs Targets for Cluster Analysis. Samples of YSOs in 22 SFiNCs SFRs used for the cluster analysis. Column 1: SFiNCs SFR \citep{Getman2017}. Column 2: Distance to SFR \citep[][their Table~1]{Getman2017}. Column 3. Total number of SFiNCs probable cluster members (SPCMs) identified by \citet{Getman2017}. Column 4: Apparent $(0.5-8)$~keV X-ray photon flux limit imposed for spatial uniformity. Columns 5-8: Numbers of SPCMs used for the cluster analysis: total number (Column 5), number of X-ray selected SPCMs that lie within the {\it Chandra}-ACIS-I field and have their X-ray photometric flux above the $F_{X,lim}$ value (Column 6), number of non-X-ray disk-bearing SPCMs that lie within the {\it Chandra}-ACIS-I field (Column 7), and number of OB-type stars within the SPCM subsample used for the cluster identification analysis (Column 8).}
 \label{tbl_cluster_sample}
 \begin{tabular}{@{\vline}c@{}c@{}c@{}c@{\vline}c@{}c@{}c@{}c@{}c@{ \vline }}
\cline{1-9}
&&&&&&&&\\
\multicolumn{3}{@{\vline}c}{SFiNCs regions} && \multicolumn{5}{c@{\vline}}{SPCM Subsample for Cluster Identification~~}\\
&&&&&&&&\\
\cline{1-3}\cline{5-9}
&&&&&&&&\\
%\hline
~Region~ & ~Distance~ & ~SPCMs~ && ~$\log F_{X,lim}$~ & ~N$_\star$~ & ~X-ray~ & ~IRE~ & ~OB~\\
& (kpc) & (stars)         && ~(ph~s$^{-1}$cm$^{-2}$)~ & (stars) & (stars) & (stars) & (stars) \\
(1)&(2)&(3)               &&(4)&(5)&(6)&(7)&(8)\\
\cline{1-9}
&&&&&&&&\\ 
Be 59 					& 0.900 & 626 &&        -6.000 &   435 &  315 &   118 &     8\\
SFO 2 					& 0.900 &    71 &&        -6.000 &    63 &     34 &     29 &     0\\
       NGC 1333 		        & 0.235 &   181 &&       -5.750 &   118 &    55 &     62 &     4\\
         IC 348 			& 0.300 &   396 &&       -5.750 &   224 &  162 &     62 &     1\\
       LkH$\alpha$ 101 			& 0.510 &   250 &&       -5.875 &   149 &    99 &     48 &     4\\
  NGC 2068-2071 			& 0.414 &   387 &&       -5.750 &   234 &  120 &   113 &     3\\
   ONC Flank S 				& 0.414 &   386 &&       -5.375 &   237 &  133 &   104 &     1\\
   ONC Flank N 				& 0.414 &   327 &&       -5.875 &   217 &  151 &     64 &     4\\
        OMC 2-3 			& 0.414 &   530 &&       -5.250 &   238 &  144 &     91 &     5\\
         Mon R2 			& 0.830 &   652 &&       -5.750 &   280 &  134 &   144 &     5\\
      GGD 12-15 			& 0.830 &   222 &&       -5.875 &   147 &    72 &     75 &     2\\
        RCW 120 			& 1.350 &   420 &&       -6.250 &   278 &  157 &   121 &     1\\
   Serpens Main 			& 0.415 &   159 &&       -6.125 &   105 &    55 &     50 &     0\\
  Serpens South 			& 0.415 &   645 &&       -6.250 &   288 &    56 &   232 &     0\\
IRAS 20050+2720 			& 0.700 &   380 &&       -6.250 &   281 &  121 &   160 &     0\\
       Sh 2-106 			& 1.400 &   264 &&       -6.125 &   221 &  123 &     98 &     1\\
        IC 5146 			& 0.800 &   256 &&       -6.250 &   232 &  141 &     90 &     6\\
       NGC 7160 			& 0.870 &   143 &&       -6.000 &     93 &    86 &       2 &     8\\
      LDN 1251B 			& 0.300 &     49 &&       -5.500 &     31 &    21 &     10 &     0\\
       Cep OB3b		 		& 0.700 & 1636 &&       -5.875 & 1019 &  551 &   465 &     9\\
          Cep A 			& 0.700 &   335 &&       -5.750 &   164 &    81 &     83 &     0\\
          Cep C 			& 0.700 &   177 &&       -5.750 &   132 &    52 &     80 &     0\\
&&&&&&&&\\ 
\cline{1-9} 
\end{tabular}
\end{minipage}
\end{table*}
%\clearpage
%\newpage

\citet[][their Appendix]{Feigelson2013} discuss issues of incompleteness and bias in samples derived in this fashion.  Most importantly, due to a well-known X-ray/mass correlation, the X-ray-selected SPCM subsamples are approximately complete above mass limits around $0.1-0.3$~M$_{\odot}$ for the range of $ F_{X,lim}$ values. For the IRE subsamples in most of the SFiNCs SFRs, the histograms of the [3.6]-magnitude for the SFiNCs {\it Spitzer} point sources peak near 17~mag equivalent to $\la 0.1$~M$_{\odot}$ sensitivity limits \citep[][their Figure 3]{Getman2017}. The IRE sub-sample is further limited by the 2MASS sensitivity limit of $K_s \sim 14.3$~mag, which translates to $\sim 0.1-0.3$~M$_{\odot}$ PMS star at distances of $300-900$~pc.  A future SFiNCs paper will compensate for these incompleteness effects using the X-ray luminosity function and initial mass function following the procedure of  \citet{Kuhn2015a}.  

The detection of highly absorbed clusters relies heavily on the catalog of the {\it Spitzer} selected disk-bearing YSOs.   But, in the cluster centers of five SFiNCs SFRs (LkH$\alpha$~101, Mon~R2, RCW~120, Sh~2-106, and Cep~A), the IRAC point source sensitivity is reduced by the high background nebular emission from heated dust \citep[][their Figure Set 6]{Getman2017}. Unlike in the case of the X-ray sample, application of a uniform MIR flux limit seems infeasible here as it would leave only a handful of bright IRAC sources. For the detection of these clusters, our procedures depend mainly on the catalog of the {\it Chandra} selected YSOs.

\section{Clusters in SFiNCs Clouds} \label{clusters_subsection}

The fitting of the isothermal ellipsoid mixture model to the ``flattened'' sample of SPCMs yields 52 clusters across the 22 SFiNCs SFRs. Table \ref{tbl_cluster_morphology} lists their morphological properties: the celestial coordinates of the cluster center, core radius, ellipticity, orientation, and total number of YSOs estimated by integrating the cluster model across an ellipse four times the size of the core. The estimated uncertainties on these cluster parameters are explained in \S \ref{error_analysis}.  Also listed is a flag indicating the association with a molecular cloud (\S \ref{cloud_subsection}). Note that $4 \times R_c$ corresponds roughly to the projected half-mass radius for a cluster with an outer truncation radius of $\sim 17 \times R_c$  \citep{Kuhn2015b}. This somewhat arbitrary decision of $4 \times R_c$, as an integration radius, was the result of our visual inspection of trial star assignments among MYStIX \citep{Kuhn2014} and SFiNCs clusters, taking into consideration two main factors, the typical separation between adjacent clusters and typical size of field of view. The numbers of stars within a specific radius follow the Equation 3 in \citet{Kuhn2014}, assuming that the model provides an accurate description of the distribution of stars. According to this equation, the number of stars in a cluster increases by a factor 1.6 when changing the integration radius from, say, $4 \times R_c$ to $10 \times R_c$.

\subsection{Morphology of Clustered Star Formation} \label{surface_density_maps_subsection}

\begin{figure}
\centering
\includegraphics[angle=0.,width=70mm]{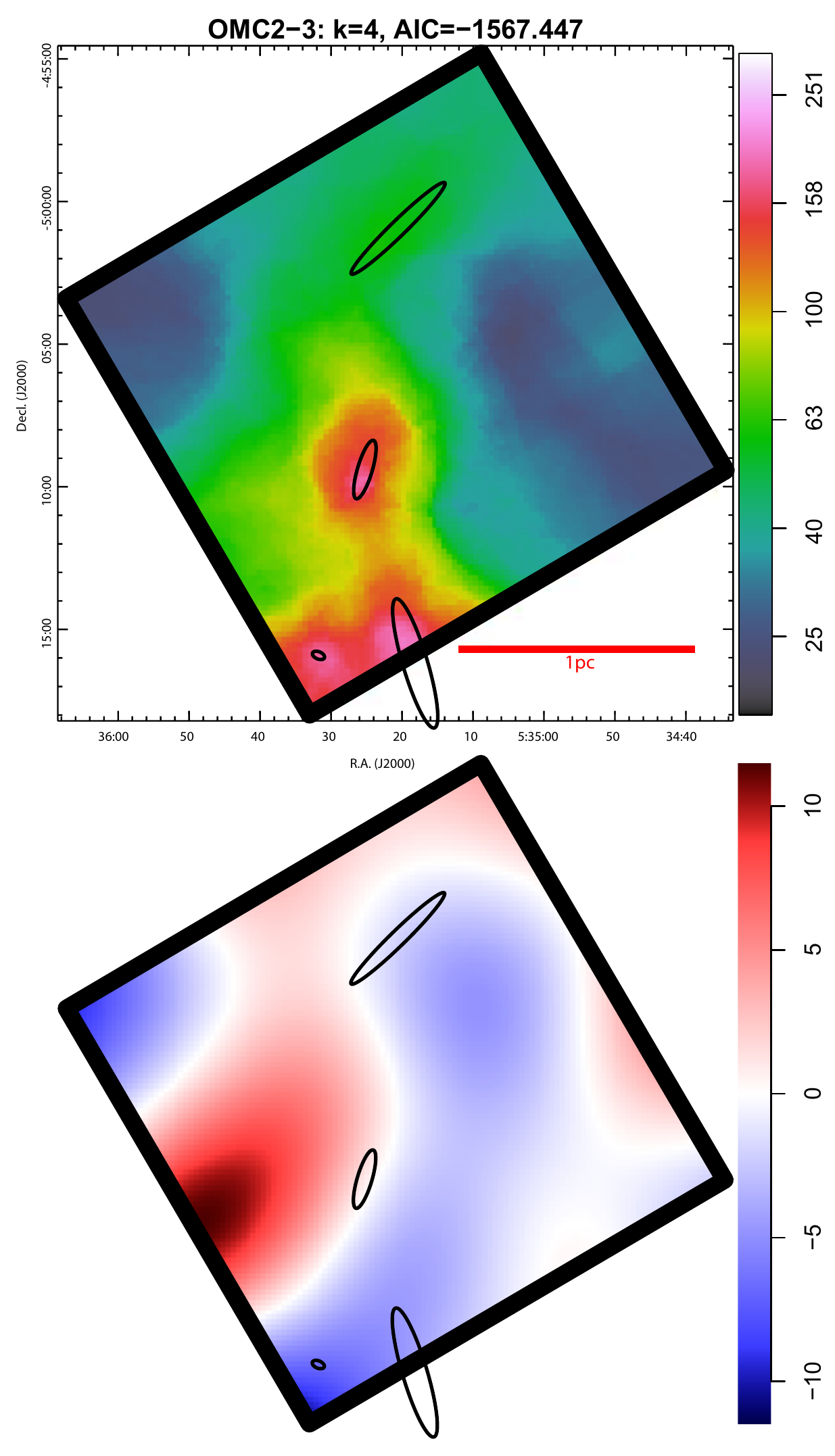}
\caption{Identification of SFiNCs clusters through the mixture model analysis, performed on the ``flattened'' SPCM samples within the {\it Chandra} ACIS fields.  An example is given for the OMC~2-3 SFR, and the full set of panels for the 22 SFiNCs fields is available in the Supplementary Material.  The upper panel shows the smoothed projected stellar surface density with a color-bar in units of observed stars per pc$^{-2}$ on a logarithmic scale. The figure title gives the name of a SFiNCs SFR, the number of identified clusters, and the final value of the Akaike Information Criterion. The lower panel shows smoothed map of residuals between the data and the model with a color bar in units of stars per pc$^{-2}$ on a linear scale. On both panels, the core radii of the identified SFiNCs clusters are outlined by the black ellipses. \label{fig_cluster_identification}}
\end{figure}

We study the spatial structure of clustering in SFiNCs molecular clouds using adaptively smoothed maps of stellar surface density.  The upper panel of Figure \ref{fig_cluster_identification} shows the map for the ``flattened'' sample for the OMC 2-3 field.  Similar maps for the other 21 SFiNCs fields are provided in the Supplementary Materials.  These maps are constructed using Voronoi tessellations, where the estimated intensity in each tile is the reciprocal of the tile area, implemented in function {\it adaptive.density} from the {\it spatstat} package \citep{Baddeley2015}.  The color scale is in units of observed stars per pc$^{-2}$.  Cluster cores are outlined by the black ellipses.  

In two cases the AIC minimization led to the acceptance of very small clusters ($R_c < 0.01$~pc) with few stellar members: Cluster A in IRAS~20050+2720, and the cluster D in Cep~OB3b. This arises when the sparse subcluster is very compact.  The mixture model has no rigid threshold on the number of points in a cluster.

As with MYStIX, SFiNCs shows wide diversity of stellar structures. For the majority of the SFiNCs SFRs (Be~59, SFO~2, IC~348, LkH$\alpha$~101, GGD~12-15, Serpens Main, Serpens South, IRAS~20050+2720, Sh~2-106, IC~5146, LDN 1251B, Cep A, Cep C), a single main compact ($R_c<0.5$~pc) cluster is detected within the {\it Chandra}-ACIS-I field.  For a few SFRs (NGC~7160 and ONC Flanking Fields), a single but rather loose ($R_c>0.5$~pc) stellar structure is found. 

\newpage
\begin{table*}\small
\centering
 \begin{minipage}{180mm}
 \caption{SFiNCs Clusters from Mixture Model. Fifty two ellipsoid cluster components from the best fit mixture models for the 22 SFiNCs SFRs.  Highly uncertain parameter values are appended with ``:''.  Column 1: The cluster component name, labeled from west to east. Columns 2-3: Celestial coordinates (J2000) for the ellipsoid center. Column 4: Positional uncertainty as 68\% error circle, in arc-seconds. Column 5: Core radius as an average between the semimajor and semiminor axes of the ellipsoid component, in parsecs. Column 6: Fractional statistical error (68\% confidence interval).  Columns 7-8: Ellipticity and its fractional statistical error. Column  9: Orientation angle of the ellipse in degrees east from north. Columns 10-11. Number of stars estimated by integrating the model component out to four times the size of the core; and its fractional statistical error. Column 12. Relation to molecular clouds, based on a visual inspection of far-infrared images discussed in \S \ref{cloud_subsection} and prior studies of the SFiNCs SFRs: ``C'' - likely embedded in a cloud;  ``R'' - likely revealed; ``...'' - unclear case.}
 \label{tbl_cluster_morphology}
 \begin{tabular}{@{\vline }c@{ \vline }c@{ \vline }c@{ \vline }c@{ \vline }c@{ \vline }c@{ \vline }c@{ \vline }c@{ \vline }c@{ \vline }c@{ \vline }c@{ \vline }c@{ \vline }}
\cline{1-12}
&&&&&&&&&&&\\ 
%\hline
Cluster & ~R.A.(J2000)~ & ~Decl.(J2000)~ & ~PosErr~ & ~~~$R_c$~~~ &$\sigma R_c/R_c$ & ~~~~$\epsilon$~~~~ & ~$\sigma \epsilon/\epsilon$~ & ~~$\phi$~~ & ~$N_{4,model}$~ & $~\sigma N/N$~ &      ~Cloud~\\
 & (deg) & (deg) & (\arcsec) & (pc) &  &  &   & (deg) & (stars)  &  & \\
   (1)         & (2)         &  (3)         & (4)     &  (5) &   (6)   &   (7)      &   (8)       &    (9) & (10) & (11) & (12)\\
\cline{1-12}
&&&&&&&&&&&\\
Be 59 A  &     0.508713 &    67.520259 &  701 & 1.320 & 0.25 & 0.39 & 0.37 &  99 &  486 &   0.37 &  R\\
Be 59 B  &     0.562134 &    67.418764 &   1 & 0.210 & 0.20 & 0.15 & 0.46 &   8 &  152 &   0.23 &  R\\
SFO 2 A  &     1.018036 &    68.565945 &   5 & 0.076 & 0.51 & 0.61 & 0.18 & 155 &   23 &   0.19 &  C\\
NGC 1333 A  &    52.276137 &    31.280885 &  70 & 0.100 & 0.55 & 0.55 & 0.24 &  48 &   41 &   0.61 &  C\\
NGC 1333 B  &    52.279194 &    31.364583 &  44 & 0.100 & 0.44 & 0.72 & 0.12 &  45 &   54 &   0.47 &  C\\
IC 348 B  &    56.141167 &    32.158825 &  31 & 0.200 & 0.21 & 0.39 & 0.25 &   6 &  156 &   0.17 &  R\\
IC 348 A  &    55.999718 &    32.031989 &  18 & 0.040 & 0.49 & 0.60 & 0.24 & 119 &   10 &   0.28 &  C\\
LkHa101 A  &    67.542392 &    35.268025 &  17 & 0.210 & 0.17 & 0.11 & 0.52 & 177 &  108 &   0.12 &  R\\
NGC 2068-2071 A  &    86.543905: &    -0.138717: & ... & 0.110: & ... & 0.96: & ... &  12: &   16: & ... &  C\\
NGC 2068-2071 B  &    86.666392 &     0.088074 &  46 & 0.290 & 0.43 & 0.50 & 0.29 & 129 &   68 &   0.28 &  R\\
NGC 2068-2071 C  &    86.775230 &     0.375235 &  17 & 0.060 & 0.49 & 0.31 & 0.38 & 159 &   25 &   0.24 &  C\\
NGC 2068-2071 D  &    86.808844 &     0.316468 &  23 & 0.110 & 0.50 & 0.86 & 0.05 &  76 &   28 &   0.21 &  C\\
ONC Flank S A  &    83.862602 &    -5.479723 & 167 & 1.230 & 0.49 & 0.75 & 0.10 &  28 & 1521 &   0.37 &  ...\\
ONC Flank N A  &    83.819526 &    -4.845647 &  90 & 0.804 & 0.21 & 0.32 & 0.36 &  15 &  578 &   0.28 &  R\\
OMC 2-3 A  &    83.825638: &    -5.271570: & ... & 0.120: & ... & 0.82: & ... &  17: &   65: & ... &  C\\
OMC 2-3 B  &    83.835852 &    -5.012671 &  34 & 0.100 & 0.77 & 0.88 & 0.06 & 134 &   14 &   0.45 &  C\\
OMC 2-3 C  &    83.855702 &    -5.156237 &   10 & 0.060 & 0.44 & 0.76 & 0.08 & 163 &   23 &   0.28 &  C\\
OMC 2-3 D &    83.883413: &    -5.266815: & ... & 0.020: & ... & 0.43: & ... &  67: &   10: & ... &  ...\\
Mon R2 A  &    91.937730 &    -6.344872 &  65 & 0.200 & 0.45 & 0.77 & 0.11 & 173 &   55 &   0.34 &  C\\
Mon R2 B  &    91.948257 &    -6.378512 &  11 & 0.100 & 0.47 & 0.77 & 0.10 &  29 &   33 &   0.47 &  C\\
Mon R2 C  &    91.961306 &    -6.428142 &   6 & 0.070 & 0.95 & 0.35 & 0.38 & 114 &   19 &   0.30 &  C\\
GGD 12-15 A  &    92.710405 &    -6.194814 &  11 & 0.140 & 0.19 & 0.46 & 0.16 &  68 &   77 &   0.13 &  C\\
RCW 120 A  &   258.037282 &   -38.516262 &   7 & 0.070 & 0.37 & 0.45 & 0.33 & 144 &   15 &   0.18 &  C\\
RCW 120 B  &   258.099513 &   -38.487682 &  15 & 0.290 & 0.25 & 0.52 & 0.17 &  31 &   60 &   0.19 &  R\\
RCW 120 C  &   258.165621 &   -38.451639 &  20 & 0.240 & 0.40 & 0.48 & 0.26 & 162 &   29 &   0.33 &  C\\
RCW 120 D  &   258.179085 &   -38.376186 &  40 & 0.110 & 1.25 & 0.86 & 0.06 & 155 &   12 &   0.43 &  C\\
Serpens Main A  &   277.464566 &     1.267515 &  35 & 0.040 & 0.66 & 0.71 & 0.19 & 126 &   12 &   0.54 &  C\\
Serpens Main B  &   277.492003 &     1.216660 &  15 & 0.060 & 0.35 & 0.48 & 0.22 &   5 &   38 &   0.25 &  C\\
Serpens South A  &   277.468911: &    -1.962172: & ... & 0.024: & ... & 0.45: & ... & 105: &   10: & ... &  C\\
Serpens South B  &   277.492366: &    -2.138106: & ... & 0.022: & ... & 0.87: & ... &  81: &    4: & ... &  C\\
Serpens South C  &   277.511870 &    -2.048380 &   10 & 0.056 & 0.25 & 0.65 & 0.08 & 168 &   58 & 0.13 &  C\\
Serpens South D  &   277.574336: &    -2.148296: & ... & 0.022: & ... & 0.31: & ... &   2: &    7: & ... &  C\\
IRAS 20050+2720 A  &   301.713709: &    27.343918: & ... & 0.007: & ... & 0.71: & ... &  17: &    4: & ... &  ...\\
IRAS 20050+2720 B &   301.741427: &    27.559032: & ... & 0.171: & ... & 0.23: & ... &  35: &    8: & ... &  C\\
IRAS 20050+2720 C  &   301.742363: &    27.511739: & ... & 0.051: & ... & 0.65: & ... & 145: &   20: & ... &  C\\
IRAS 20050+2720 D  &   301.772854 &    27.487538 &  13 & 0.081 & 0.57 & 0.56 & 0.25 & 178 &   63 &   0.40 &  C\\
IRAS 20050+2720 E  &   301.915203: &    27.562103: & ... & 0.075: & ... & 0.26: & ... &  39: &   33: & ... &  ...\\
Sh 2-106 A  &   306.812857 &    37.461188 &  31 & 0.300 & 0.35 & 0.58 & 0.25 & 104 &   15 &   0.39 &  R\\
Sh 2-106 B  &   306.823055 &    37.376261 &  11 & 0.110 & 0.45 & 0.11 & 0.50 &  56 &   16 &   0.34 &  C\\
Sh 2-106 C  &   306.853428: &    37.293251: & ... & 0.030: & ... & 0.39: & ... &  20: &   10: & ... &  C\\
Sh 2-106 D  &   306.860795 &    37.382119 &   6 & 0.090 & 0.31 & 0.48 & 0.23 & 120 &   41 &   0.14 &  C\\
IC 5146 A  &   328.140096 &    47.228756 &  16 & 0.090 & 0.67 & 0.51 & 0.31 & 165 &   12 &   0.32 &  C\\
IC 5146 B  &   328.381131 &    47.265152 &   7 & 0.170 & 0.16 & 0.25 & 0.34 &  17 &  115 &   0.08 &  R\\
NGC 7160 A  &   328.443151 &    62.585448 &  29 & 0.720 & 0.24 & 0.30 & 0.35 & 146 &  108 &   0.18 &  R\\
LDN 1251B A  &   339.696683 &    75.193642 &  19 & 0.030 & 0.60 & 0.41 & 0.38 &  73 &   15 &   0.34 &  C\\
Cep OB3b A  &   343.446066 &    62.596289 &  24 & 0.450 & 0.20 & 0.44 & 0.21 &  52 &  304 &   0.15 &  R\\
Cep OB3b B  &   343.743449 &    62.569278 &  20 & 0.090 & 1.00 & 0.48 & 0.22 &  66 &   30 &   0.28 &  ...\\
Cep OB3b C  &   344.167892 &    62.701711 &  22 & 0.580 & 0.14 & 0.18 & 0.44 &   1 &  520 &   0.13 &  R\\
Cep OB3b D &   344.279145: &    62.641799: & ... & 0.001: & ... & 0.53: & ... &  86: &    7: & ... &  C\\
Cep A A  &   344.073598 &    62.030905 &  14 & 0.210 & 0.20 & 0.54 & 0.12 & 112 &   87 &   0.14 &  C\\
Cep C A  &   346.441016 &    62.502827 &  23 & 0.180 & 0.25 & 0.53 & 0.16 &  50 &   64 &   0.14 &  C\\
Cep C B  &   346.654364: &    62.530596: & ... & 0.020: & ... & 0.25: & ... &  34: &    5: & ... &  C\\
&&&&&&&&&&&\\ 
\cline{1-12} 
\end{tabular}
\end{minipage}
\end{table*}
\clearpage
\newpage

In some SFRs (IC~348, Serpens Main, Serpens South, IRAS 20050+2720, Sh~2-106, Cep C), their main clusters are accompanied by minor siblings on $\sim 1$~pc scales. A few fields (OMC~2-3, Mon~R2, RCW~120) harbor linear chains of clusters on $2-3$~pc scales. For the NGC~2068-2071 and Cep~OB3b SFRs with wider angular coverage, multiple stellar structures are seen on $5$~pc scales. 

In MYStIX, \citet{Kuhn2014,Kuhn2015b} define the following four morphological classes of clusters: isolated clusters, core-halo structures, clumpy structures, and linear chains of clusters. The SFiNCs' single compact and loose clusters are morphological analogues to the isolated MYStIX clusters. Meanwhile, SFiNCs lacks structures similar to the MYStIX clumpy structures that are seen on large spatial scales of $\ga 5-10$~pc in the rich M~17, Lagoon, and Eagle SFRs.  SFiNCs' linear chain clusters, observed on a $2-3$~pc scale, are much smaller than the MYStIX chain structures observed on $10-15$~pc scales (DR~21, NGC~2264, NGC~6334, and NGC~1893). The differences between the SFiNCs and MYStIX cluster morphologies could have several causes:  (1) the intrinsic spatial scale of the sample (MYStIX SFRs are chosen to be giant molecular clouds); (2) observational fields of view (MYStIX fields are generally more distant with more mosaicked {\it Chandra} exposures) allowing larger-scale clumpy or linear structures; and/or (3) intrinsic cluster richness (MYStIX clusters are richer allowing the algorithm to find more secondary structures).  

In Appendix, we summarize the YSO cluster distributions in a multi-wavelength astronomical context such as molecular cloud and other interstellar features.  

\subsection{Model Validation} \label{model_validation}

As discussed in the Appendix, parametric modeling such as our mixture model has advantages over nonparametric clustering methods  in identifying optimal number of clusters with clear physical properties.  However, these advantages accrue only if the parametric assumptions hold; in statistical parlance, the model must be correctly specified.  It is therefore important to examine whether the mixture of two-dimensional isothermal ellipsoid models fits the stellar distribution in SFiNCs clusters. We perform here two model validation tests.  

First, we show maps of the residuals between the data and model in the lower panels of Figure~\ref{fig_cluster_identification} computed using the $R$ function {\it diagnose.ppm} from the {\it spatstat} package.  As discussed by \citet{Kuhn2014} and \citet{Baddeley2015}, residual maps for spatial point processes offer direct information about where the model and the data agree or disagree.  The residuals are shown smoothed by a Gaussian kernel width of 0.4~pc in units of stars per pc$^{-2}$; the blue and red colors indicate negative (model $>$ data) and positive (data $>$ model) residuals, respectively.  

Examining the relative surface density scales of the original and residual maps, one sees for nearly all of the SFiNCs regions that the peak residuals are less than $\sim 10$\% of the original peak levels indicating good matches between the data and the model. On occasion, positive residuals exceeding $30$\% are present such as residual hotspots in SFO 2, RCW 120, IRAS 20050+2720, and Cep~C. Some of these coincide with dusty cloudlets seen in far-infrared images of the regions and could be small embedded star groups whose contributions to the model likelihood were too small to be accepted as new clusters.  For a few other cases such as RCW 120 (clusters A and C), Sh 2-106 (A and C), and IRAS 20050+2720 (B), the residuals exceed $30$\%, indicating that the observed stellar distribution is not well fit with equilibrium elliptical models.  Evaluation of individual residual maps appears in Appendix \ref{sec_individual_subclusters}.  Overall, we conclude that the residual maps show mostly random noise and, in most SFiNCs fields, the mixture model did not either create false or miss true clusters in an obvious fashion.  
\begin{figure}
\centering
\includegraphics[angle=0.,width=70mm]{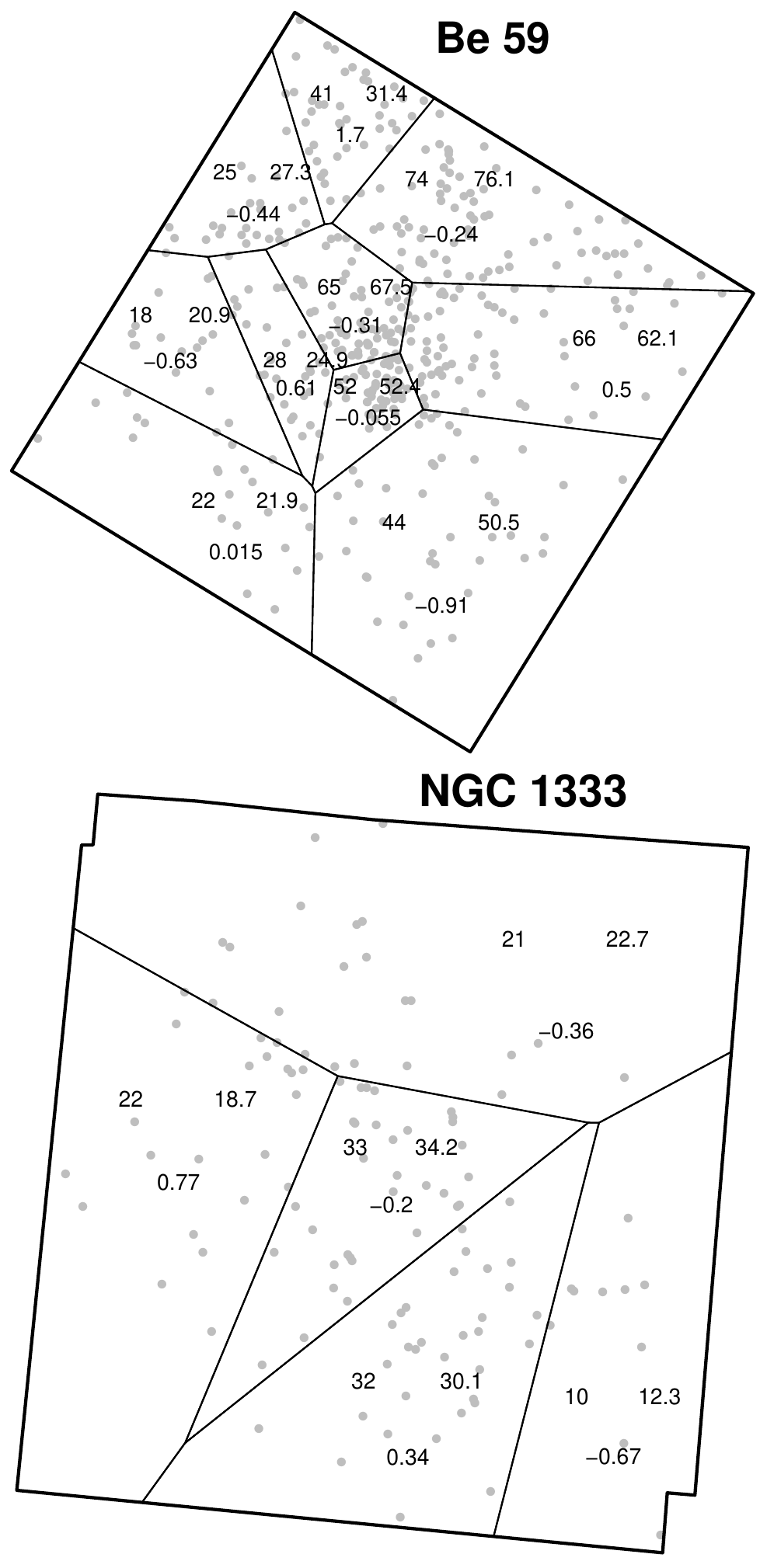}
\caption{Validation of the mixture models with random Voronoi tesselations.  Two SFiNCs fields are shown here; similar panels for the full SFiNCs sample are presented in the Supplementary Materials. YSOs from the ``flattened'' SPCM sample are shown as gray points. Three numbers are given in each tile: the number of observed YSOs ($N_{observed}$; upper left), the number expected from the best fit mixture model ($N_{expected}$; upper right); and Pearson residual or `sigma' deviation (bottom). \label{fig_chi2_global}}
\end{figure}

Second, we compare observed and model star counts in polygonal tiles obtained from Voronoi tesselation using Poisson statistics.  This method is used in astronomy \citep{Schmeja11} and is a variant of the quadrat counting test widely used in other fields \citep{Baddeley2015}. Goodness of fit can be estimated from Pearson's $X^2$ statistic 
\begin{equation}
X^{2} = \sum_{i=1}^{n} \frac{(N_{i,observed}-N_{i,expected})^2}{N_{i,expected}},
\end{equation}
where $n$ is the number of tiles, and $N_{i,observed}$ and $N_{i,expected}$ are the numbers of observed and expected YSOs in an $i$ tile, respectively. This statistic is compared to the $\chi^{2}$ distribution with $n-1$ degrees of freedom to evaluate the probability for the null hypothesis that the data are drawn from the point process model distribution. The expected values are obtained from an inhomogeneous Poisson process based on the mixture model clusters.  The test is computed using the {\it quadrat.test} and {\it dirichlet} functions in {\it spatstat} \citep{Baddeley2015}.    
\begin{table}\tiny
\centering
 \begin{minipage}{80mm}
 \caption{Mixture Model Goodness of Fit. Pearson's chi-squared as a goodness of fit measure for the best fit mixture models given in Table \ref{tbl_cluster_morphology} based on the random tessellation given in Figure \ref{fig_chi2_global}.  Column 1: Star forming region. Column 2: Number of clusters in the model. Columns 3-4: Pearson's $X^2$ value and degrees of freedom.  Column 5: Two-sided $p$-value for the null hypothesis that the observed YSO distribution is drawn from the model. If $P_{\chi^2} < 0.01$ then the data are likely not drawn from the model.}
 \label{tbl_chi2_global}
 \begin{tabular}{@{\vline }c@{ \vline }c@{ \vline }c@{ \vline }c@{ \vline }c@{ \vline }}
\cline{1-5}
&&&&\\ 
%\hline
Region & ~$N_{subclust}$~ & ~~$X^2$~~ & ~~dof~~ & ~~$P_{\chi^2}$~~\\
   (1)         & (2)         &  (3)         & (4)     &  (5)\\
\cline{1-5}
&&&&\\
Be 59  &      2& 5.1 &  9 &0.35\\
SFO 2   &     1& 1.6 &  2 &0.89\\
NGC 1333 &    2& 1.3 &  4 &0.29\\
IC 348   &    2& 5.1 & 7 &0.71\\
LkH$\alpha$ 101 &    1& 7.9 &  5 &0.33\\
NGC 2068-2071 &    4&13.8 &10 &0.36\\
ONC Flank S  &    1& 4.9 & 9 &0.32\\
ONC Flank N  &    1& 8.5 & 8 &0.76\\
OMC 2-3  &    4& 4.5 & 7 &0.56\\
Mon R2   &    3& 7.0 & 9 &0.73\\
GGD 12-15 &   1& 2.6 & 4 &0.75\\
RCW 120   &   4& 4.8 & 7 &0.63\\
Serpens Main & 2& 3.4 &  5 &0.73\\
Serpens South & 4&10.0 & 12 &0.77\\
~IRAS 20050+2720~   & 5&13.6 &  9 &0.27\\
Sh 2-106     & 4&13.0 &  8 &0.23\\
IC 5146      & 2&12.8 & 7 &0.16\\
NGC 7160     & 1&5.0  &4 &0.57\\
LDN 1251B    & 1&5.0  &2 &0.16\\
Cep OB3b     & 4&30.8 &30 &0.85\\
Cep A        & 1&10.0 & 6 &0.25\\
Cep C        & 2&4.9  &5 &0.85\\
&&&&\\ 
\cline{1-5} 
\end{tabular}
\end{minipage}
\end{table}
%\clearpage
%\newpage

Figure \ref{fig_chi2_global} shows randomly tessellated stellar spatial distributions for two SFiNCs fields; the complete SFiNCs sample is shown in the Supplementary Materials. The number of points in a tile is drawn from the Poisson distribution, but the calculation of $\chi^{2}$ assumes that the distribution of points is normal. To allow the normal approximation to Poisson distribution, only SFiNCs tessellations with the number of sources in a tile $\geq 10$ are considered. Pearson residuals $X$ are shown as labels to each tile. Out of $>$190 tiles across 22 SFiNCs SFRs, only 3 tiles exhibit Pearson residuals above 2 in absolute value, an indicator of a likely departure from the fitted model. The high fraction of tiles with low Pearson residuals is a clear evidence that the SFiNCs YSOs distributions are generally well fit with the isothermal elliptical models.   Table \ref{tbl_chi2_global} shows high probability values ($P_{\chi^2} >> 0.01$) that the data are successfully drawn from the model  for the full SFiNCs sample.

\subsection{Cluster Parameter Uncertainties} \label{error_analysis}

\begin{figure}
\centering
\includegraphics[angle=0.,width=80mm]{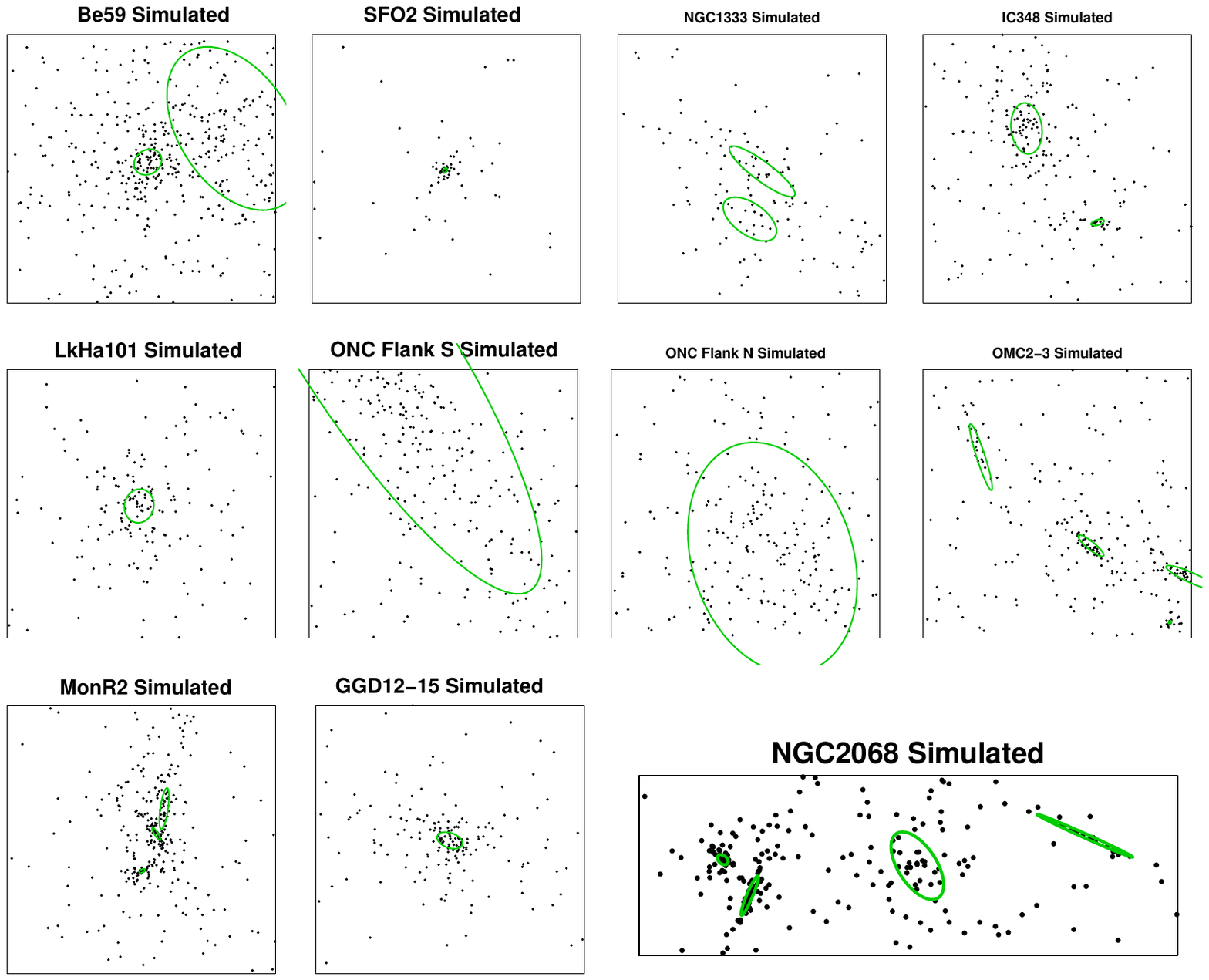}
\caption{Error analysis of the mixture models with simulated stellar spatial distributions (black points).  Six SFiNCs fields are shown here; similar panels for the remaining SFiNCs regions are presented in the Supplementary Materials.  The core radii of the simulated isothermal ellipsoid clusters are marked by the green ellipses. \label{fig_mma_simulations}}
\end{figure}

In order to estimate statistical errors on the SFiNCs cluster parameters derived in \S \ref{clusters_subsection}, we conduct Monte Carlo simulations. For each of the SFiNCs SFRs, we simulate 100 random sets of the spatial distributions of their ``flattened'' stellar samples; these distributions follow isothermal ellipsoid models with individual parameters taken from Table \ref{tbl_cluster_morphology}. This analysis was performed using function {\it simulate.ppm} in {\it spatstat}, as well as the CRAN  {\it spatgraphs} package \citep{Baddeley2015,Rajala2015}. Figure \ref{fig_mma_simulations} shows individual examples of the simulated SFiNCs data.

For the model fitting of the simulated data, the ``initial guess'' stage is imitated by choosing the positions of initial cluster models to match the positions of the simulated clusters. Likelihood maximization is further mimicked by running the model refinement stage via Nelder-Mead \citep[Appendix of][]{Kuhn2014}  several times. For each of the cluster parameters, the 68\% confidence interval is derived from a sample distribution of 100 simulated values. The inferred confidence intervals for the cluster's position, core radius, ellipticity, and modeled number of stars within the area four times the size of the core are reported in Table \ref{tbl_cluster_morphology}.

\begin{figure}
\centering
\includegraphics[angle=0.,width=80mm]{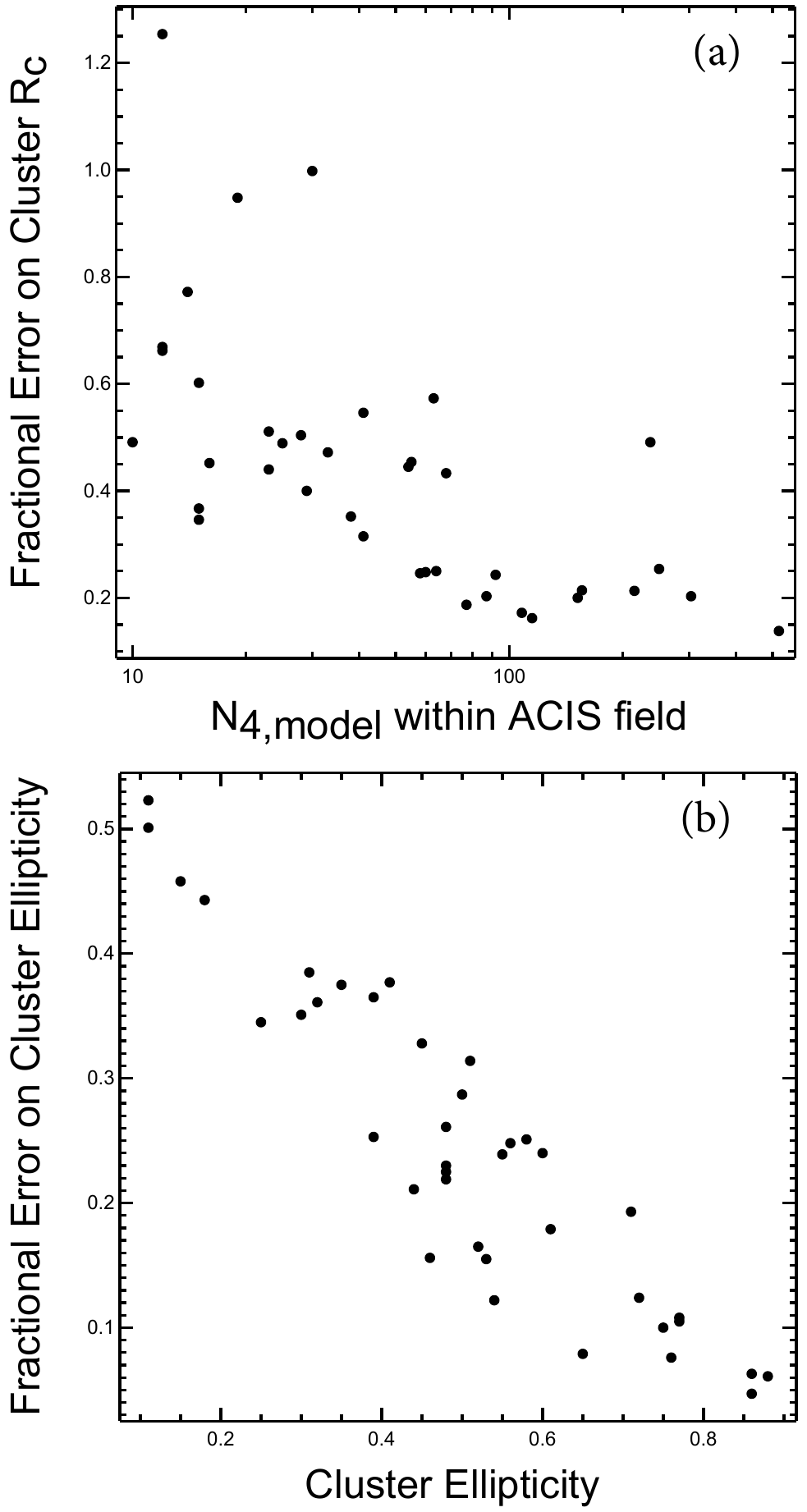}
\caption{Uncertainties of cluster core radii and ellipticities inferred from the simulations of SFiNCs clusters (\S \ref{error_analysis}). (a) Fractional error of cluster core radius versus number of stars estimated by integrating the model component out to four times the size of the cluster core.  (b) Fractional error of cluster ellipticity versus cluster ellipticity. \label{fig_simulation_errors}}
\end{figure}

Two cluster quantities of interest are the core radius and ellipticity. The fractional statistical error on the core radius ranges from 30-60\% for weak clusters with $20<N_{4,model} < 70$ stars to $<25$\% for more populous clusters (Figure \ref{fig_simulation_errors}a). The fractional statistical error on cluster ellipticity does not correlate with cluster population but rather depends on the ellipticity value itself (Figure \ref{fig_simulation_errors}b). The ellipticity errors are typically $>$30\% for clusters with $\epsilon < 0.4$ and $<$15\% for extremely elongated clusters.

For SFiNCs clusters that are either sparse ($N_{4,model} < 20$ stars) and/or strongly affected by nearby clusters, parameter values are poorly constrained by the simulations.  In Table \ref{tbl_cluster_morphology}, the reported parameter values for these clusters are appended by the warning sign ``:'', and they are omitted from the multivariate analyses in \S \ref{ma_section} and \ref{cloud_subsection}.

\begin{table*}\footnotesize
\centering
\begin{minipage}{180mm}
 \caption{Membership of SFiNCs Clusters. This table is available in its entirety (8,492 SFiNCs Probable Complex Members across the 22 SFiNCs SFRs) in the Supplementary Materials.   Column 1: SFiNCs SFR name. Column 2: Source IAU designation. Columns 3-4: Right ascension and declination in decimal degrees (J2000). Column 5: A three digit flag. The first digit indicates whether the source is located inside (1) or outside (0) the {\it Chandra} ACIS-I field. The second digit indicates whether the source is an X-ray object (1) or not (0). The third digit indicates whether the source is an OB-type star (1) or not (0). Column 6: X-ray median energy in the $(0.5-8)$~keV band. Columns 7-8: 2MASS photometry in the $J$ and $H$-bands. Column 9: Apparent spectral energy distribution slope $\alpha_{IRAC} = d \log(\lambda F_{\lambda})/d \log(\lambda)$ measured in the IRAC wavelength range from 3.6 to 8.0~$\mu$m. Column 10: Intrinsic X-ray luminosity in the $(0.5-8)$~keV band. Column 11. Stellar age estimate using the method of \citet{Getman2014a}. Column 12: Cluster assignment: ``A''-``E'' indicates the assigned SFiNCs cluster; ``U'' indicates the unclustered stellar population; ``X'' indicates uncertain assignment; and ``...'' indicates that no assignment is made for sources located outside the {\it Chandra} ACIS-I field.}
 \label{tbl_cluster_membership}
 \begin{tabular}{@{\vline }c@{ \vline }c@{ \vline }c@{ \vline }c@{ \vline }c@{ \vline }c@{ \vline }c@{ \vline }c@{ \vline }c@{ \vline }c@{ \vline }c@{ \vline }c@{ \vline }}
\cline{1-12}
&&&&&&&&&&&\\ 
%\hline
~Region~ & Desig & ~R.A.(J2000)~ & ~Decl.(J2000)~ & ~Flag~ & ~$ME$~ & ~~~~~~~$J$~~~~~~~ & ~~~~~~~$H$~~~~~~~ & ~~~~$\alpha_{IRAC}$~~~~ & ~$\log L_{X,tc}$~ & ~$Age_{JX}$~ &      ~Clus~\\
  &   & (deg) & (deg) &   & (keV) & (mag) & (mag) &   & (erg~s$^{-1}$) & (Myr) &  \\
   (1)         & (2)         &  (3)         & (4)     &  (5) &   (6)   &   (7)      &   (8)       &    (9) & (10) & (11) & (12)\\
\cline{1-12}
&&&&&&&&&&&\\
Be59 & ~000033.87+672446.2~ &     0.141150 &    67.412846 & 110 & 2.62 & $15.19\pm 0.05$ & $13.58\pm 0.04$ & $ -2.44\pm  0.04$ & 30.94 &  ... & A\\
Be59 & ~000036.43+672658.5~ &     0.151798 &    67.449596 & 110 & 1.88 & $13.68\pm 0.03$ & $12.64\pm 0.03$ & $ -2.69\pm  0.02$ & 30.79 &  ... & A\\
Be59 & ~000045.20+672805.8~ &     0.188345 &    67.468297 & 110 & 1.66 & $13.90\pm 0.03$ & $12.85\pm 0.03$ & $ -2.60\pm  0.04$ &   ... &  ... & A\\
Be59 & ~000046.19+672358.2~ &     0.192477 &    67.399503 & 110 & 1.58 & $14.08\pm $... & $13.01\pm 0.04$ & $ -2.77\pm  0.11$ & 30.25 &  ... & A\\
Be59 & ~000050.10+672721.4~ &     0.208781 &    67.455954 & 110 & 1.77 & $15.09\pm 0.05$ & $13.89\pm 0.04$ & $ -2.58\pm  0.05$ & 29.90 &  2.5 & A\\
Be59 & ~000051.39+672648.8~ &     0.214161 &    67.446914 & 110 & 2.12 & $14.66\pm 0.04$ & $13.33\pm 0.04$ & $ -2.31\pm  0.04$ & 30.63 &  ... & A\\
Be59 & ~000053.45+672615.0~ &     0.222725 &    67.437501 & 110 & 2.37 & $15.43\pm 0.06$ & $13.99\pm 0.05$ & $ -2.63\pm  0.10$ & 30.82 &  ... & A\\
Be59 & ~000054.01+672119.8~ &     0.225079 &    67.355504 & 110 & 2.37 & $15.41\pm 0.05$ & $14.03\pm 0.04$ & $ -2.40\pm  0.04$ & 30.81 &  ... & A\\
Be59 & ~000055.58+672647.8~ &     0.231621 &    67.446638 & 110 & 2.08 & $15.30\pm 0.05$ & $13.96\pm 0.04$ & $ -2.54\pm  0.09$ & 30.21 &  4.3 & A\\
Be59 & ~000056.24+672835.1~ &     0.234343 &    67.476426 & 110 & 3.00 & $16.19\pm 0.09$ & $15.00\pm 0.08$ & $ -2.68\pm  0.13$ & 31.25 &  ... & A\\
&&&&&&&&&&&\\ 
\cline{1-12} 
\end{tabular}
\end{minipage}
\end{table*}
%\clearpage
%\newpage

\subsection{Cluster Membership} \label{membership_subsection}

\begin{figure}
\centering
\includegraphics[angle=0.,width=100mm]{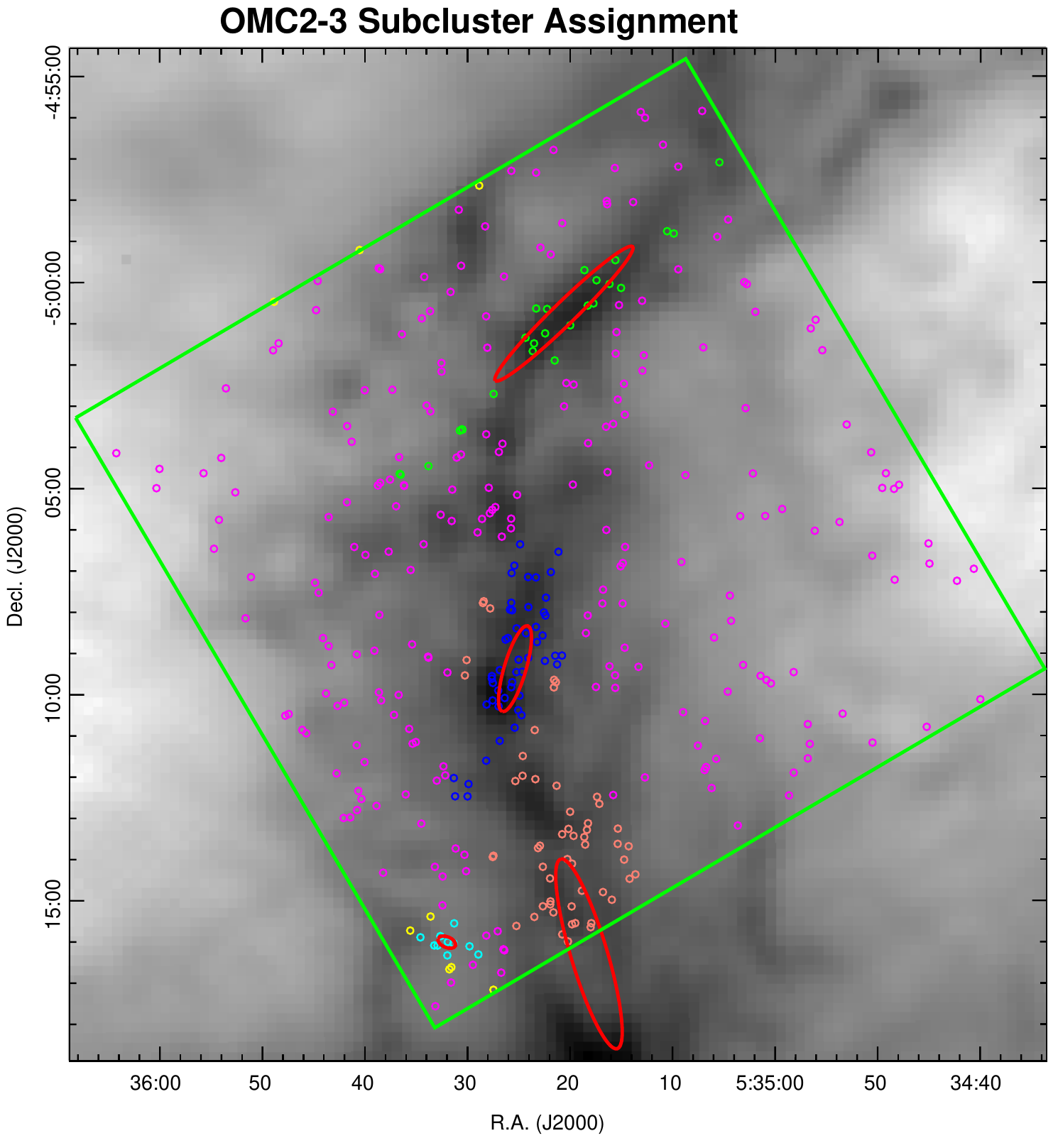}
\caption{Cluster assignments in SFiNCs, shown for the full SFiNCs Probable Complex Member sample within the {\it Chandra} ACIS field for the OMC~2-3 star forming region;  similar panels for the full SFiNCs sample are presented in the Supplementary Materials.  SPCMs are superimposed on gray-scale far-IR images taken by $Herschel$-SPIRE at 500~$\mu$m or (for NGC~7822, IRAS~00013+6817, IRAS~20050+2720, NGC~7160, CepOB3b) $AKARI$-FIS at 160~$\mu$m. These images trace the locations of the SFiNCs molecular clouds shown on a logarithmic grayscale where denser clouds appear darker.  The red ellipses show the best fit mixture model and the stars assigned to each cluster are color coded. Unclustered members are shown in magenta and unassigned stars in yellow.  The {\it Chandra}-ACIS field of view is outlined by the green polygon. \label{fig_cluster_assignment_maps}}
\end{figure}

As in MYStIX \citep{Kuhn2014}, the mixture model is also used here as a ``soft classifier'' that, with additional decision rules, allows individual YSOs to be assigned to the clusters or to the unclustered component. For each YSO, the probability of membership is calculated based on the relative contribution of the different cluster model components to the stellar density at the location of a YSO. The following MYStIX decision rules are further adopted for the cluster membership assignment of the SFiNCs YSOs: the probability for the assigned cluster must exceed $30$\% and the cluster members must lie within an ellipse four times the size of the cluster core.  Stars that fail these rules have ``uncertain'' membership.  The result of this membership assignment procedure is shown in Figure~\ref{fig_cluster_assignment_maps}.

For the entire SPCM sample (8,492 stars), Table \ref{tbl_cluster_membership} lists the cluster assignments along with several X-ray, NIR, and MIR source properties.  For the SPCM sources located outside the {\it Chandra} ACIS-I fields their cluster assignments are unknown (indicated as ``...'' in Column~12). Inside the ACIS-I fields, the individual SPCMs are either assigned to specific clusters (``A''-``E'' in Column~12) or unclustered population (``U'') or marked as objects with uncertain membership (``X''). The difference between the YSOs with uncertain membership (``X'') and unclustered YSOs (``U'') is that the former fail to meet the assignment criteria for any model component while the latter are assigned successfully to the unclustered model component. Out of 8,492 SPCMs, 5,214 are clustered (``A''-``E''), 1,931 are unclustered (``U''), 252 have uncertain membership (``X''), and 1,095 lie outside the ACIS-I fields and have unknown membership (``...''). 

Source properties included in Table \ref{tbl_cluster_membership} are taken from \citet{Getman2017}: X-ray median energy ($ME$) that measures absorption to the star, X-ray luminosity corrected for this absorption, NIR 2MASS $J$ and $H$-band magnitudes, slope of the $3.6-8$~$\mu$m spectral energy distribution $\alpha_{IRAC} = d \log(\lambda F_{\lambda})/d \log(\lambda)$, and stellar age, $Age_{JX}$.  SFiNCs $Age_{JX}$ values are calculated by \citet{Getman2017} following the methodology of \citet{Getman2014a}. This age estimator is based on an empirical X-ray luminosity-mass relation calibrated to well studied Taurus PMS stars and to theoretical evolutionary tracks of \citet{Siess2000}.  Below, median values of these properties for member stars will be used to characterize each cluster. 

\subsection{Comparison with other astronomical studies}

In the Appendix \S\ref{sec_individual_subclusters}, we discuss each cluster found above with respect to molecular cloud maps and previous studies of young stellar clustering.  Most previous studies use nonparametric techniques based either on the Minimal Spanning Tree (MST) or $k$-nearest neighbor analysis.  In Appendix \S\ref{sec_sfincs_vs_g09} we compare in detail our parametric mixture model method and the nonparametric MST-based procedure of \citet{Gutermuth2009}.  The latter exhibits significant deficiencies called 'chaining' and 'fragmentation'.  This has been well-established in the statistical literature (\S\ref{sec.stat.bkgd}), appears in some simple simulations we perform (\S\ref{sec_mst_simulations}), and in detailed applications to our SFiNCs star distributions (\S\ref{sec_mma_mst_sfincs}).     

The main results of these comparisons can be summarized as follows:
\begin{enumerate}

\item For most of the richer SFiNCs clusters, the data-minus-model residual map values are small (typically $<10$\%) indicating that the isothermal ellipsoid models provide good fits. Our second validation technique, the quadrat counting test presented in \S \ref{model_validation}, shows similar results.  

\item The majority of the SFiNCs clusters are associated with clumpy and/or filamentary dusty structures seen in the far-IR images of the SFiNCs SFRs.

\item In many cases, cluster identification with nonparametric Minimal Spanning Tree (MST), either by us or by \citet{Gutermuth2009}, ``chain'' multiple SFiNCs clusters into unified structures.  Examples include: Be 59 (A+B), NGC 1333 (A+B), NGC 2068-2071 (C+D), Mon R2 (A+B+C), RCW 120 (A+B), Serpens Main (A+B), IRAS 20050+2720 (C+D), and Sh 2-106 (D+B). Independent information on associated molecular cloud structures, when  available (NGC 1333, RCW 120, Serpens Main, Sh 2-106), suggests that the MST chaining is physically unreasonable.

\item In other cases, the MST procedure fragments the SFiNCs clusters into multiple structures and/or fractionates the outer regions of larger clusters.  Examples include:  Be 59 (A) into 4 fragments, IC 348 (B) into 3 fragments, LkH$\alpha$~101 (A) into 3 fragments, RCW 120 (D) into 2 fragments, Cep OB3b (A) into several fragments, and Cep C (A) into 2 fragments.  Most of the MST stellar structures resulted from such fragmentations have no associations with any cloud structures. 
\clearpage
\begin{figure*}
\centering
\includegraphics[angle=0.,width=180mm]{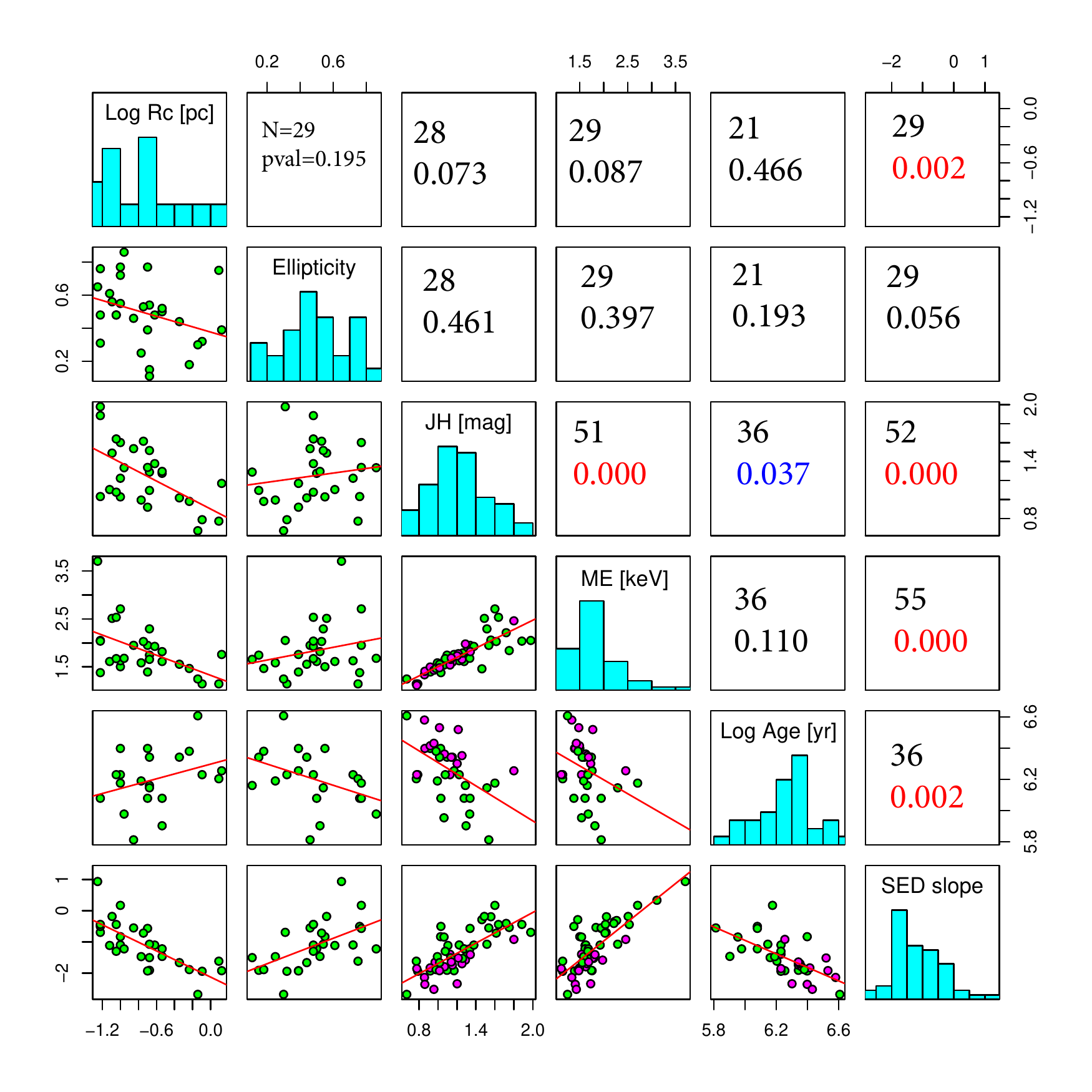}
\caption{Pairs plot showing bivariate relationships between properties of SFiNCs clusters (green) and SFiNCs unclustered stellar structures (magenta).  Bivariate scatter plots with linear fits (red) are below the diagonal, and univariate histograms on the diagonal. Above the diagonal, panels give the number of plotted stellar structures and $p-$values of Kendall's $\tau$ correlation coefficient adjusted for multiple tests.  Marginally ($0.003 < p < 0.05$) and strongly ($p \leq 0.003$) statistically significant correlations have $p$-values in blue and red, respectively. \label{fig_pairs_plot_sfincs}}
\end{figure*}

\begin{figure*}
\centering
\includegraphics[angle=0.,width=180mm]{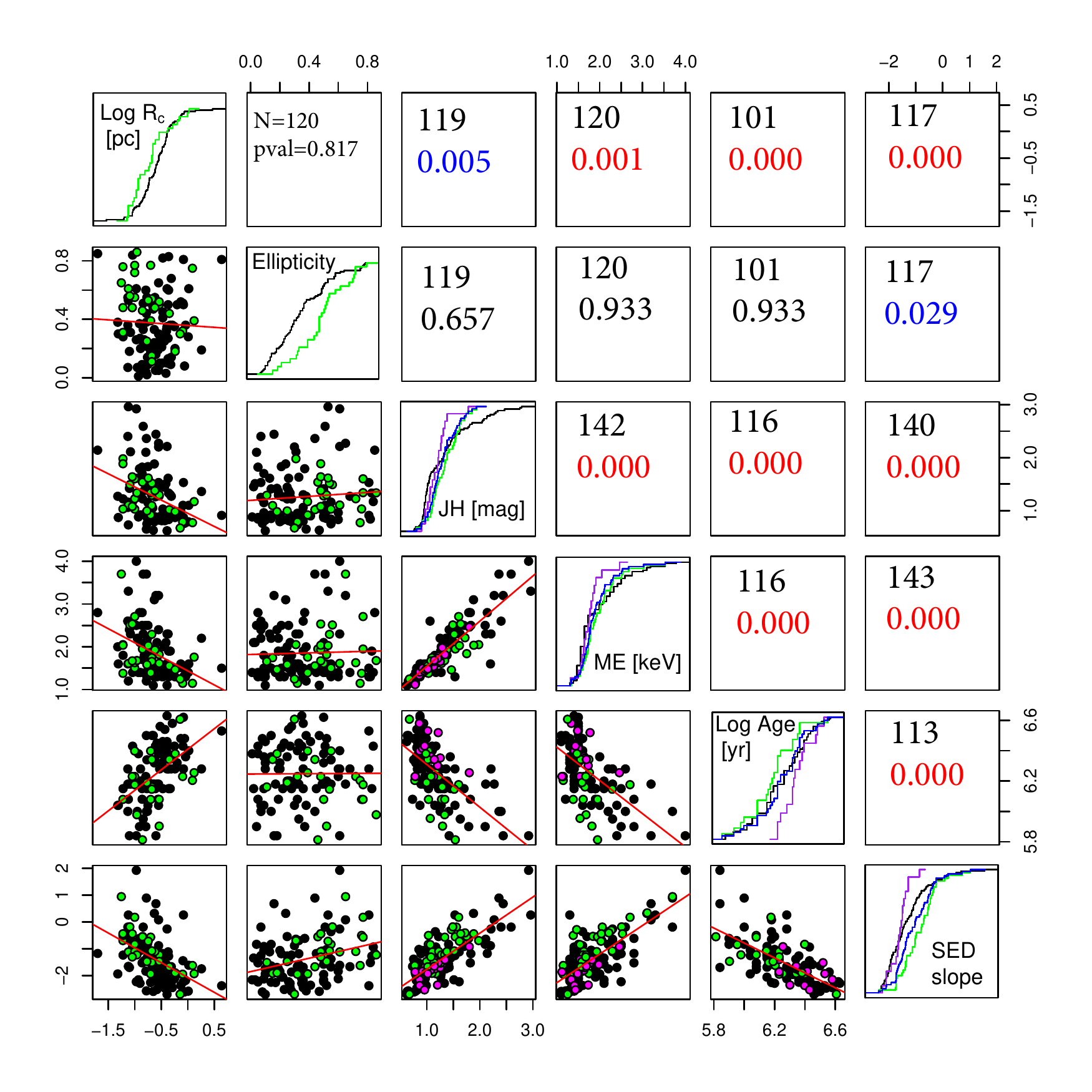}
\caption{Pairs plot for the combined samples of MYStIX clusters (black), SFiNCs clusters (green), and SFiNCs unclustered samples (magenta).  The panels are similar to Figure~\ref{fig_pairs_plot_sfincs} except that the diagonal shows cumulative distribution functions for four samples: MYStIX clusters (black), SFiNCs clusters (green), SFiNCs unclustered samples (magenta), and a merger of the SFiNCs clustered and unclustered samples (blue). \label{fig_pairs_plot_mystixsfincs}}
\end{figure*}

\begin{table*}\scriptsize
\centering
 \begin{minipage}{180mm}
 \caption{Properties of SFiNCs Clusters. Column 1: Cluster component name. In addition to the cluster components from Table \ref{tbl_cluster_morphology} (A, B, C, D, and E), the unclustered components (UnCl) are listed.  Columns 2: Number of observed SPCMs. Column 3: Number of observed SPCMs with available reliable $J-H$ color.  Column 4: Median $J-H$ color. Column 5: Number of observed SPCMs with available X-ray median energy ($ME$). Column 6: Median $ME$. Column 7: Number of observed SPCMs with available IRAC spectral energy distribution slope ($\alpha_{IRAC}$). Column 8: Median SED slope. Column 9. Number of observed SPCMs with available age estimates ($Age_{JX}$). Column 10: Median age.  In Columns (2, 3, 5, 7, 9), the numbers are given for the SPCMs located within an $\sim 4 \times R_c$ region around the center of the A$-$E components.  In Columns (4, 6, 8, and 10), the reported uncertainties on the medians are derived using the bootstrap approach given in \citet{Getman2014a}.}
 \label{tbl_cluster_other_props}
 \begin{tabular}{@{\vline }c@{ \vline }c@{ \vline }c@{ \vline }c@{ \vline }c@{ \vline }c@{ \vline }c@{ \vline }c@{ \vline }c@{ \vline }c@{ \vline }}
\cline{1-10}
&&&&&&&&&\\ 
%\hline
Cluster & 
~$N_{4,data}$~ &
~$N_{JH}$~ &
~$J-H$~ &
~$N_{ME}$~ &
$ME$ &
~$N_{\alpha_{IRAC}}$~ &
$\alpha_{IRAC}$ &
~$N_{Age_{JX}}$~ &
$Age_{JX}$ \\
  & (stars) & (stars) & (mag) & (stars) & (keV) & (stars) &   & (stars) & (Myr)\\
   (1)         & (2)         &  (3)         & (4)     &  (5) &   (6)   &   (7)      &   (8)       &    (9) & (10)\\
\cline{1-10}
&&&&&&&&&\\
Be 59 A &  321 & 301 & ~$1.17\pm0.01$~ & 248 & ~$1.76\pm0.02$~ & 319 & ~$-2.31\pm 0.11$~ &  92 & ~$1.80\pm0.23$~\\
Be 59 B &  220 & 189 & $1.10\pm0.02$ & 180 & $1.74\pm0.03$ & 198 & $-2.30\pm 0.11$ &  39 & $2.20\pm0.37$\\
Be 59 UnCl &    1 & ... & ... & ... & ... & ... & ... & ... & ...\\
SFO 2 A &   66 &  51 & $1.11\pm0.07$ &  37 & $1.61\pm0.18$ &  66 & $-1.11\pm 0.15$ &   5 & $1.80\pm1.81$\\
SFO 2 UnCl &    5 &   4 & $0.85\pm0.09$ &   5 & $1.54\pm0.09$ &   5 & $-2.76\pm 0.08$ &   3 & $1.90\pm1.63$\\
NGC 1333 A &   60 &  39 & $1.03\pm0.16$ &  36 & $1.50\pm0.17$ &  60 & $-0.83\pm 0.27$ &  13 & $2.50\pm1.18$\\
NGC 1333 B &  101 &  67 & $1.22\pm0.07$ &  71 & $1.63\pm0.08$ &  97 & $-1.10\pm 0.14$ &  23 & $1.70\pm0.34$\\
NGC 1333 UnCl &   15 &  11 & $0.97\pm0.16$ &   9 & $1.34\pm0.24$ &  15 & $-1.65\pm 0.44$ &   4 & $2.55\pm0.80$\\
IC 348 B &  280 & 259 & $0.92\pm0.01$ & 245 & $1.39\pm0.03$ & 279 & $-2.34\pm 0.06$ & 109 & $2.50\pm0.14$\\
IC 348 A &   28 &  13 & $1.56\pm0.33$ &  14 & $2.06\pm0.46$ &  27 & $-0.41\pm 0.23$ &   6 & $2.30\pm0.78$\\
IC 348 UnCl &   45 &  44 & $0.86\pm0.05$ &  35 & $1.34\pm0.07$ &  45 & $-2.47\pm 0.08$ &  15 & $3.80\pm0.37$\\
LkH$\alpha$ 101 A &  182 & 156 & $1.29\pm0.03$ & 152 & $1.66\pm0.05$ & 167 & $-1.50\pm 0.19$ &  54 & $1.45\pm0.36$\\
LkH$\alpha$ 101 UnCl &   63 &  53 & $1.15\pm0.06$ &  43 & $1.60\pm0.08$ &  56 & $-1.74\pm 0.17$ &  25 & $2.20\pm0.61$\\
NGC 2068-2071 A &   23 &  15 & $1.75\pm0.18$ &  14 & $1.84\pm0.54$ &  23 & $-0.53\pm 0.63$ &   4 & $0.45\pm0.21$\\
NGC 2068-2071 B &  116 & 101 & $1.28\pm0.06$ &  79 & $1.61\pm0.06$ & 116 & $-1.22\pm 0.13$ &  41 & $1.20\pm0.18$\\
NGC 2068-2071 C &   33 &  18 & $1.98\pm0.16$ &  19 & $2.05\pm0.12$ &  30 & $-0.69\pm 0.40$ &   7 & $0.60\pm0.18$\\
NGC 2068-2071 D &   44 &  34 & $1.34\pm0.13$ &  28 & $1.68\pm0.09$ &  44 & $-1.22\pm 0.23$ &  10 & $0.95\pm0.29$\\
NGC 2068-2071 UnCl &  117 &  90 & $1.07\pm0.06$ &  85 & $1.63\pm0.06$ & 114 & $-1.80\pm 0.17$ &  37 & $2.30\pm0.44$\\
ONC Flank S A &  325 & 286 & $0.77\pm0.02$ & 222 & $1.15\pm0.02$ & 316 & $-1.62\pm 0.08$ & 109 & $1.60\pm0.18$\\
ONC Flank N A &  260 & 242 & $0.79\pm0.02$ & 195 & $1.15\pm0.03$ & 250 & $-1.94\pm 0.09$ &  79 & $1.70\pm0.22$\\
OMC 2-3 A &   54 &  44 & $1.06\pm0.13$ &  35 & $1.54\pm0.18$ &  52 & $-0.84\pm 0.18$ &  18 & $0.90\pm0.25$\\
OMC 2-3 B &   25 &   9 & $2.15\pm0.41$ &  18 & $2.23\pm0.43$ &  23 & $-0.34\pm 0.43$ &   3 & $2.10\pm1.45$\\
OMC 2-3 C &   50 &  30 & $1.03\pm0.16$ &  35 & $1.38\pm0.16$ &  50 & $-0.50\pm 0.18$ &  13 & $1.20\pm0.43$\\
OMC 2-3 D &   10 &   9 & $1.02\pm0.12$ &   9 & $1.34\pm0.21$ &   9 & $-2.03\pm 0.76$ &   6 & $2.05\pm0.73$\\
OMC 2-3 UnCl &  234 & 203 & $0.78\pm0.02$ & 187 & $1.12\pm0.04$ & 230 & $-1.86\pm 0.09$ &  89 & $1.70\pm0.19$\\
Mon R2 A &  134 &  85 & $1.34\pm0.07$ & 107 & $1.95\pm0.07$ & 119 & $-0.56\pm 0.10$ &  26 & $1.20\pm0.13$\\
Mon R2 B &  127 &  51 & $1.60\pm0.08$ & 101 & $2.71\pm0.22$ &  87 & $ 0.18\pm 0.15$ &  11 & $1.50\pm0.41$\\
Mon R2 C &   32 &  19 & $1.46\pm0.20$ &  16 & $1.46\pm0.21$ &  32 & $-0.28\pm 0.27$ &   5 & $0.80\pm0.20$\\
Mon R2 UnCl &  247 & 214 & $1.13\pm0.02$ & 175 & $1.54\pm0.04$ & 247 & $-1.40\pm 0.10$ &  80 & $1.70\pm0.18$\\
GGD 12-15 A &  108 &  56 & $1.54\pm0.13$ &  59 & $1.95\pm0.21$ & 105 & $-0.55\pm 0.10$ &  12 & $0.65\pm0.69$\\
GGD 12-15 UnCl &  104 &  90 & $0.86\pm0.03$ &  79 & $1.41\pm0.03$ & 104 & $-2.36\pm 0.20$ &  33 & $2.50\pm0.51$\\
RCW 120 A &   30 &   5 & $1.72\pm0.35$ &  22 & $3.11\pm0.30$ &  24 & $ 0.34\pm 0.30$ & ... & ...\\
RCW 120 B &  110 &  77 & $1.30\pm0.06$ &  98 & $1.82\pm0.05$ & 103 & $-1.46\pm 0.19$ &  19 & $0.80\pm0.20$\\
RCW 120 C &   56 &  41 & $1.38\pm0.07$ &  40 & $1.93\pm0.10$ &  56 & $-1.10\pm 0.15$ &   7 & $0.70\pm0.42$\\
RCW 120 D &   17 &   5 & $1.68\pm0.11$ &   6 & $2.71\pm0.21$ &  17 & $-0.73\pm 0.24$ & ... & ...\\
RCW 120 UnCl &  157 &  71 & $1.29\pm0.10$ &  86 & $1.98\pm0.27$ & 156 & $-1.58\pm 0.15$ &   8 & $1.25\pm0.25$\\
Serpens Main A &   14 & ... & ... &   7 & $3.91\pm0.44$ &  13 & $ 1.30\pm 0.66$ & ... & ...\\
Serpens Main B &   61 &  27 & $1.88\pm0.30$ &  39 & $2.04\pm0.28$ &  59 & $-0.44\pm 0.16$ &   9 & $0.60\pm0.76$\\
Serpens Main UnCl &   65 &  45 & $1.25\pm0.07$ &  45 & $1.66\pm0.10$ &  65 & $-1.70\pm 0.16$ &  16 & $2.25\pm0.51$\\
Serpens South A &   12 & ... & ... & ... & ... &  12 & $ 1.28\pm 0.56$ & ... & ...\\
Serpens South B &    6 & ... & ... & ... & ... &   6 & $ 0.04\pm 0.39$ & ... & ...\\
Serpens South C &   74 &   4 & $1.29\pm0.51$ &  33 & $3.70\pm0.19$ &  70 & $ 0.94\pm 0.19$ & ... & ...\\
Serpens South D &    7 & ... & ... & ... & ... &   7 & $ 0.97\pm 0.60$ & ... & ...\\
Serpens South UnCl &  199 &  31 & $1.80\pm0.21$ &  44 & $2.46\pm0.24$ & 198 & $-0.92\pm 0.14$ &  13 & $1.80\pm0.84$\\
IRAS 20050+2720 A &    0 & ... & ... & ... & ... & ... & ... & ... & ...\\
IRAS 20050+2720 B &   14 &   5 & $1.26\pm0.14$ &   7 & $1.79\pm0.08$ &  14 & $-0.73\pm 0.31$ &   4 & $2.70\pm0.48$\\
IRAS 20050+2720 C &   29 &   8 & $1.49\pm0.06$ &  13 & $2.27\pm0.18$ &  29 & $-0.31\pm 0.35$ &   3 & $1.60\pm0.57$\\
IRAS 20050+2720 D &  111 &  26 & $1.49\pm0.08$ &  64 & $2.51\pm0.11$ & 106 & $-0.18\pm 0.14$ &   9 & $1.90\pm0.69$\\
IRAS 20050+2720 E &   31 &  25 & $1.26\pm0.04$ &  20 & $1.79\pm0.12$ &  31 & $-1.53\pm 0.14$ &   6 & $4.00\pm0.18$\\
~IRAS 20050+2720 UnCl~ &  130 &  82 & $1.21\pm0.04$ &  85 & $1.77\pm0.09$ & 127 & $-1.52\pm 0.10$ &  25 & $3.30\pm0.41$\\
Sh 2-106 A &   25 &  19 & $1.13\pm0.15$ &  15 & $1.66\pm0.20$ &  24 & $-1.51\pm 0.36$ & ... & ...\\
Sh 2-106 B &   24 &  13 & $1.72\pm0.14$ &  11 & $2.21\pm0.40$ &  24 & $-0.40\pm 0.22$ & ... & ...\\
Sh 2-106 C &    5 &   4 & $1.60\pm0.29$ &   3 & $2.15\pm0.06$ &   5 & $-1.68\pm 0.78$ & ... & ...\\
Sh 2-106 D &   53 &  16 & $1.64\pm0.11$ &  51 & $2.53\pm0.20$ &  23 & $-0.44\pm 0.57$ & ... & ...\\
Sh 2-106 UnCl &  144 &  98 & $1.34\pm0.06$ &  76 & $1.82\pm0.12$ & 143 & $-1.41\pm 0.14$ &   4 & $0.80\pm0.36$\\
IC 5146 A &   10 &   9 & $1.01\pm0.08$ &   9 & $1.13\pm0.13$ &  10 & $-1.05\pm 0.22$ & ... & ...\\
IC 5146 B &  142 & 129 & $0.99\pm0.03$ &  85 & $1.58\pm0.05$ & 140 & $-1.47\pm 0.07$ &  32 & $1.55\pm0.18$\\
IC 5146 UnCl &   93 &  87 & $0.92\pm0.04$ &  62 & $1.50\pm0.06$ &  93 & $-1.83\pm 0.19$ &  23 & $2.60\pm0.49$\\
NGC 7160 A &  141 & 135 & $0.67\pm0.02$ & 134 & $1.25\pm0.02$ & 140 & $-2.68\pm 0.01$ &  28 & $4.05\pm0.44$\\
LDN 1251B A &   14 &   8 & $1.01\pm0.29$ &   8 & $1.31\pm0.99$ &  12 & $-0.80\pm 0.85$ & ... & ...\\
LDN 1251B UnCl &   34 &  29 & $0.96\pm0.08$ &  30 & $1.43\pm0.10$ &  34 & $-2.52\pm 0.22$ &  13 & $2.70\pm0.83$\\
Cep OB3b A &  508 & 420 & $1.02\pm0.01$ & 344 & $1.55\pm0.02$ & 502 & $-1.66\pm 0.06$ & 161 & $2.20\pm0.21$\\
Cep OB3b B &   42 &  37 & $1.08\pm0.03$ &  25 & $1.67\pm0.06$ &  42 & $-1.30\pm 0.07$ &  10 & $1.70\pm0.32$\\
Cep OB3b C &  817 & 725 & $0.98\pm0.01$ & 586 & $1.47\pm0.01$ & 810 & $-1.89\pm 0.08$ & 281 & $2.40\pm0.14$\\
Cep OB3b D &    0 & ... & ... & ... & ... & ... & ... & ... & ...\\
Cep OB3b UnCl &   98 &  84 & $1.02\pm0.02$ &  57 & $1.48\pm0.04$ &  98 & $-1.92\pm 0.18$ &  35 & $3.40\pm0.43$\\
Cep A A &  172 &  82 & $1.52\pm0.09$ & 120 & $2.29\pm0.17$ & 159 & $-1.07\pm 0.11$ &  29 & $1.40\pm0.28$\\
Cep A UnCl &   98 &  89 & $1.20\pm0.02$ &  70 & $1.73\pm0.07$ &  98 & $-2.34\pm 0.15$ &  48 & $2.00\pm0.26$\\
Cep C A &   86 &  43 & $1.61\pm0.13$ &  36 & $2.03\pm0.14$ &  84 & $-0.72\pm 0.13$ &   9 & $0.80\pm0.28$\\
Cep C B &    4 & ... & ... &   4 & $2.88\pm0.65$ &   4 & $-0.97\pm 0.61$ & ... & ...\\
Cep C UnCl &   82 &  66 & $1.14\pm0.03$ &  53 & $1.69\pm0.06$ &  76 & $-1.64\pm 0.33$ &  27 & $2.20\pm0.87$\\
&&&&&&&&&\\ 
\cline{1-10} 
\end{tabular}
\end{minipage}
\end{table*}
\clearpage
\newpage

\begin{table*}\small
\centering
 \begin{minipage}{180mm}
 \caption{Properties of MYStIX Clusters. A few rows of this table are shown here; the full table of 91 MYStIX clusters (that have $N_{4,model} \geq 20$ stars) is available in the Supplementary Materials. Column 1: MYStIX cluster component name. Column 2. Number of observed stars estimated by integrating the model component out to four times the size of the core. Column 3. Logarithm of cluser core radius. Column 4. Cluster ellipticity. Columns 5-8: Median $J-H$, $ME$, age (in log), and SED slope of a cluster. Column 9: Relation to molecular clouds, based on a visual inspection of Figure Set 7 in \citet{Getman2014a}: ``C'' - likely associated with a cloud, i.e., embedded in or emerging from a cloud; ``R'' - likely revealed, i.e., already emerged from a cloud; ``...'' - unclear case.}
 \label{tbl_mystix_clusters}
 \begin{tabular}{@{\vline }c@{ \vline }c@{ \vline }c@{ \vline }c@{ \vline }c@{ \vline }c@{ \vline }c@{ \vline }c@{ \vline }c@{ \vline }}
\cline{1-9}
&&&&&&&&\\ 
%\hline
Cluster &
~$N_{4,model}$~ & 
~$\log(R_c)$~ &
~~~$\epsilon$~~~ & 
~$J-H$~ &
~$ME$~ &
~$\log(Age_{JX})$~ &
~$\alpha_{IRAC}$~ &
~Cloud~\\
  & (stars) & (pc) &   & (mag) & (keV) & (yr) &   &  \\
   (1)         & (2)         &  (3)         & (4)     &  (5) &   (6)   &   (7)      &   (8)       &    (9)\\
\cline{1-9}
&&&&&&&&\\
Orion B &    73 & -1.31 & 0.30 & 0.87 & 1.6 & 6.04 & ... & R\\
Orion C &   834 & -0.66 & 0.49 & 1.05 & 1.6 & 6.18 & -1.08 & R\\
Orion D &    48 & -1.04 & 0.84 & 1.17 & 1.4 & 6.43 & -0.93 & C\\
Flame A &   219 & -0.91 & 0.37 & 1.79 & 2.8 & 5.90 & -0.61 & C\\
W40 A &   187 & -0.79 & 0.04 & 2.10 & 2.5 & 5.90 & -0.83 & R\\
~RCW36 A~ &   196 & -0.84 & 0.33 & 1.63 & 2.3 & 5.95 & -0.69 & C\\
&&&&&&&&\\ 
\cline{1-9} 
\end{tabular}
\end{minipage}
\end{table*}
%\clearpage
%\newpage

\item In several cases, very small and sparse clusters are identified by the SFiNCs mixture modeling procedure with $N_{4,data} < 10$.  Examples include: Serpens South (B and D), IRAS 20050+2720 (A), Sh 2-106 (C), Cep OB3b (D), and Cep C (B).  Based on the presence of cloud counterparts and/or independent identification with MST method, these seem to be real stellar groupings; MST analysis often adds nearby YSOs to these clusters.  However, the SFiNCs model parameters (core radius, ellipticity, etc.) of these structures are undoubtedly unreliable. 

\end{enumerate}

\section{Properties of SFiNCs and MYStIX Clusters} \label{ma_section}

\subsection{Combined SFiNCs and MYStIX Sample} \label{sample_subsection}

In Table \ref{tbl_cluster_other_props}, we provide a homogeneous set of median properties for 71 SFiNCs stellar structures: 52 SFiNCs clusters and 19 unclustered stellar structures. The median values of the $J-H$, $ME$, $\alpha_{IRAC}$, and $Age_{JX}$ quantities and their bootstrap uncertainties are computed by averaging over individual stellar members of these groups listed in Table \ref{tbl_cluster_membership}. The table also provides $N_{4,data}$, the estimated number of stars assigned to each cluster.  In a future SFiNCs paper, the full intrinsic stellar populations will be estimated by correcting for incompleteness at low masses following the procedure of \citep{Kuhn2015a}.  

Some median values are omitted due to small sample limitations.  For highly absorbed clusters, the measurements reported in Table \ref{tbl_cluster_other_props} might not be representative of their true, intrinsic property values due to the lack of the measurements for most of their extremely absorbed stellar members.  For the multivariate cluster analysis given below the median values of various properties are omitted when less than 10 members are available. We also omit the ellipticity and core radius values for weak SFiNCs clusters where $N_{4,data} < 30$ stars (equivalent to $N_{4,model} < 20$ stars)\footnote{Recall that $N_{4,model}$ and $N_{4,data}$ differ mainly due to the differences between the ``flattened'' and entire stellar samples.}. 

In Table \ref{tbl_mystix_clusters} we provide a similar list of properties for 91 MYStIX clusters.  Data are obtained from \citet{Kuhn2014,Kuhn2015b} except for $\alpha_{IRAC}$ that we calculate here.  As with  the SFiNCs sample, the MYStIX dataset is culled of weak ($N_{4,model} < 20$ stars) clusters with poorly constrained cluster parameters. 

\subsection{Relationships Between Cluster Properties} \label{mv_presentation_subsection}

Figures \ref{fig_pairs_plot_sfincs} and \ref{fig_pairs_plot_mystixsfincs} summarize univariate distributions and bivariate relationships for the SFiNCs and SFiNCs$+$MYStIX samples, respectively. Similar plots for the MYStIX sample alone, but totaling 141 clusters including weak clusters ($N_{4,model} < 20$ stars), appear in \citet{Kuhn2015b}.  These plots were created using functions  {\it pairs.panels} and {\it corr.test} in CRAN package $psych$ \citep{Revelle17}. To guide the eye for identification of possible trends, a least squares linear fit (controlled by {\it pairs.panels}) is added to each of the bivariate scatter plots (red line) obtained using $R$'s {\it lm} function.   

\begin{figure}
\centering
\includegraphics[angle=0.,width=80mm]{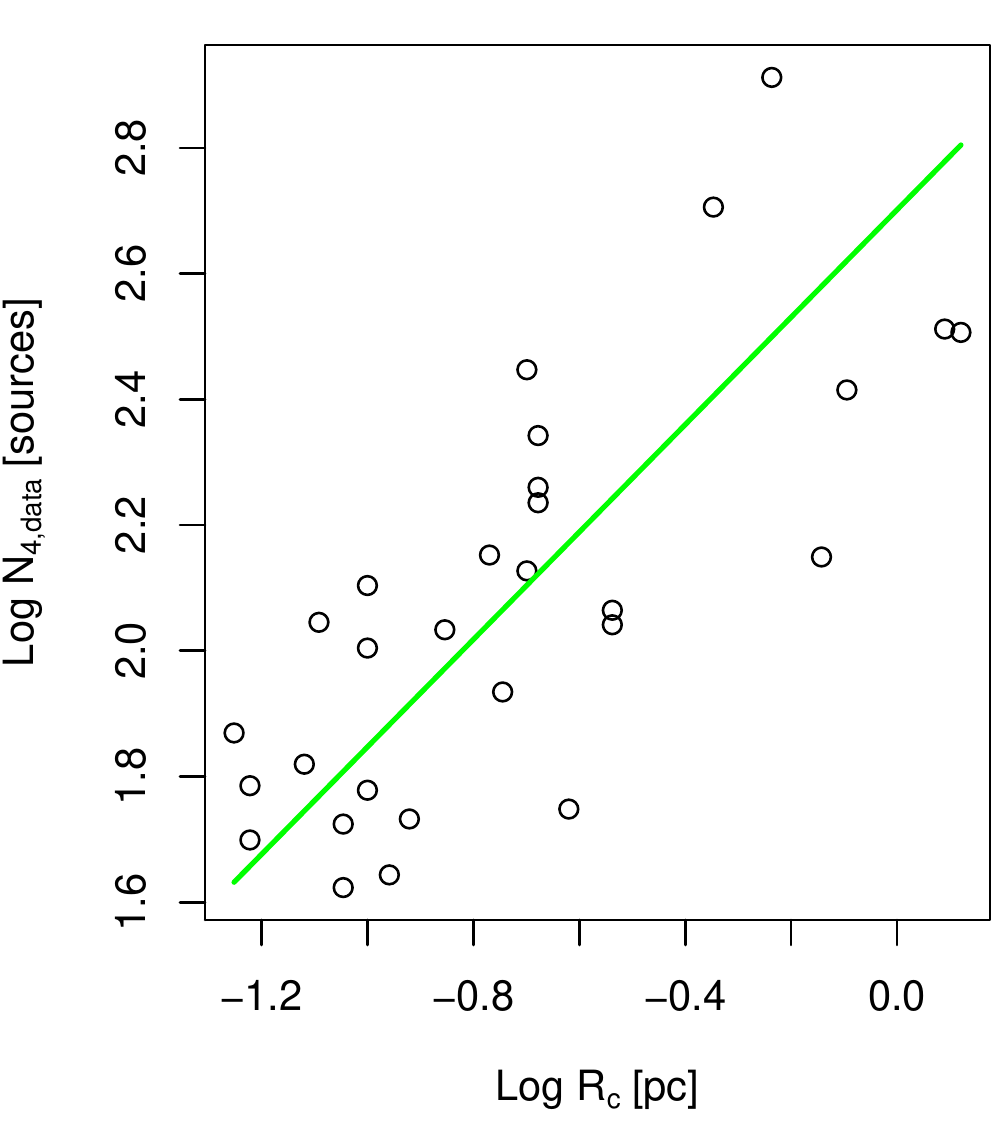}
\caption{Number of observed (not corrected for the observation incompleteness at low masses) SPCMs in a SFiNCs cluster as a function of cluster size.  A symmetrical linear regression fit is plotted in green. \label{fig_logn_logrc}}
\end{figure}

Figure \ref{fig_logn_logrc} highlights an additional relationship between the total apparent number of SPCMs in SFiNCs clusters ($N_{4,data}$) and cluster size ($R_c$).  The correlation has a significance level of $p=0.002$ using Kendall's $\tau$ statistic.  The fitted line, computed with a standard major axis procedure that treats the variables symmetrically \citep{Legendre1998}, has slope $0.85$.  However, we recall that the $N_{4,data}$ measurement is subject to strong observational selection biases involving the distance, absorption and instrumental exposure sensitivities.  In a future study, the $N_{4,data}$ quantity will be corrected for the incompleteness at low masses  using the well established X-ray luminosity function and initial mass function analyses following \citet{Kuhn2015a}.

For each of the pairs, the statistical significance of their correlations were evaluated by testing the null hypothesis that the Kendall's $\tau$ coefficient is equal to zero. Here we adopt $p-$values of $0.003<p\leq0.05$ and $p\leq0.003$ as indicators of marginally and strong statistical significant correlations, respectively.  The {\it pairs.panels} function offers a control against Type I errors (the null hypothesis is true, but is rejected) when multiple comparisons are under consideration.  As we computed each correlation for three cases (SFiNCs alone, MYStIX alone (not shown), and SFiNCs+MYStIX) we adjust $p$-values for 3 tests using the False Discovery Rate procedure of \citet{Benjamini1995}. 

The important relationships between cluster properties emerging from these plots are: 
\begin{enumerate}

\item {\it Cluster core radius:} Core radii are clearly linked to four interrelated properties: anti-correlation with X-ray median energy $ME$ and $J-H$ color, two measures of cloud absorption; anti-correlation with {\it SED slope}, a measure of inner protoplanetary disk; and correlation with $Age_{JX}$, a measure of stellar age.  Together, these empirical relationships provide strong evidence that clusters expand (i.e. $R_c$ increases) as they age.  This result was first reported by \citet{Kuhn2015b} for the MYStIX sample alone and is discussed further below (\S \ref{expansion_subsection}).  For the merged MYStIX$+$SFiNCs sample, a linear regression fit that treats variables symmetrically (obtained using $R$'s $lmodel2$ function) gives the relationship 
\begin{equation}
\log R_c = -12.4(\pm 2.0) + 1.9(\pm 0.3) \times \log Age_{JX}~{\rm pc} 
\end{equation}
over the approximate ranges $0.08 < R_c < 1$~pc and $1 < Age_{JX} < 3.7$~Myr.

There is also a hint that the SFiNCs clusters have systematically smaller sizes (median core radius = 0.18~pc) compared to those of MYStIX (0.22~pc). An Anderson-Darling two-sample test on the core radii distributions indicates this is only a possibility ($p_{AD} = 0.04$). 

\item{\it Cluster ellipticity:} The ellipticities of the MYStIX+SFiNCs clusters show no correlations with any of the cluster properties.  However, there is a significant difference between the two samples with MYStIX clusters being rounder than SFiNCs clusters ($p_{AD} = 0.001$). The median ellipticities are 0.48 and 0.30 for the SFiNCs and MYStIX clusters, respectively.  

\item{\it Extinction:} Since both the NIR $J-H$ and X-ray $ME$ quantities serve as surrogates for extinction, there is a strong and tight correlation between these two variables. For the merged MYStIX$+$SFiNCs sample, $J-H$ and $ME$ strongly correlate with core radius, age, and SED slope.

\item{\it Age:} As mentioned in item 1 above, the ages of SFiNCs$+$MYStIX clusters are significantly correlated with cluster radius and inversely correlated with  extinction indicators (negative), and SED slope.

The SFiNCs and MYStIX clusters have statistically indistinguishable age ECDFs with median of 1.7~Myr.  However the unclustered SFiNCs stars are systematically older than the clustered stars with median 2.3~Myr ($p_{AD} = 0.001$).  This echoes a similar result for clustered vs. unclustered populations in the MYStIX sample reported by \citet{Getman2014a} (see \S \ref{db_populations_subsection} for further discussion).  

\item{\it SED Slope:} This slope is a well-established surrogate for disk-bearing stars \citep{Richert2018}.  As mentioned in item 1, in the SFiNCs$+$MYStIX cluster sample, it is strongly correlated with core radius, extinction\footnote{Since the SFiNCs and MYStIX $\alpha_{IRAC}$ quantities are observed (not corrected for extinction) slopes, they may overestimate intrinsic SED slopes \citep[][their Table 2]{Lada2006} by $<3$\% ($<15$\%) for YSOs that are subject to source extinction of $A_V\leq5$~mag ($A_V \leq 10$~mag).}, and age. 

The SED slopes of the SFiNCs unclustered structures are systematically lower than those of the SFiNCs clusters ($p_{AD} < 0.001$).  This indicates that spatial gradients of apparent disk fraction are generally present, where the disk fraction is decreasing from the cluster centers towards the peripheries of the SFiNCs SFRs (see \S \ref{db_populations_subsection} below).

\end{enumerate}

\section{Association With Molecular Clouds} \label{cloud_subsection}

Figure~\ref{fig_cluster_assignment_maps} shows all SPCMs within the {\it Chandra} ACIS-I fields superimposed on the {\it Herschel} and {\it AKARI} far-IR images that trace the SFiNCs molecular clouds. For MYStIX, similar maps appear in Figure 7 of \citet{Getman2014a}.  SPCM stars are color-coded according to their cluster assignments.  We visually identify clusters that closely associated with clouds, either embedded within a cloud or revealed and emerging from a cloud. The results are given in the last column of Tables~\ref{tbl_cluster_morphology} and \ref{tbl_mystix_clusters}. 

Nearly all of the SFiNCs clusters are associated with molecular clouds; of the 52 clusters, 35 appear embedded and 12 appear revealed.  These include all of the clusters in the SFiNCs SFRs close to the Sun such as NGC~1333, IC~348, NGC~2068-2071, OMC~2-3, GGD~12-15, Serpens Main, Serpens South, and Cep~C.  In most cases, the clusters are positioned and elongated along $\ga 1$~pc-long molecular filaments that are clearly distinguished by eye on the far-IR images. Some clusters are associated with large ($\ga 1$~pc) molecular clump and/or hub-filament systems such as Cep A, Sh~2-106, Mon~R2, and IRAS~20050+2720. Other star-cloud configurations are present: for instance, the three minor clusters in RCW~120 lie projected along a molecular shell, and the main cluster in SFO~2 is embedded at the tip of a bright-rimmed cloud. All of the aforementioned clusters are marked by the flag ``C'' in Table \ref{tbl_cluster_morphology}.
\begin{figure}
\centering
\includegraphics[angle=0.,width=90mm]{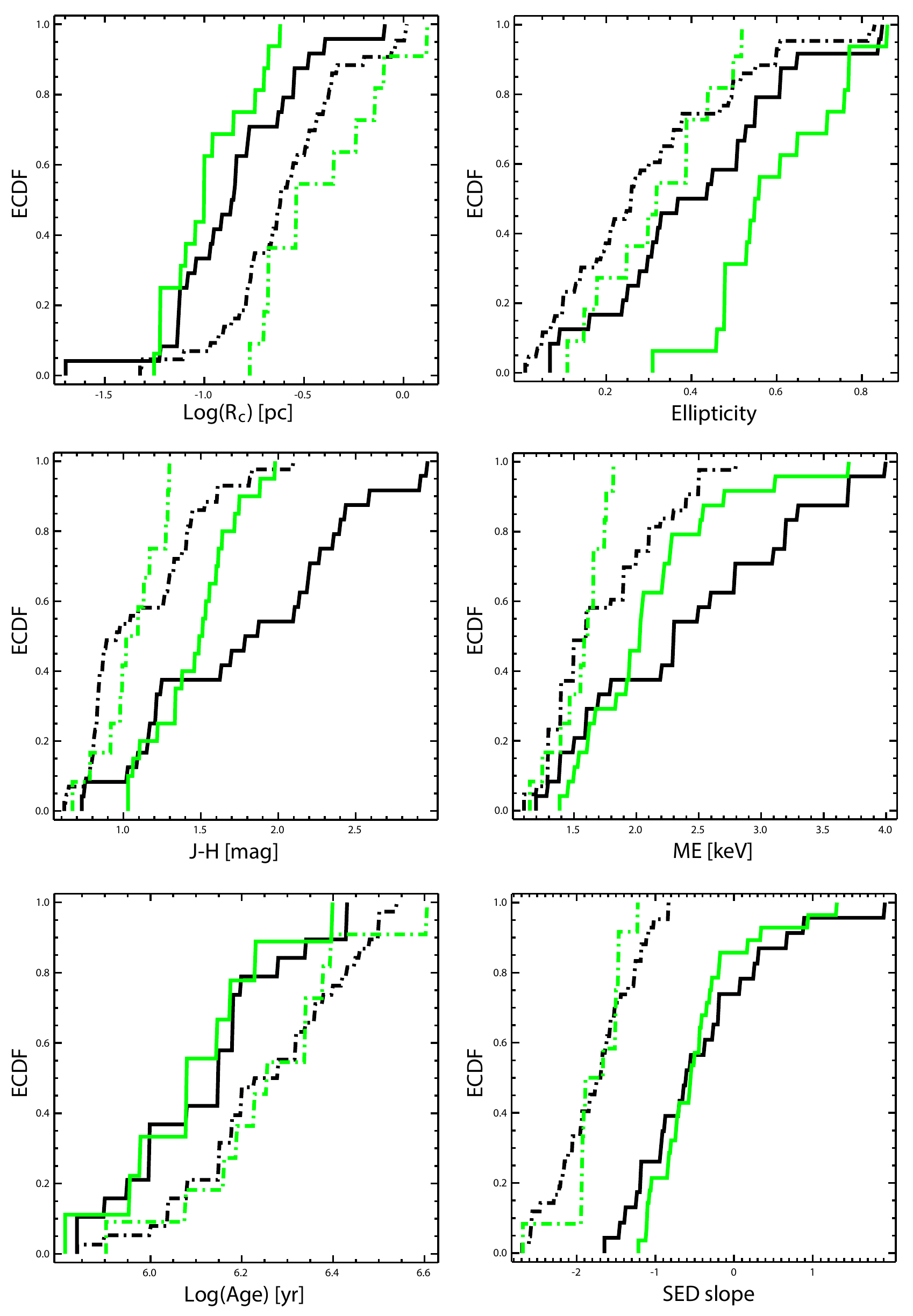}
\caption{Cluster properties stratified by the presence or absence of molecular cloud material. Empirical cumulative distribution functions of cluster core radius, ellipticity, $J-H$ and $ME$ (as surrogates for absorption), age, and SED slope. SFiNCs clusters are in green and MYStIX clusters are in black. Clusters closely associated with molecular clouds (i.e., fully or partially embedded in molecular clouds) are shown with solid lines.  Clusters emerged (revealed) from clouds are shown with dash-dotted lines. \label{fig_ecdfs_rev_vs_cloud}}
\end{figure}

About a quarter of the SFiNCs clusters are revealed, near but not embedded in clouds. For instance, the three main clusters in IC~348, NGC~2068-2071, and Be~59 lie in regions of dispersed molecular material. The main clusters in ONC Flank N, RCW~120, and IC~5146 lie mainly inside ionized molecular bubbles. The two main clusters in Cep~OB3b lie outside a giant molecular cloud. And no molecular material left in the vicinity of the oldest cluster, NGC~7160.  These clusters are marked as ``R'' in Table \ref{tbl_cluster_morphology}.

Figure~\ref{fig_ecdfs_rev_vs_cloud} shows various physical properties derived from the best-fit mixture models for SFiNCs clusters stratified by their location with respect to clouds.  Quantitative comparisons between these univariate distributions for SFiNCs and MYStIX clusters, stratified by cloud proximity, appear in Table~\ref{tbl_rev_cloud_univariate}. 

The following results emerge from the cluster-cloud associations:
\begin{enumerate}

\item A clear difference in cloud location is seen between SFiNCs and MYStIX clusters: about two-thirds of SFiNCs sample are closely associated with clouds (probably embedded) while 60\% of the MYStIX clusters are revealed.   

\item For the merged SFiNCs+MYStIX sample, the revealed clusters are substantially larger (median core radius $R_c = 0.25 \pm 0.03$~pc) than the cloud-associated clusters ($0.11 \pm 0.02$~pc).

\item The cloud-associated (embedded) clusters are significantly more elongated (median $\epsilon = 0.51 \pm 0.03$) than the revealed clusters ($0.27 \pm 0.03$). Ellipticities of SFiNCs cloud-associated clusters appear systematically higher than those from the  MYStIX sample ($p_{AD} = 0.009$).

\item The cloud-associated clusters are more absorbed (median $A_V$ of $8-10$~mag\footnote{Using the conversion from $J-H$ and $ME$ to the $V$-band extinction $A_V$ based on the SFiNCs data tabulated in \citet{Getman2017}.}) than the revealed clusters ($2-4$~mag).

\item The cloud-associated clusters are systematically younger (median $Age = 1.4 \pm 0.1$~Myr) than the revealed clusters ($1.8 \pm 0.2$~Myr).

\item The cloud-associated clusters have systematically much higher SED slope (median $\alpha_{IRAC} = -0.6 \pm 0.1$) than the revealed clusters ($-1.7 \pm 0.1$).

\end{enumerate}

\begin{table*}\small
\centering
 \begin{minipage}{180mm}
 \caption{Properties of Cloud-Associated $vs.$ Revealed MYStIX+SFINCs Clusters. Comparisons of univariate distributions of physical properties for SFiNCs and MYStIX cloud-associated and revealed clusters; see Figure \ref{fig_ecdfs_rev_vs_cloud}. Column 1: Property. Column 2: Cluster sample: SFiNCs cloud-associated (``Sc-C''), SFiNCs revealed (``Sc-R''), MYStIX cloud-associated (``Mc-C''), MYStIX revealed (``Mc-R''), MYStIX$+$SFiNCs cloud-associated (``McSc-C''), and MYStIX$+$SFiNCs revealed (``McSc-R''). Column 3. Number of clusters in the sample. Column 4: Median value of the univariate distribution and its bootstrap error. All variables are on a linear scale. Columns 5-9: $p$-values reported by the two-sample Anderson-Darling test. $p$-values $\la 0.003$ indicate that the null hypothesis that the two samples are drawn from the same distribution can be confidently rejected (strong significance).  $p$-values between $0.003 < p \leq 0.05$ indicate a marginal significance. $p$-values $>0.05$ indicate similar distributions (not significant).}
 \label{tbl_rev_cloud_univariate}
 \begin{tabular}{@{\vline}c@{}c@{}c@{}c@{}c@{}c@{}c@{}c@{}c@{ \vline }}
\cline{1-9}
&&&&&&&&\\
\multicolumn{4}{@{\vline}c}{} & \multicolumn{5}{c@{\vline}}{p-values From AD Test}\\
%\cline{5-9}
&&&&&&&&\\
\cline{5-9}
&&&&&&&&\\
%\hline
Property & 
Sample &
N & 
Median &
~~Sc-R~~ &
~~Mc-C~~ &
~~Mc-R~~ &
~~McSc-C~~ &
~~McSc-R~~ \\
  &   & (clusters) & (Value) &   &   &   &   &   \\
(1)&(2)&(3)               &(4)&(5)&(6)&(7)&(8)&(9)\\
\cline{1-9}
&&&&&&&&\\
$R_c$ [pc]	& Sc-C & 16 & $0.10\pm0.02$ & $0.000$ & $0.127$ & $0.000$ & ... & ...\\
			& Sc-R & 11 & $0.29\pm0.15$ & ... & $0.001$ & $0.166$ & ... & ...\\
			& Mc-C & 24 & $0.14\pm0.02$ & ... & ... & $0.001$ & ... & ...\\
			& Mc-R & 43 & $0.25\pm0.03$ & ... & ... & ... & ... & ...\\
			& McSc-C & 40 & $0.11\pm0.02$ & ... & ... & ... & ... & $0.000$\\
			& McSc-R & 54 & $0.25\pm0.03$ & ... & ... & ... & $0.000$ & ...\\
\cline{1-9}
Ellipticity 	& Sc-C & 16 & $0.56\pm0.05$ & $0.000$ & $0.009$ & $0.000$ & ... & ...\\
			& Sc-R & 11 & $0.32\pm0.06$ & ... & $0.233$ & $0.382$ & ... & ...\\
			& Mc-C & 24 & $0.41\pm0.07$ & ... & ... & $0.039$ & ... & ...\\
			& Mc-R & 43 & $0.26\pm0.04$ & ... & ... & ... & ... & ...\\
			& McSc-C & 40 & $0.51\pm0.03$ & ... & ... & ... & ... & $0.000$\\
			& McSc-R & 54 & $0.27\pm0.03$ & ... & ... & ... & $0.000$ & ...\\
\cline{1-9}
$J-H$ [mag]	& Sc-C & 20 & $1.50\pm0.07$ & $0.000$ & $0.015$ & $0.000$ & ... & ...\\
			& Sc-R & 12 & $1.06\pm0.07$ & ... & $0.001$ & $0.156$ & ... & ...\\
			& Mc-C & 24 & $1.83\pm0.27$ & ... & ... &  $0.000$ & ... & ...\\
			& Mc-R & 43 & $0.96\pm0.14$ & ... &  ... & ... & ... & ...\\
			& McSc-C & 44 & $1.58\pm0.09$ & ... & ... & ... & ... & $0.000$\\
			& McSc-R & 55 & $0.99\pm0.09$ & ... & ... & ... & $0.000$ & ...\\
\cline{1-9}
$ME$ [keV]	& Sc-C & 24 & $2.03\pm0.10$ & $0.001$ & $0.092$ & $0.001$ & ... & ...\\
			& Sc-R & 12 & $1.60\pm0.06$ & ... & $0.006$ & $0.203$ & ... & ...\\
			& Mc-C & 24 & $2.30\pm0.26$ & ...& ... & $0.000$ & ... & ...\\
			& Mc-R & 43 & $1.60\pm0.12$ & ... & ... & ... & ... & ...\\
			& McSc-C & 48 & $2.13\pm0.13$ & ... & ... & ... & ... & $0.000$\\
			& McSc-R & 55 & $1.60\pm0.06$ & ... & ... & ... & $0.000$ & ...\\
\cline{1-9}
$Age$ [Myr] & Sc-C &  9 & $1.20\pm0.19$ & $0.063$ & $0.916$ & $0.028$ & ... & ...\\
			& Sc-R & 11 & $1.80\pm0.28$ & ... & $0.035$ & $0.833$ & ... & ...\\
			& Mc-C & 19 & $1.41\pm0.17$ & ... & ... & $0.012$ & ... & ...\\
			& Mc-R & 38 & $1.80\pm0.22$ & ... & ... & ... & ... & ...\\
			& McSc-C & 28 & $1.41\pm0.13$ & ... & ... & ... & ... & $0.001$\\
			& McSc-R & 49 & $1.80\pm0.21$ & ... & ... & ... & $0.001$ & ...\\
\cline{1-9}
SED Slope 	& Sc-C & 28 & $-0.54\pm0.10$ & $0.000$ & $0.338$ & $0.000$ & ... & ...\\
			& Sc-R & 12 & $-1.78\pm0.16$ & ... & $0.000$ & $0.502$ & ... & ...\\
			& Mc-C & 23 & $-0.61\pm0.21$ & ... & ... & $0.000$ & ... & ...\\
			& Mc-R & 42 & $-1.71\pm0.11$ & ... & ... & ... & ... & ...\\
			& McSc-C & 51 & $-0.55\pm0.09$ & ... & ... & ... & ... & $0.000$\\
			& McSc-R & 54 & $-1.71\pm0.10$ & ... & ... & ... & $0.000$ & ...\\ 
&&&&&&&&\\ 
\cline{1-9} 
\end{tabular}
\end{minipage}
\end{table*}
%\clearpage
%\newpage

\section{Discussion} \label{discussion_sec}

\subsection{Bias in the SFiNCs Cluster Catalog} \label{bias_subsection}

It is important to recall a few biases pertinent to the current study of the SFiNCs clusters.  Similar issues are discussed for the MYStIX survey in the Appendix of \citet{Feigelson2013}.  

First, our methodology (\S\ref{model_section}) is biased against identifying poor clusters. The AIC model selection statistic requires that a putative cluster significantly improve the model likelihood for the entire region (\S\ref{model_section}).  Thus a real cluster with only a few members may not be discriminated from the unclustered component, or a nearby cluster, in a field with hundreds of SFiNCs Probable Complex Members.  Even fairly rich deeply embedded clusters may be missed because only a small fraction of the population is detected by $Chandra$ due to X-ray absorption by cloud gas.  Any clustering finding procedure will encounter similar difficulties when using the SPCM sample. 

Second, for highly absorbed SFiNCs clusters, due to the lack of the $J-H$, $ME$, $\alpha_{IRAC}$, and $Age_{JX}$ estimates for the vast majority of their absorbed cluster core members, the values reported in Table \ref{tbl_cluster_other_props} might not be representative of the intrinsic properties of these clusters. We seek to mitigate this problem by restricting the analysis in \S\S\ref{ma_section} and \ref{cloud_subsection} to cluster sub-samples, that have at least 10 cluster members with reliable measurements of $J-H$, $ME$, $Age_{JX}$, and SED slope. With respect to the $R_c$ and $\epsilon$ properties, the sub-samples are restricted to richer clusters ($N_{4,model} \geq 20$ stars). A result of this decision is that for instance the $\log R_{c} - \epsilon$ and $\log Age_{JX} - \alpha_{IRAC}$ relationships for the SFiNCs clusters (Figure \ref{fig_pairs_plot_sfincs}) employ unequal numbers of clusters. 

Third, the $N_{4,data}$ and $N_{4,model}$ quantities are not reliable measures of the total intrinsic stellar populations due to different distances, $Chandra$ and $Spitzer$ exposure times, and intervening absorption. A future SFiNCs study employing the analyss of X-ray luminosity functions and Initial Mass Functions \citep{Kuhn2015a} will give intrinsic population estimates. 

\subsection{Cluster Elongations in SFiNCs and MYStIX} \label{elongation_subsection}

Shells, bubbles, and filamentary molecular cloud structures are ubiquitous in the Galaxy \citep{Churchwell2006,Andre2014} and are often sites of star formation. Hierarchical fragmentation in molecular cloud filaments is often observed on scales ranging from several parsecs to $\leqslant 0.1$pc.  Different physical mechanisms appear to trigger the cloud fragmentation including: gravitational collapse; thermal, turbulent, and magnetic pressures; angular momentum; and dynamical feedback from young stellar outflows, winds and radiation pressure.  

Observations suggest that different mechanisms could dominate the cloud fragmentation in different situations \citep{Takahashi2013, Contreras2016, Teixeira2016}.  It is also possible that filamentary clouds are made up of collections of velocity-coherent subfilaments \citep{Hacar2013}. Turbulent energy cascades are proposed to play a major role in the formation of both the subfilaments and integrated filaments \citep{Smith2016}. Dense, gravitationally bound prestellar cores then form by cloud fragmentation along the densest filaments; core growth through filamentary accretion is also reported \citep{Andre2014}. Small star clusters can then emerge in these cores through star formation mediated by turbulent core accretion \citep{McKeeTan2003}, competitive accretion \citep{Bonnell2001,Wang2010}, and/or stellar  mergers \citep{BonnellBate2005}. Molecular gas can then be expelled by feedback effects of young stars including OB ionizing radiation and winds, supernovae, protostellar accretion heating, protostellar jets and outflows \citep{Dale2015}.

We therefore expect embedded SFiNCs clusters to inherit morphological characteristics from these star forming processes.   Perhaps most interesting is our finding that the SFiNCs clusters are both (a) more elongated than the MYStIX clusters (\S \ref{mv_presentation_subsection}) and (b) more closely associated with molecular filaments and clumps (\S \ref{cloud_subsection}).  Furthermore, examination of Figure \ref{fig_cluster_assignment_maps} shows that, in most cases, cluster elongations are oriented along the axes of their filamentary molecular clouds.  

The higher elongations of the SFiNCs cloud-associated clusters could be due to their more tranquil environments lacking numerous OB-type stars.  O-stars, if present, dominate the stellar feedback \citep{Dale2015}; a single O7 star may be capable of photoionising and dispersing a $10^{4}$~M$_{\odot}$ molecular cloud in $1-2$~Myr \citep{Walch2012}.  Since the SFiNCs environments harbor fewer and less massive OB-type stars than MYStIX regions, their gas removal timescales could be longer, allowing SFiNCs clusters to remain bound to and retain the morphological imprints of their parental molecular gas for a longer period of time. 

We warn, however, that elongated cluster shapes can also result from dynamical cluster mergers and be unrelated to the original cloud morphology \citep{Maschberger2010, Bate2012}.  And the lower ellipticities of the more populous MYStIX clusters may reflect the mergers of numerous smaller SFiNCs-like clusters that would reduce the ellipticity of the merger product.  We will examine this hypothesis in a future paper, where the total intrinsic stellar populations could be compared to the cluster ellipticity for the combined SFiNCs+MYStIX cloud-associated samples.

\subsection{Cluster Sizes in SFiNCs and MYStIX} \label{sizes_subsection}

For young stellar clusters in the solar neighborhood with a wide range of masses, from 30~M$_{\odot}$ to $>10^{4}$~M$_{\odot}$, \citet{Kuhn2015b} and \citet{Pfalzner2016} report a clear correlation between the cluster mass and cluster radius following $M_c \propto R_c^{1.7}$. A similar slope of $\sim 1.7$ is found in the relationship of mass and radius for the sample of a thousand massive star-forming molecular clumps across the inner Galaxy \citep{Urquhart2014}.  These findings point to the uniform star formation efficiency for stellar clusters with drastically different sizes and masses \citep{Pfalzner2016}. The mass-radius cluster relation could be a result from cluster growth, either through star formation or subcluster mergers \citep{Pfalzner2011,Kuhn2015b} and/or from an initial cloud mass-radius relation \citep{Pfalzner2016}.  

Our finding of a clear trend of increasing cluster's apparent population size ($N_{4,data}$) with increasing cluster radius ($R_c$) in SFiNCs (Figure \ref{fig_logn_logrc}) is consistent with the trend of the positive correlation $M_c-R_c$ reported by \citet{Kuhn2015a} and \citet{Pfalzner2016}. The SFiNCs relationship has a shallower slope ($N_{4,data} \propto R_c^{0.85}$); however, at this moment it is unclear if the slope difference is due to an astrophysical effect or simply due to the imperfection of $N_{4,data}$ as being an apparent rather than an intrinsic quantity (\S \ref{bias_subsection}). 

Since the SFiNCs targets are generally closer and less populous SFRs than MYStIX, the hint of smaller cluster sizes for SFiNCs, compared to those of MYStIX (\S \ref{mv_presentation_subsection}), is in line with the presence of the aforementioned correlations ($N_{4,data}-R_c$ and $M_c-R_c$). This is also tightly linked to our findings that the revealed (older and less absorbed) clusters appear to have much larger core radii (median $R_c \sim 0.25$~pc) than the cloud-associated (younger and more absorbed) clusters (0.11~pc).  This can be astrophysically linked to the effect of cluster expansion (\S \ref{expansion_subsection}).

It is also important to note that the median value of $R_c \sim 0.11$~pc found in both the SFiNCs and MYStIX cloud-associated cluster samples is very similar to the typical inner width of the molecular filaments ($\sim 0.1$~pc) found in the nearby clouds of the Gould Belt \citep{Andre2014}. This provides an indirect link between cloud-associated clusters and their natal molecular filaments.

\subsection{Cluster Expansion} \label{expansion_subsection}

It is well established that upon the removal of the residual molecular gas via the feedback of newly born stars, the gravitational potential of the molecular gas weakens causing young cluster expansion and, in most cases, eventual dispersal \citep[e.g.,][]{Tutukov1978,Moeckel2010,Banerjee2017}. In addition to the gas loss, expansion can arise from other causes: two-body relaxation and binary heating \citep{Moeckel2012,Parker2014,Banerjee2017}; hierarchical cluster merging \citep{Maschberger2010, Banerjee2015};  and mass loss via winds of massive stars and supernova explosions \citep{Banerjee2017}.

For the MYStIX clusters, the $R_c - Age$, $R_c - \rho_0$ ($\rho_0$ is the volumetric stellar density), and $\rho_0 - Age$ correlations provide direct empirical evidence for cluster expansion \citep{Kuhn2015b}. For the sample of embedded clusters provided by \citet{Lada2003}, \citet{Pfalzner2011} reports a clear radius-density correlation. Both the Kuhn et al. and Pfalzner et al. radius-density relationships are flatter that that expected from a pure isomorphic cluster expansion. The cause is uncertain but may arise from a non-uniform initial cluster state via inside-out star formation \citep{Pfalzner2011} or a cluster growth process involving hierarchical cluster mergers \citep{Kuhn2015b}.  We note that \citet{Getman2014b} and \citet{Getman2018} find a radial age gradient opposite from what is expected in the inside-out scenario of Pfalzner.  

Figure \ref{fig_pairs_plot_mystixsfincs} shows that the SFiNCs and MYStIX clusters occupy the same locii on the diagrams of $R_c -$absorption, $R_c - Age$, and $R_c - \alpha_{IRAC}$ (\S\ref{mv_presentation_subsection}).  The cluster radii are strongly correlated with $ME$, age, and SED slope, as well as a marginally significant correlation with $J-H$. Considering that the absorption and SED slope are excellent surrogates for age, these correlations give a strong empirical evidence of SFiNCs+MYStIX cluster expansion.  This echoes the MYStIX-only result of \citet{Kuhn2015b}.  
 
For the combined MYStIX+SFiNCs cluster sample, the $\log(R_c)-\log(Age)$ and $\log(R_c)-\alpha_{IRAC}$ regression fits (Figure~\ref{fig_pairs_plot_mystixsfincs}) indicate that the clusters with a core radius of $R_c \sim 0.08$~pc (0.9~pc) have a typical age of 1~Myr (3.5~Myr). Assuming that the projected half-mass radius for a cluster is roughly $4 \times R_c$ \citep{Kuhn2015b}, the MYStIX+SFiNCs clusters might grow from 0.3~pc to 3.6~pc half-mass radii over 2.5~Myr. 

It remains unclear what fraction of stars remains in bound clusters after cluster expansion and ultimately at the end of star formation \citep{Lada2003,Kruijssen2012}. A number of physical parameters are proposed to control the fraction of stars that remain bound in an expanding and re-virialized cluster, such as the star formation efficiency (SFE), gas removal time ($\tau _{gr}$), clusters's initial dynamical state, mass, and size \citep{Lada1984,Goodwin2009,Brinkmann2016}. For instance, the N-body simulations of relatively small ($<100$~M$_{\odot}$) and large ($>10^4$M$_{\odot}$) clusters predict that the bound fraction of the stellar cluster increases with a higher SFE, longer  $\tau_{gr}$, smaller initial cluster size, and larger initial cluster mass. A cluster with an initial sub-virial dynamical state might have higher chances to remain bound after gas expulsion. 

In the past, these physical parameters have not been well constrained by empirical data. Here, for the cloud-associated (i.e., fully or partially embedded in a cloud) SFiNCs+MYStIX clusters, we can employ our age estimates to provide empirical constraints on the gas removal timescale parameter $\tau_{gr}$.  Figures \ref{fig_ecdfs_rev_vs_cloud} and Table \ref{tbl_rev_cloud_univariate} show that for the SFiNCs and MYStIX cloud-associated samples, the median ages are of $1.2-1.4$~Myr, meaning that for at least half of the SFiNCs/MYStIX cloud-associated clusters $\tau _{gr} > 1.2$~Myr. For the sub-samples with reliable SED slope measurements, their median SED slope is of $\alpha_{IRAC}(SFiNCs+MYStIX) = -0.55$. Employing the SFiNCs$+$MYtIX regression line\footnote{This regression line (obtained using $R$'s {\it lmodel2} function) is analogous to the red line shown in Figure~\ref{fig_pairs_plot_mystixsfincs}, but treating variables symmetrically.} $\alpha_{IRAC} = 24.8(\pm3.7) - 4.2 (\pm 0.6) \times \log(Age_{JX})$, the corresponding age values are of $1.1$~Myr. Therefore, we conclude that for at least about half of the SFiNCs and MYStIX cloud-associated clusters, the gas removal timescale is longer than $1$~Myr. 

We thus emerge with two estimates of the cluster expansion timescale: core radii increase by an order of magnitude over $1-3.5$~Myr period, and the molecular gas is removed over $>1$~Myr. This timescale does not naturally emerge from theoretical models of star cluster evolution.  Some astrophysical calculations have assumed the gas expulsion and cluster expansion occurs within $\sim 10^5$ yr \citep[e.g.,][]{Lada1984, Baumgardt2007}.  These appear to be excluded by our findings.  Note however that our timescales are tied to $Age_{JX}$ estimates of \citet{Getman2014a} which in turn are calibrated to the pre-main sequence evolutionary tracks of \citet{Siess2000}.  If, for example,  we were to use the more recent tracks of \citet{Feiden2016} that treat magnetic pressure in the stellar interior, the timescales would be a factor of two or more longer than the estimates here.  See our study \citet{Richert2018} for the effects of evolutionary tracks on pre-main sequence timescales.

\subsection{SFiNCs Distributed Populations} \label{db_populations_subsection}

In the MYStIX SFRs, \citet{Kuhn2014} and \citet{Getman2014a} show that dispersed young stellar populations that surround the compact clusters and molecular clouds are ubiquitous on spatial scales of $5-20$~pc. In the Carina Complex, where a very large {\it Chandra} mosaic survey is available, the population size of the dispersed stellar sample is comparable to that of the clustered stellar sample \citep{Townsley2011,Feigelson2011}.  In the smaller MYStIX fields, typically $10-20$\% of the YSOs are in widely distributed populations \citep{Kuhn2015a}. \citet{Getman2014a} find that these distributed populations are nearly always have older ages than the principal MYStIX clusters. These results demonstrate that massive molecular clouds with current star formation have had (continuous or episodic) star formation for many millions of years in the past.

We find here that, for most SFiNCs SFRs (exceptions are Be~59, SFO~2, NGC~1333, ONC Flanking Fields, and NGC~7160) relatively rich populations of distributed young stars are identified (Figure \ref{fig_cluster_assignment_maps}). Across all 22 SFiNCs SFRs, the relative fractions of the observed (not intrinsic) clustered and distributed YSOs within the {\it Chandra} fields of view are 70\% and 26\%, respectively. (The remaining 4\% are unassigned).  In several fields (OMC~2-3, Serpens South, Sh~2-106, and LDN~1251B), the apparent numbers of the distributed YSOs exceed those of the clustered YSOs.

Figures \ref{fig_pairs_plot_sfincs} and \ref{fig_pairs_plot_mystixsfincs} show that across all SFiNCs SFRs the SFiNCs cluster sample (median age of 1.6~Myr) is significantly younger than the sample of SFiNCs distributed populations (median age of 2.3~Myr) with the difference at the $p<0.001$ significance level. These older dispersed populations are present on the spatial scales of $\gtrsim 2-3$~pc (in IC~348, Serpens Main, Cep~A, Cep~C, GGD~12-15, and Mon~R2) and $\gtrsim 5-7$~pc (in NGC~2068, IC~5146, Cep~OB3b, and RCW~120).  Similarly, in \S \ref{mv_presentation_subsection}, we find evidence for spatial gradients of the IRAC SED slope (a surrogate for age), with the slope decreasing from the cluster centers towards the peripheries of the SFiNCs SFRs. 

Two interpretative issues arise.  First, the reported star membership assignments in the clustered versus unclustered MYStIX/SFiNCs mixture model components are approximate and can be changed, to some degree, by varying the membership probability threshold and the limiting cluster size (\S \ref{membership_subsection}).  Second, the stars assigned to the SFiNCs/MYStIX ``unclustered'' components (and referred here as stars belonging to distributed populations) may have different astrophysical origins. Some may be ejected members of nearby clusters while others may belong to earlier generations of star formation.  In cases where the distributed stars are a large fraction of the total YSO population, the interpretation that SFiNCs SFRs have star formation enduring for millions of years seems reasonable. 

\section{Conclusions} \label{summary_sec}

The SFiNCs (Star Formation in Nearby Clouds) project is aimed at providing detailed study of the young stellar populations and star cluster formation in nearby 22 star forming regions.  The input lists of young stars, SFiNCs Probable Complex Members, were obtained in our previous study \citep{Getman2017}.  This study complements and extends our earlier MYStIX survey of richer, more distant clusters. Both efforts share consistent data sets, data reduction procedures, and cluster identification methods.  The latter are based on maximum likelihood parametric mixture models (\S \ref{model_section}) which differs from the nonparametric procedures used in most previous studies.  Appendix gives a detailed comparison of the methodological approaches.  

We identify 52 SFiNCs clusters and 19 unclustered stellar components in the 22 SFiNCs  star forming regions (\S \ref{clusters_subsection}).  The clusters include both recently formed embedded structures and somewhat older revealed clusters.  The unclustered components represent a distributed stellar populations (\S \ref{db_populations_subsection}). Model validation analyses show that the SFiNCs YSOs spatial distributions are generally well-fit with isothermal elliptical models (\S \ref{model_validation}); our parametric modeling procedures are thus self-consistent.  

Our parametric mixture model results include the number of significant clusters and a homogeneous suite of cluster physical parameters  (Tables \ref{tbl_cluster_morphology} and \ref{tbl_cluster_other_props}).  These include: cluster celestial location, core radius ($R_c$), ellipticity ($\epsilon$), observed number of YSO members ($N_{4,data}$), association to molecular clouds, interstellar absorption (based on $J-H$ color and X-ray median energy $ME$), age \citep[the $Age_{JX}$ estimate derived by][]{Getman2014a}, and a circumstellar disk measure ($\alpha_{IRAC}$).  Similar properties are compiled for the MYStIX clusters from our previous studies.  These cluster properties are mostly median values obtained on the largest available samples of YSOs derived in a uniform fashion. 

Together, these cluster characteristics can powerfully aid our understanding of clustered star formation.  The multivariate analyses of the univariate distributions and bivariate relationships of the merged SFiNCs and MYStIX cluster samples  are presented in \S\S \ref{ma_section} and \ref{cloud_subsection}. We emerge with the following main science results.
\begin{enumerate}

\item The SFiNCs sample is dominated ($75$\%) by cloud-associated clusters that are fully or partially embedded in molecular clouds. In contrast, the majority ($60$\%) of the MYStIX clusters are already emerged from their natal clouds.

\item The cloud-associated clusters are found to be on average younger and more absorbed than the revealed clusters (\S \ref{cloud_subsection}).  This was previously reported for the MYStIX-only sample by \citet{Getman2014a}. 

\item The SFiNCs cloud-associated clusters are on average more elongated than the revealed SFiNCs/MYStIX and cloud-associated MYStIX clusters. Their major axes are generally aligned with the host molecular filaments. Therefore their high ellipticity is probably inherited from the morphology of their parental molecular filaments (\S \ref{elongation_subsection}). 

\item The cloud-associated clusters are considerably smaller than the revealed clusters. In part, this is a consequence of the effect of cluster expansion that is clearly evident from the strong $R_c - Age$, $R_c -$absorption, and $R_c - \alpha_{IRAC}$ correlations for the combined MYStIX+SFiNCs cluster sample (\S\S \ref{sizes_subsection} and  \ref{expansion_subsection}). Core radii increase dramatically from $\sim0.08$ to $\sim0.9$~pc over the age range $1-3.5$~Myr. These confirm and extend previously reported MYStIX-only results by \citet{Kuhn2015b}.   

\item For at least about half of the SFiNCs and MYStIX cloud-associated clusters, the estimated gas removal timescale responsible for cluster expansion is longer than $\sim 1$~Myr (assuming \citet{Siess2000} timescale; \S \ref{expansion_subsection}). This gives an important constraint on early star cluster evolution.   

\item For the majority of the SFiNCs SFRs, relatively rich populations of distributed YSOs are identified. These probably  represent early generations of star formation (\S \ref{db_populations_subsection}). For the MYStIX SFRs, similar findings on the presence of older distributed stellar populations were previously reported \citep{Feigelson2013, Kuhn2014,Getman2014a,Kuhn2015a}.

\end{enumerate}

A number of studies using the combined SFiNCs and MYStIX cluster results are planned: 
\begin{itemize}
\item[] \citet{Richert2018} reexamines the longevity distribution of inner protoplanetary disks using this large sample of young clusters, updating results of \citet{Haisch2001} and others based on smaller, less homogenous cluster samples.
\item[] \citet{Getman2018} use the $Age_{JX}$ chronometer to show that age spreads are common within young stellar clusters with a particular spatial age gradient: stars in cluster cores appear younger (formed later) than in cluster peripheries.  This extends the earlier result on only two clusters by \citet{Getman2014b}.  
\item[] A study is planned that will correct the observed cluster populations ($N_{4,data}$) to intrinsic populations for the full Initial Mass Function, following the procedure of \citet{Kuhn2015a}. This will allow a number of astrophysical issues concerning cluster formation to be addressed with more assurance than possible in the present paper. 
\item[] Additional possible efforts include dynamical modeling of SFiNCs-like clusters (analogous to the simulation studies of \citet{Bate2009,Bate2012}), measurements of gas-to-dust ratios in the molecular clouds following \citet{Getman2017}, and others.
\end{itemize}

\section*{Acknowledgements}

We thank the anonymous referee for helpful comments. The MYStIX project is now supported by the {\it Chandra} archive grant AR7-18002X. The SFiNCs project is supported at Penn State by NASA grant NNX15AF42G, {\it Chandra} GO grant SAO AR5-16001X, {\it Chandra} GO grant GO2-13012X, {\it Chandra} GO grant GO3-14004X, {\it Chandra} GO grant GO4-15013X, and the {\it Chandra} ACIS Team contract SV474018 (G. Garmire \& L. Townsley, Principal Investigators), issued by the {\it Chandra} X-ray Center, which is operated by the Smithsonian Astrophysical Observatory for and on behalf of NASA under contract NAS8-03060. The Guaranteed Time Observations (GTO) data used here were selected by the ACIS Instrument Principal Investigator, Gordon P. Garmire, of the Huntingdon Institute for X-ray Astronomy, LLC, which is under contract to the Smithsonian Astrophysical Observatory; Contract SV2-82024. This research made use of data products from the {\it Chandra} Data Archive and the {\it Spitzer Space Telescope}, which is operated by the Jet Propulsion Laboratory (California Institute of Technology) under a contract with NASA. This research used data products from the Two Micron All Sky Survey, which is a joint project of the University of Massachusetts and the Infrared Processing and Analysis Center/California Institute of Technology, funded by the National Aeronautics and Space Administration and the National Science Foundation. This research has also made use of NASA's Astrophysics Data System Bibliographic Services and SAOImage DS9 software developed by Smithsonian Astrophysical Observatory.

%%%%%%%%%%%%%%%%%%%%%%%%%%%%%%%%%%%%%%%%%%%%%%%%%%



%\bibliographystyle{mnras}


%%%%%%%%%%%%%%%%%%%%%%%%%%%%%%%%%%%%%%%%%%%%%%%%%%
\clearpage
\newpage

%%%%%%%%%%%%%%%%%%%%%%%%%%%%%%%%%%%%%%%%%%%%%%%%%%
%%%%%%%%%%%%%%%%% APPENDICES %%%%%%%%%%%%%%%%%%%%%

\appendix
\section{Comparison of Clusters Identified with Mixture Model and MST Analysis Methods} \label{sec_sfincs_vs_g09}

We compare our isothermal ellipsoid mixture model analysis (MMA) with the cluster identification procedure of  \citet[][hereafter G09]{Gutermuth2009} based on the Minimum Spanning Tree (MST).  Following a brief review of statistical issues (\S\ref{sec.stat.bkgd}), we apply the two methods to  SFiNCs ``flattened'' YSO samples (\S\S \ref{sec_mst_to_sfincs} and \ref{sec_mma_mst_sfincs}) and to a simulated multi-cluster region (\S\ref{sec_mst_simulations}). Detail comparisons of the methods for individual SFiNCs regions are provided in \S\ref{sec_individual_subclusters}. 

\subsection{Statistical Background} \label{sec.stat.bkgd}

The use of the pruned MST for cluster identification is a well-known technique of nonparametric clustering for multivariate data.  Though operationally based on the MST, it has been reinvented many times in different forms since the 1950s \citep{Rohlf1982}.  It is mathematically identical to the astronomers'  friends-of-friends percolation algorithm \citep{Turner1976} and the statisticians' single-linkage agglomerative clustering algorithm.  It is important to recognize that the method has several limitations:
\begin{enumerate}  
\item Like all nonparametric clustering methods, there is no obvious quantity (such as likelihood) to maximize and thus no clear way to choose the number of clusters present in an objective and reproducible manner. 
\item There are no theorems to guide the choice of algorithm.  There is no criterion to choose between different hierarchical clustering procedures (single linkage, average linkage, complete linkage, and Ward's distances are most commonly used) or between agglomerative hierarchical and other approaches (such as $k$-means and other partitioning procedures).   
\item When clustering methods are compared in simulation, single linkage agglomeration produces dendrograms and clusters that can differ widely from all other common methods \citep[e.g.][]{Jain2004, Izenman2008}.  
\item Single linkage clustering has the particular problem of `chaining' together unrelated clusters in the presence of noise.  
\end{enumerate} 
The strong limitations of this approach are summarized in the most widely read statistical textbook in the field by \citet{Everitt11} (see also \citet{Aggarwal14}): 
\begin{quote}
``It has to be recognized that hierarchical clustering methods may give very different results on the same data, and empirical studies are rarely conclusive. ... Single linkage, which has satisfactory mathematical properties and is also easy to program and apply to large datasets, tends to be less satisfactory than other methods because of `chaining'; this is the phenomenon in which separate clusters with `noise' point in between tend to be joined together."
\end{quote}

\subsection{Application of MST to SFiNCs Fields} \label{sec_mst_to_sfincs}

\begin{figure*}
\centering
\includegraphics[angle=0.,width=6.5in]{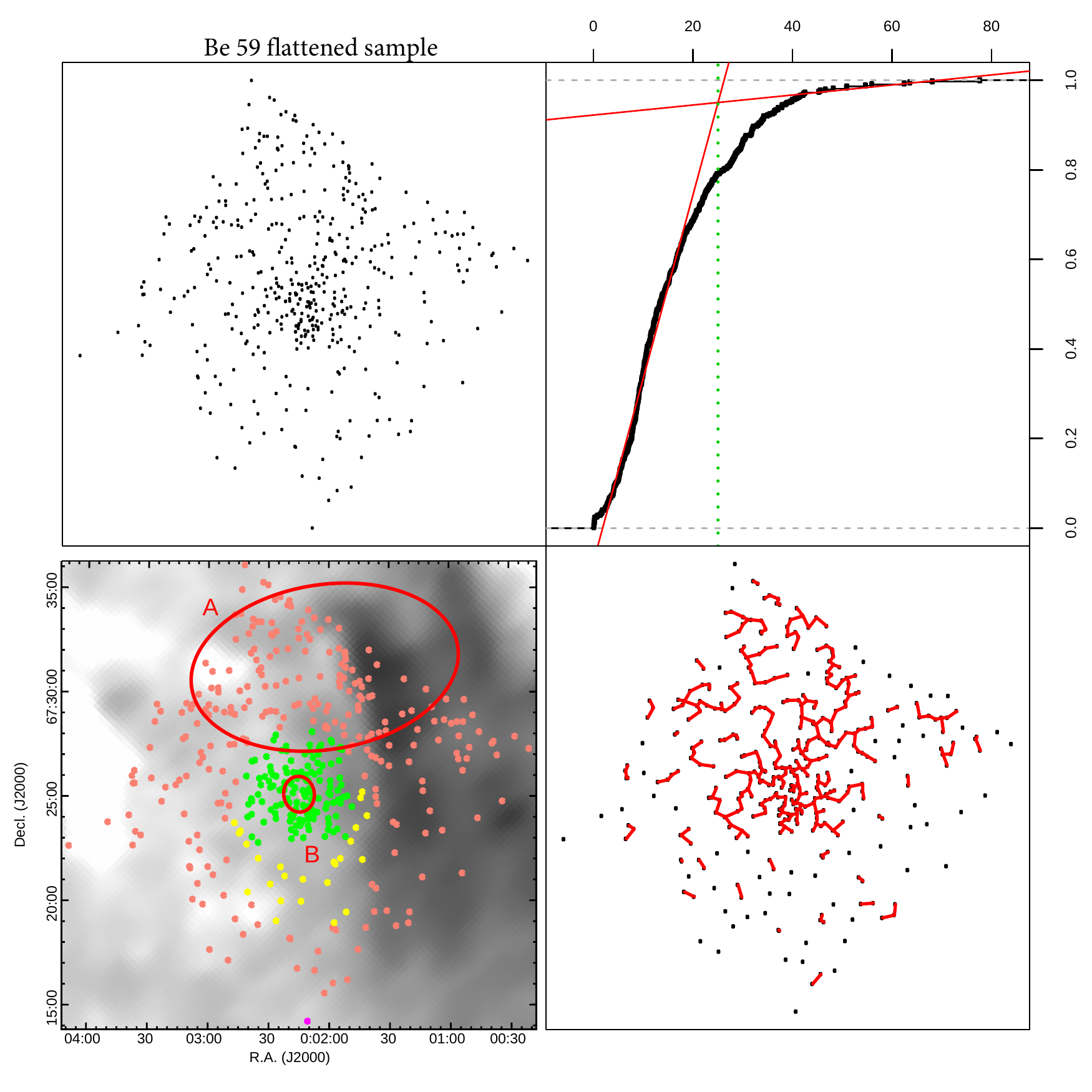}
\caption{Comparison of the SFiNCs cluster identification results between our mixture model and MST methods. The Be 59 SFR is shown here; figures for the full SFiNCs sample are given in the Supplementary Materials.  The upper left panel shows the spatial distribution of SPCMs from the ``flattened'' sample. The upper right panel shows the cumulative distribution function of the resulting MST branch lengths. The lengths are given in physical (x,y) scales proportional to parsec scales. The green dotted line indicates the threshold for pruning the MST branch lengths to isolate local stellar groups/clusters, following the procedure of \citet{Gutermuth2009}. The lower left panel shows the results of our mixture model analysis, similar to that of Figure Set \ref{fig_cluster_assignment_maps}. The SPCM ``flattened'' sample with cluster membership is coded by color: clusters A (salmon), B (green), C (blue), D (cyan), and E (pink), unclustered members (magenta), and unassigned stars (yellow). The elliptical contours (red) show the core radii of the isothermal cluster structures. SPCMs are superimposed on the FIR Herschel-SPIRE or AKARI-FIS images tracing the locations of molecular clouds. The lower right panel shows the resulting pruned MST tree (red). \label{fig_mma_vs_mst}}
\end{figure*}

The MST analysis was performed using the physical (x,y) coordinates of the SFiNCs YSOs that are proportional to physical parsec scales, projected on the sky.  Briefly stated, the MST procedure: constructs the unique Minimal Spanning Tree for the stellar two-dimensional spatial distribution; plots the cumulative distribution function of the MST branches; chooses a critical branch length at the intersection of linear regressions to the upper and lower portions of this function; removes all tree branches longer than this critical length; defines cluster members as contiguous linked data points.   These steps are visualized for the SFiNCs stellar distributions in the right-hand panels of Figure \ref{fig_mma_vs_mst} and its associated Supplementary Materials.  The MMA result is summarized in the lower-left panel. For many regions, the SFiNCs MST results based on the X-ray/IR datasets resemble the MST results of G09 that are based on purely IR stellar samples. Detailed comparison of the MMA and MST results for individual SFiNCs regions (depicted in Figure \ref{fig_mma_vs_mst}) are further provided in \S\ref{sec_individual_subclusters}.   

\subsection{MMA and G09-MST Comparison for SFiNCs Fields} \label{sec_mma_mst_sfincs}

G09 used {\it Spitzer} data to provide a homogeneous set of disk-bearing YSO populations across 36 nearby star forming regions. Using the MST method, they identify 39 cluster cores and derive some of their basic properties, such as cluster position, size, aspect ratio, stellar density, and extinction. Fourteen G09 regions are in common with SFiNCs. For these clusters, we compare the cluster sizes and stellar densities inferred by our MMA procedure to those derived in G09 in Table \ref{tbl_sfincs_vs_g09}. 

\begin{figure*}
\centering
\includegraphics[angle=0.,width=5.6in]{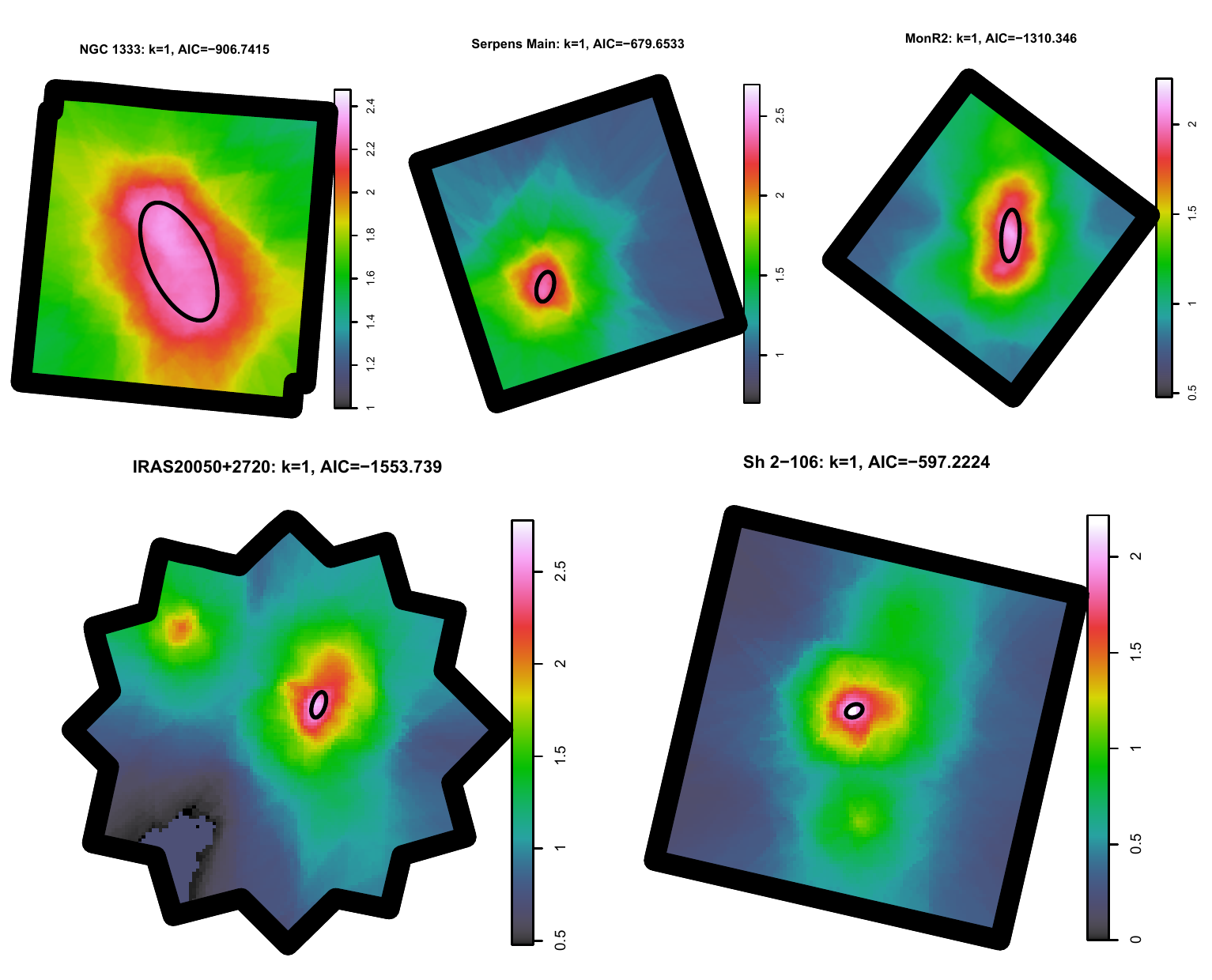}
\caption{For the cases where a single G09 cluster spans multiple SFiNCs clusters (NGC 1333, Serpens Main, Mon R2, IRAS 20050+2720, Sh 2-106), the SFiNCs YSO spatial distribution is re-fitted with a single-cluster component model ($k=1$). The panels show smoothed projected stellar surface densities of the ``flattened'' SPCM sample with a color-bar in units of observed stars per pc$^{-2}$ (on a logarithmic scale). The figure title gives the name of a SFiNCs SFR, the number of identified clusters, and the final value of AIC. The core radii ($R_c$) of the identified isothermal SFiNCs clusters are marked by the black ellipses. The sizes and apparent stellar densities for these clusters are listed in Table \ref{tbl_sfincs_vs_g09}. \label{fig_new_fits}}
\end{figure*}

In five SFiNCs SFRs (NGC 1333, Mon R2, Serpens Main, IRAS 20050+2720, Sh 2-106), our mixture model procedure  find multiple clusters within the locations of single G09 clusters. To facilitate the comparison with MST method, the SFiNCs YSO spatial distributions in these SFRs are re-fitted here with mixture models composed of a single cluster component ($k=1$) plus a single unclustered component. These fits are shown in Figure \ref{fig_new_fits}. Notice that the systematically higher AIC values, compared to the original multi-cluster fits given in Figure \ref{fig_cluster_identification}, indicate poorer fits.

\begin{table*}\small
\centering
 \begin{minipage}{180mm}
 \caption{Clusters in Common Between SFiNCs and Gutermuth et al. Column 1: Star forming region. Column 2: SFiNCs cluster component name. Columns 3-5: Cluster size, observed number of YSOs, and the mean stellar density derived from SFiNCs model. The reported size is a cluster radius four times the size of the cluster core ($R_c$) given in Table \ref{tbl_cluster_morphology}. The mean density is estimated as $N_{4,data}/\pi (4 R_c)^2$, where the total number of all SFiNCs YSOs observed within the cluster ($N_{4,data}$) is taken from Table \ref{tbl_cluster_other_props}.  Column 6: Cluster name from G09. Columns 7 and 8: Cluster radius and mean stellar density inferred from the analysis of G09 (see their Table 8). For SFO 2, NGC 1333, IC 348, LkH$\alpha$ 101, Serpens Main, Sh 2-106, Cep A, and Cep C the G09 values are adjusted to match the distances adopted in SFiNCs; the distance values adopted in SFiNCs and G09 can be found in Table \ref{tbl_cluster_sample} here and Table 1 of G09, respectively.}
 \label{tbl_sfincs_vs_g09}
 \begin{tabular}{@{\extracolsep{4pt}}@{\vline}c@{}c@{}c@{}c@{}c@{}c@{}c@{}c@{}c@{\vline}}
\cline{1-9}
&&&&&&&&\\
SFR & \multicolumn{4}{c}{SFiNCs} && \multicolumn{3}{c@{\vline}}{Gutermuth et al.}\\
\cline{2-5} \cline{7-9}
&&&&&&&&\\
&
~Cluster~ &  
~Size~ &
~Number~ &
~$\Sigma_{mean}$~~~~~ &&
~Name~ &
~$R_{hull}$~ &  
~$\sigma_{mean}$~~~\\
& & 
(pc) &
(stars) &
(pc$^{-2}$) && 
&
(pc) & 
(pc$^{-2}$)\\
(1) & (2) & (3) & (4) & (5) && (6) & (7) & (8) \\
\cline{1-9}
&&&&&&&&\\
SFO 2 & A & 0.32            &66~~ & 205~~ && Core & 0.34 &   89~~\\
NGC 1333 & AB & 0.76        &176~~ & 97~~ && Core & 0.37 &  219~~\\
IC 348 & A & 0.16           &28~~ & 348~~ && Core-2 & 0.23 &  148~~\\
IC 348 & B & 0.80           &280~~ & 139~~ && Core-1 & 0.39 &  114~~\\
LkH$\alpha$ 101 & A & 0.84  &182~~ &  82~~ && Core & 0.67 &   40~~\\
Mon R2 & ABC & 1.12         &385~~ & 98~~ && Core & 0.73 &   79~~\\
GGD 12-15 & A & 0.56        &108~~ & 110~~ && Core & 0.50 &   99~~\\
Serpens Main & AB & 0.36    &66~~ & 162~~ && Core & 0.35 &  141~~\\
~IRAS 20050+2720~ & CD & 0.44 &155~~ & 255~~ && Core-1 & 0.36 &  200~~\\
Sh 2-106 & BD & 0.60        &63~~ &  56~~ && Core & 0.79 &   18~~\\
IC 5146 & B & 0.68          &142~~ &  98~~ && Core & 0.68 &   66~~\\
Cep A & A & 0.84            &172~~ &  78~~ && Core-1 & 0.53 &   60~~\\
Cep C & A & 0.72            &86~~ &  53~~ && Core-1 & 0.40 &   77~~\\
Cep C & B & 0.08            &4~~ & 199~~ && Core-2 & 0.20 &   90~~\\
&&&&&&&&\\ 
\cline{1-9} 
\end{tabular}
\end{minipage}
\end{table*}

\begin{figure*}
\centering
\includegraphics[angle=0.,width=5.5in]{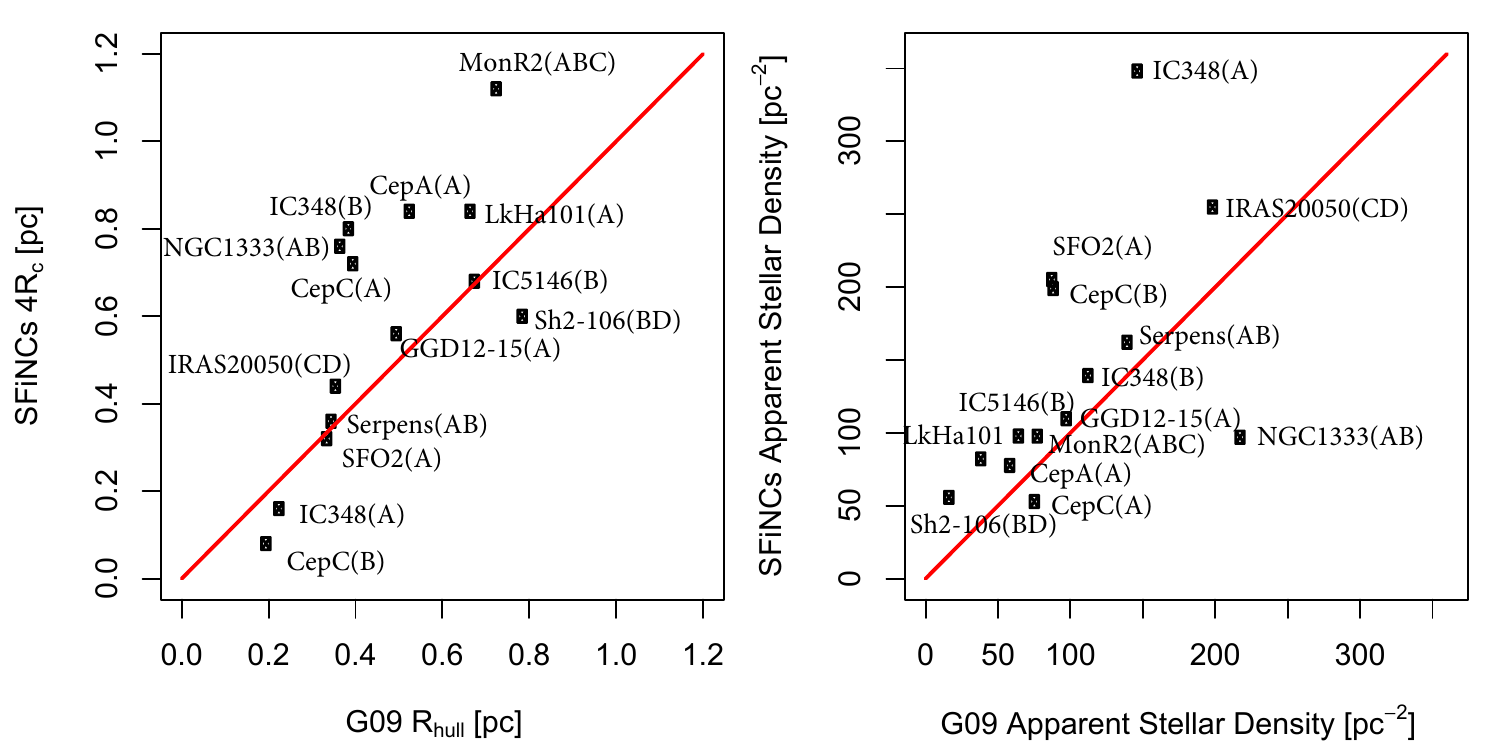}
\caption{Comparison of cluster sizes (left panel) and mean stellar densities (right panel) between SFiNCs and G09. SFiNCs properties are derived here based on the mixture model analysis of X-ray/IR data. G09 properties are derived in \citet{Gutermuth2009} based on the application of the MST analysis to IR data. The clusters are shown as black points with cluster name labels. Unity lines are shown in red. \label{fig_sfincs_vs_g09}}
\end{figure*}

We summarize two of the cluster properties, cluster size and apparent stellar surface density, found using the SFiNCs and G09 methods in Figure \ref{fig_sfincs_vs_g09}. The two methods give correlated values but with wide scatter. 

Some of the discrepancies may have identifiable causes. Mon R2 and Cep A are subject to bright mid-IR nebular background emission; the smaller cluster sizes inferred by G09 may arise from an IR catalog bias.  In IC 348, G09 split the main cluster (SFiNCs cluster B) into two subclusters, Core 1 and Core 3, which leads to a truncated size of the main cluster.  For NGC 1333, the MST fragmentation of the northern region results in a smaller cluster size.  In the stellar density plot, the SFiNCs stellar densities are systematically higher than those of G09; this is expected due to the inclusion of X-ray selected disk-free YSOs as well as IR excess YSOs.  

\subsection{Simulation of mixture model and MST procedures} \label{sec_mst_simulations}

\begin{figure*}
\centering
\includegraphics[angle=0.,width=6.5in]{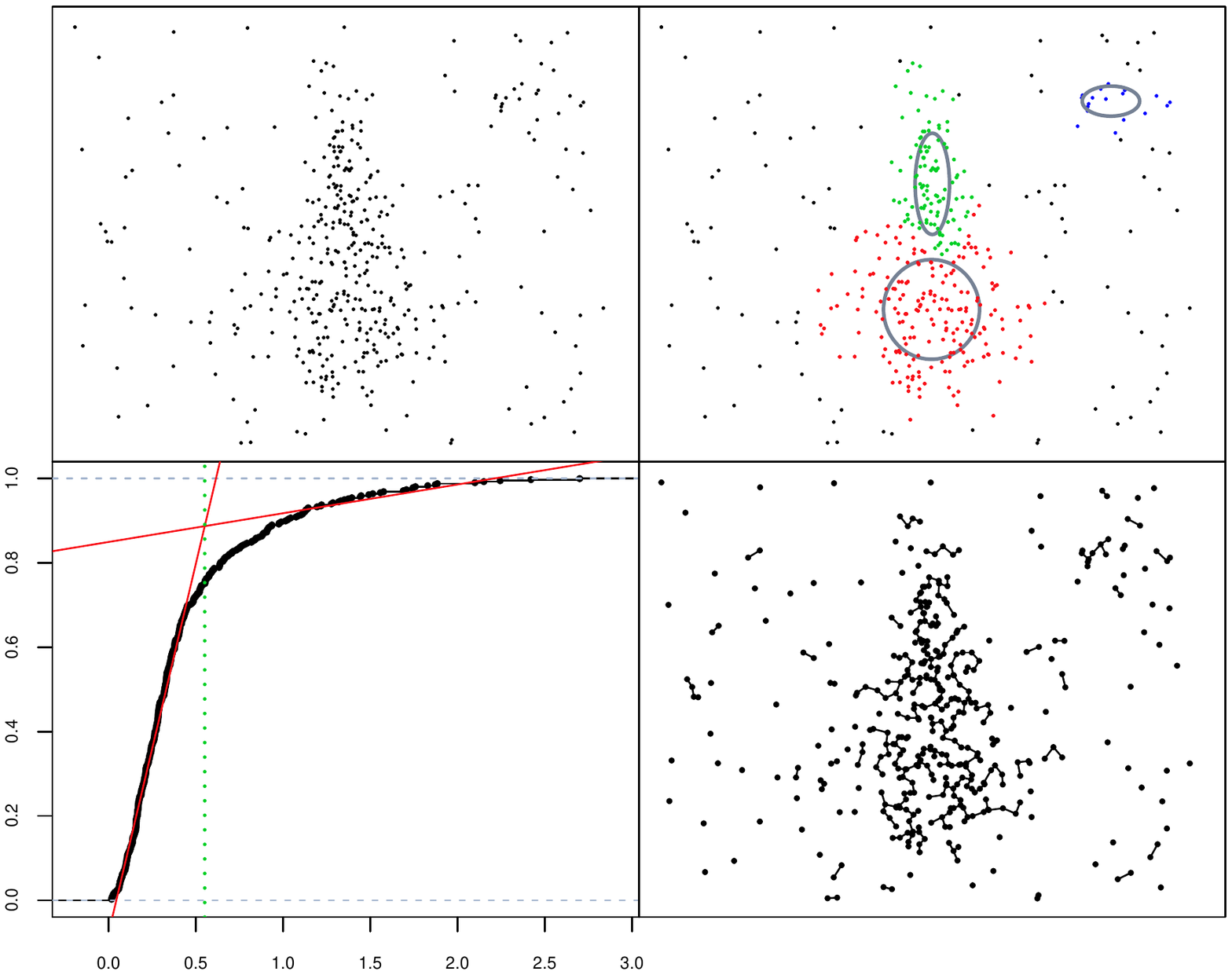}
\caption{Simulated application of the MST method. (Upper left panel) The spatial distribution of three simulated Gaussian-shaped clusters merged with an unclustered component. (Upper right panel) Clustering result from the application of the Gaussian mixture model. Cluster membership is coded by color: the main round cluster of 200 stars (red), the secondary highly elongated cluster with 100 stars (green), and a sparser cluster of 20 stars (blue). (Lower left panel) The cumulative distribution function of the MST branch lengths based on point locations. The green dotted line indicates the threshold for pruning the MST branch lengths to isolate local stellar groups/clusters based on two linear regressions, following the procedure of \citet{Gutermuth2009}. (Lower right panel) Resulting pruned MST tree. \label{fig_mst_simulations}}  
\end{figure*}

To illustrate the difficulties of MST-derived clusters and the effectiveness of mixture modeling under some circumstances, we conducted a single series of simulations.  It is a Gaussian mixture model (GMM) with a central round cluster of 200 stars, an overlapping elongated cluster with $4:1$ axis ratio of 100 stars, a sparser well-separated  cluster of 20 stars, and 100 uniformly distributed unclustered stars. These are placed in a dimensionless $10 \times 10$ square window.  Figure~\ref{fig_mst_simulations} (upper left panel) shows a typical example of the simulated star distribution.   

A standard maximum likelihood GMM fitting procedure, {\it R} function  {\it Mclust} \citep{Fraley2002},  fits maximum likelihood Gaussian clusters with $1 < k < 10$, choosing the best-fit $k$ with the Bayesian Information Criterion.  The methods has no user-supplied parameters.  For an ensemble of 100 simulations, the GMM separates the central and elongated clusters in all of the simulations, and identified sparse 20-point cluster in 3/4 of the simulations.  The upper right panel of Figure~\ref{fig_mst_simulations} gives a typical example.  This failure to uniformly capture sparse clusters is consistent with the more extensive simulations of \citet{Lee2009}. In no case is the main cluster subdivided into clusters.  

Each simulated star distribution was then subject to the MST procedures described by G09\footnote{This analysis was performed using {\it R}'s {\it spatstat} and {\it spatgraphs} packages \citep{Baddeley2015,Rajala2015}. Useful functions include {\it ppp},  {\it psp}, {\it edgeLengths} and {\it spatgraph}.}.  The branch length distribution of the MST of the typical example is displayed in the lower left panel of Figure~\ref{fig_mst_simulations}. Two user-supplied parameters are needed.  First, one must define how the linear regressions of the MST branch length distribution are calculated.  Here we choose the lower 40\% of the distribution to fit one line and the upper 10\%  to fit the other line (red lines).  The lines intersect around branch length $0.5-0.6$ (green line) which serves as the threshold for pruning the MST.  A third user-supplied parameter is then needed to exclude small fragments of two or more points.  We choose a threshold that  clusters must have at least 10 members.  The clusters can then be identified. 

The MST-based procedure was much less successful than the GMM procedure in recovering the simulated clusters. The elongated cluster was discriminated from the main cluster in only 5\% of the cases, and the sparse separated cluster was recovered in 25\% of the cases.  In a third of the simulations, the main cluster was erroneously subdivided into subclusters.  Different choices of the user-supplied parameters would not give much better solutions.  A longer branch threshold improves the detection of sparse clusters but worsens the ability to discriminate close or overlapping clusters. A smaller cluster membership threshold improves the detection of the sparse cluster but increases the erroneous fractionation of the main cluster. In real astronomical situations, there is insufficient knowledge to tune the user-supplied parameters.

The simulation study here is limited and perhaps tuned to the success of GMMs as the clusters were assumed to have Gaussian shapes. In the SFiNCs analysis (\S\ref{model_section}), we use isothermal ellipsoids rather than multivariate Gaussians as the model, and then validate that this model accurately fits the resulting clusters (\S\ref{model_validation}).  The MST method shows similar problems of chaining and fragmentation in the real SFiNCs datasets (see \S\ref{sec_individual_subclusters} below) as seen in the simulations here.  Overall, the simulation here shows that that nonparametric clustering procedures like the pruned MST can be substantially less successful than parametric mixture model procedures in recovering star clusters in patterns that resemble the spatial distribution of SFiNCs membership samples (\S \ref{sec_individual_subclusters}).

\section{Comments on Individual SFiNCs Regions and Clusters} \label{sec_individual_subclusters}

Here we provide information on the association of the individual SFiNCs clusters found with the mixture model analysis (MMA) with known molecular structures, previously published stellar clusters, previous searches for stellar clusters, and results of the MST analysis described in \S\ref{sec_mst_to_sfincs} and shown in Figure \ref{fig_mma_vs_mst}.   

{\it Be 59 (NGC 7822)} is the principal cluster responsible for the ionization of the nearby, $\sim 40$~pc diameter Cepheus Loop bubble \citep[][and references therein]{Kun2008}. The western part of Be 59 is bounded by the giant molecular cloud associated with the Cepheus Loop shell (Figure Set \ref{fig_cluster_assignment_maps}). Our MMA identifies two clusters (A and B): B is a rich dense central cluster and A is an excess of stars in the halo of B. Such a cluster morphology with a cluster core and an asymmetric halo, observed within the ACIS-I field on a spatial scale of a few parsecs, is reminiscent of a number of MYStIX SFRs \citep{Kuhn2014}. Low data-model residual values of $<10$\% across both clusters provide evidence for a good model fit to the data (Figure Set \ref{fig_cluster_identification}).

The MST method picks out the main rich cluster B, but treats B as an elongated structure chained to the densest part of the cluster A. MST also subdivides the sparser part of the cluster A into four sparse ($N\geq10$) groups (Figure \ref{fig_mma_vs_mst}).

{\it SFO 2 (BRC 2, S 171, IRAS 00013+6817)} As part of the aforementioned Cepheus Loop bubble, about 1$\degr$ north of Be 59 lies the bright-rimmed cloud SFO 2, surrounded by the ionized rim NGC 7822 facing Be 59 \citep[][and references therein]{Kun2008}. MMA identifies a single cluster A located at the tip of the bright-rimmed cloud.  The data-model residuals are below $15$\% at the center of the cluster (Figure \ref{fig_cluster_identification}). There is a small residual spot at the south-eastern part of the field; but this is associated with only a few points and could be a random fluctuation.  G09 applied the MST analysis to {\it Spitzer} YSO sample of this region (their target S~171) and obtained a similar result of a single cluster.  
  
{\it NGC 1333} is an SFR within the Perseus molecular cloud complex noted for its large population of protostars and young stellar outflows \citep[][and references therein]{Luhman2016}. MMA identifies two clusters (A and B), each corresponding to different filamentary parts of the molecular cloud (Figure \ref{fig_cluster_assignment_maps}). Low residuals of $<3$\% and $<10$\% at the cores and halos of the clusters, respectively, indicate good model fits to the data (Figure Set \ref{fig_cluster_identification}). 

Both the SFiNCs MST (Figure Set \ref{fig_mma_vs_mst}) and the MST analysis of G09 chain A and B into a single cluster.  However, the SFiNCs clusters A and B coincide with the double-cluster identified earlier by \citet{Gutermuth2008} who applied a nearest neighbor algorithm to their {\it Spitzer} YSO catalog. No kinematic differences between the stellar populations of the two clusters are found in the INfrared Spectra of Young Nebulous Clusters (IN-SYNC) project by \citet{Foster2015}. 

{\it IC 348} is the richest SFR in the nearby Perseus molecular cloud complex \citep[][and references therein]{Luhman2016}. MMA identifies two clusters.  The small, heavily absorbed cluster A is associated with the dense part of a molecular filament, while the main rich and lightly absorbed cluster B lies projected against the area with dispersed molecular material (Figure \ref{fig_cluster_assignment_maps}).  Model residuals across the clusters are $<10$\% (Figure \ref{fig_cluster_identification}). 

The SFiNCs MST analysis fragments the main cluster B into three components: two sparse northern groups each with $\sim 10$ stars, and a main cluster.  The MST analysis by G09 identifies the small embedded SFiNCs cluster A, but fragments cluster B into two pieces: a main rich scluster and a secondary southern sparser cluster with $\ga20$ members.  

{\it LkH$\alpha$ 101} is a rich SFR associated with the dense molecular filament L1482, part of the giant California molecular cloud \citep{Andrews2008,Lada2009}. SFiNCs MMA finds one single rich cluster A that lies projected against a dense molecular structure (Figure \ref{fig_cluster_assignment_maps}). The data-model residuals are $<10$\% (Figure \ref{fig_cluster_identification}). 

SFiNCs MST fragments the cluster A into three parts: one main cluster and two secondary sparser ($N \ga 10$) groups to the west of the main cluster (Figure \ref{fig_mma_vs_mst}). However,  the MST analysis of G09 treats the cluster A as a single cluster, consistent with the MMA result.

{\it NGC 2068-2071} NGC 2068 and NGC 2071 are SFRs associated with the northern part of the Orion B molecular cloud, also known as L1630N \citep[][and references therein]{Spezzi2015}. MMA identifies four clusters: the richest B cluster lies projected against a dispersed cloud structure in NGC 2068; the C and D clusters are associated with clumpy and filamentary molecular structures in NGC 2071; and the A cluster is associated with filamentary molecular structure in the southern part of L1630N, termed by Spezzi et al. as the HH 24-26 area (Figure \ref{fig_cluster_assignment_maps}). The data-model residuals are $\la 10$\% across the B, C, and D clusters and about 20\% around the A cluster (Figure Set \ref{fig_cluster_identification}).

SFiNCs MST chains clusters C and D together. Consistent with MMA, MST identifies the cluster B as a single cluster, but fragments the elongated cluster A into two sparse groups (Figure Set \ref{fig_mma_vs_mst}).

\citet{Spezzi2015} applied a nearest neighbor algorithm to their VISTA$+${\it Spitzer} YSO catalog of L1630N. They define NGC 2071 (SFiNCs clusters C and D) and NGC 2068 (SFiNCs cluster B) as loose stellar clusters, and HH 24-26 (SFiNCs cluster A) as a loose stellar group. \citet{Megeath2016} applied the `N10' $k$-nearest neighbor surface density estimator to the Spitzer YSO catalog of NGC 2068-2071. They identify two main clusters: their northern cluster is a composite of the SFiNCs clusters C and D; their southern cluster coincides with SFiNCs cluster B. Their surface density map shows a density increase at the location of HH 24-26 (SFiNCs cluster A) but it is not labeled as a cluster. 

{\it OMC 2-3} is associated with a part of the Orion A molecular filament that extends northward from the OMC 1/Orion Nebula region \citep{Peterson2008}. MMA identifies four clusters. Clusters A, B, and C are positioned and elongated along the main molecular filament. Cluster D represents a group of several YSOs located off the filament. The data-model residuals are $\la 5$\% around the clusters and $\sim 20$\% at the outskirts of the {\it Chandra} field. 

SFiNCs MST identifies all four SFiNCs clusters. Compared to MMA, MST chains several additional YSOs to clusters C and D. Based on their N10 surface density estimator of the Orion {\it Spitzer} YSO catalog, \citet{Megeath2016} do not report any stellar clusters/groups at the location of OMC 2-3.

{\it ONC Flank N (NGC 1977)} Located at the northern end of the Orion A molecular cloud (right to the north of OMC 2-3), this is an HII bubble ionized by a few early B-type stars \citep{Peterson2008}. Our MMA analysis identifies a rich, un-obscured cluster A that lies projected against the HII region. The data-model residuals are $\la 10$\% across the cluster and about 20\% at the outskirts of the {\it Chandra} field. 

The SFiNCs MST analysis is consistent with SFiNCs MMA at identifying this single cluster. Based on their N10 surface density estimator of the Orion Spitzer YSO catalog \citet{Megeath2016} do not report any stellar clusters at the location of NGC 1977.

{\it ONC Flank S} Located in the OMC 4 part of the Orion A molecular filament that extends southward from the OMC 1/Orion Nebula region \citep{{ODell2008}}. MMA analysis identifies a single rich, unobscured, and elongated cluster A that lies projected against the OMC 4 filament. The data-model residuals are $\la 5$\% across the cluster and  $<20$\% at the outskirts of the {\it Chandra} field. 

The SFiNCs MST analysis fragments the SFiNCs cluster A into three structures: a main ($N>100$ YSOs) and two secondary each with several-dozen YSOs. In addition, several small groups of $N \sim 10-20$ YSOs are identified in the outer region by MST. Based on their N10 surface density estimator of the Orion {\it Spitzer} YSO catalog \citet{Megeath2016} do not report any stellar clusters/groups at the location of ONC Flank S.

{\it Mon R2} is a rich SFR associated with the Mon R2 molecular core that is part of the giant Monoceros R2 molecular cloud complex \citep[][and references therein]{Pokhrel2016}. MMA identifies three star clusters. The extremely absorbed cluster B is associated with the central part of the molecular clump, and the less obscured clusters A and C lie projected against the northern and southern edges of the clump, respectively (Figure \ref{fig_cluster_assignment_maps}). The data-model residuals are $<$5\% across the clusters (Figure \ref{fig_cluster_identification}). 

Both the SFiNCs MST (Figure  \ref{fig_mma_vs_mst}) and the MST analysis of G09 chain the A, B, and C clusters into a single structure. The YSO catalog of G09  misses numerous X-ray selected YSOs at the center of the field due to the presence of high nebular MIR emission \citep[Figure 6 in][]{Getman2017}.

{\it GGD~12-15} is part of the giant Monoceros R2 molecular cloud complex located to the east of the Mon R2 SFR \citep[][and references therein]{Pokhrel2016}. MMA finds a single rich cluster A associated with a molecular clump (Figure Set \ref{fig_cluster_assignment_maps}).  The data-model residuals range from less than 5\% at the center to $<25$\% at the halo of the cluster (Figure Set \ref{fig_cluster_identification}).  

The SFiNCs MST analysis (Figure \ref{fig_mma_vs_mst}) and the MST analysis of G09 are consistent with SFiNCs MMA at identifying this single cluster A.

{\it RCW 120} is a nearby, $\sim 4$~pc diameter HII bubble \citep[][and references therein]{Figueira2017}. MMA identifies four clusters: the primary ionizing cluster B surrounded by a dusty shell, and the secondary clusters A, C, and D associated with the filamentary and clumpy parts of the dusty shell (Figure \ref{fig_cluster_assignment_maps}). The data-model residuals do not exceed 10\% at the centers of the clusters (Figure \ref{fig_cluster_identification}). However, the high ($30-50$\%) negative residuals (model $>$ data) in the halos of the embedded clusters A and C indicate poor fits with the isothermal ellipsoid models for these clusters. In addition, a few spots with positive residuals of $50-100$\% at the locations of clumpy molecular structures suggest that a few independent, possibly embedded, sparse ($N = 3-7$) stellar groups might have been missed by MMA.

SFiNCs MST chains the clusters A and B into a single cluster (Figure \ref{fig_mma_vs_mst}). MST is consistent with MMA at identifying the cluster C. MST breaks the cluster D into two sparse groups, each consisting of several YSOs. MST shows groups of $N=3-7$ YSOs at the aforementioned spots with the positive MMA residuals.

{\it Serpens Main} is one of a few young, active SFRs in the nearby Serpens/Aquila Molecular Complex \citep[][and references therein]{OrtizLeon2017}. MMA identifies two clusters associated with different molecular filamentary and clumpy structures, B being richer than A (Figure \ref{fig_cluster_assignment_maps}). Low data-model residual values of less than a few percent across both clusters show a good model fit (Figure \ref{fig_cluster_identification}). 

SFiNCs MST analysis is consistent with MMA at identifying both clusters (Figure \ref{fig_mma_vs_mst}). Unlike the SFiNCs MST and MMA analyses, the MST analysis of G09 chains both clusters into a single cluster. This result  disagrees with their own smoothed surface density map that shows two cluster cores.

{\it Serpens South} is another active SFR in the Serpens/Aquila Molecular Complex \citep[][and references therein]{OrtizLeon2017}. It is one of the youngest among nearby rich SFRs. MMA identifies four clusters associated with different filamentary molecular structures (Figure \ref{fig_cluster_assignment_maps}). The data-model residuals are less than $10$\% across the field (Figure \ref{fig_cluster_identification}).  Parameters for the sparse clusters B and D are poorly constrained due to the small samples ($N_{4,data} = 6-7$). 

SFiNCs MST chains the two main clusters A and C into a single cluster (Figure Set \ref{fig_mma_vs_mst}). In agreement with MMA, MST identifies the cluster B as a group of several YSOs. MST chains  cluster D to a dozen of additional YSOs; some of those lie projected against an associated molecular filament. At the edges of the field, MST identifies two additional possible weak clusters ($N \ga 10$) with half YSOs lying projected against molecular clumps.

{\it IRAS 20050+2720} is an active SFR in the Cygnus rift molecular complex \citep[][and references therein]{Poppenhaeger2015}. MMA identifies five clusters. Clusters B, C and D are associated with molecular clumps while A and E lie projected against diffuse molecular structures (Figure \ref{fig_cluster_assignment_maps}). The data-model residuals are less than $10$\% across the A, C, D, and E clusters (Figure \ref{fig_cluster_identification}). High positive residuals of $\ga 30$\% between the B and C clusters suggest that independent sparse stellar groups might have been missed by MMA. Parameters are poorly constrained for the sparse cluster A.

SFiNCs MST chains the main clusters C and D into a single cluster (Figure \ref{fig_mma_vs_mst}). MST identifies the sparse clusters A and B as groups of several YSOs.  In agreement with MMA, MST identifies the cluster E as a single rich stellar structure. In addition, MST suggests the presence of a new small stellar group of $N=6$ YSOs projected against a faint molecular clumpy structure  located between the A and D clusters.

As with the SFiNCs MST analysis, the MST procedure of G09 chains clusters C and D into a single cluster. This is inconsistent with their smoothed source density map that shows two cluster cores. Their MST analysis does not identify the small cluster B. Their field of view does not include the clusters A and E.

\citet{Gunther2012} apply the MST clustering analysis to their {\it Spitzer/Chandra} YSO catalog of IRAS 20050+2720. Their MST analysis finds two cluster structures: ``cluster core E'' and ``cluster core W''. Their ``cluster core E'' matches the SFiNCs cluster E. Their ``cluster core W'' chains the three SFiNCs clusters B, C, and D into a single cluster. G\"unther et al. acknowledge the limitations of MST in discriminating close clusters. They further attempt a manual separation of ``cluster core W'' into two clusters: ``If we cut the cluster core W along the dashed black line (Figure 6), we end up with two subcores.''

{\it Sh 2-106} is a rich SFR in the giant Cygnus-X molecular complex \citep[][and references therein]{Adams2015}. MMA identifies four clusters. The two main clusters B and D are associated with dense molecular clumps (Figure \ref{fig_cluster_assignment_maps}). The sparse cluster C is associated with a diffuse molecular structure, and the sparse cluster A is revealed. The data-model residuals are  $<10$\% across the main B and D clusters, but higher residuals of $>30$\% around A and C suggest poor fits with the isothermal ellipsoid models (Figure \ref{fig_cluster_identification}). Parameters for the sparse cluster C ($N_{4,data}=5$) are poorly constrained.

SFiNCs MST chains clusters B and D into a single cluster (Figure \ref{fig_mma_vs_mst}). MST identifies clusters A and C. For the latter case, MST adds many additional YSOs that are projected on a diffuse molecular structure. 

The MIR YSO catalog of G09  misses the core of the SFiNCs cluster D and the eastern portion of the cluster B due to the presence of high nebular MIR emission (Figure 6 in \citet{Getman2017}). These are recovered in SFiNCs through X-ray selection. The MST procedure of G09 identifies a single cluster that spans both main SFiNCs clusters B and D. However their clustering result is likely affected by the aforementioned bias in their data.

{\it IC 5146} is a SFR in the Cocoon Nebula that is part of the IC 5146 molecular complex \citep[][and references therein]{Johnstone2017}. MMA identifies two clusters: the main revealed cluster B surrounded by a dusty shell, and the secondary cluster A associated with molecular clumpy structures (Figure \ref{fig_cluster_assignment_maps}). The data-model residuals are less than 2\% at the core of the cluster B, but reach $\la 20$\% at the two halo spots where the cluster crosses parts of the surrounding dusty shell, suggesting possible contribution from YSOs embedded in the shell (Figure \ref{fig_cluster_identification}). Higher residuals of $>30$\% across the weak cluster A indicate a poor fit with the isothermal ellipsoid model for that cluster. 

Consistent with MMA, SFiNCs MST identifies both of these clusters (Figure \ref{fig_mma_vs_mst}). Consistent with SFiNCs MST and MMA, the MST analysis of G09 identifies the main cluster B. A group of several YSOs chained together by their MST method is located at the position of the cluster A. Their smoothed surface density map also shows two clusters at the locations of A and B.

{\it NGC 7160} is part of the Cepheus OB2 association located inside a 100~pc diameter dusty shell, named the Cepheus Bubble \citep[][and references therein]{Kun2008}. The NGC 7160 region is relatively old and free of molecular material. Both MMA and MST identify a single revealed stellar cluster in this field (Figure \ref{fig_cluster_assignment_maps}, Figure \ref{fig_mma_vs_mst}). The high data-model residuals of over 30\% across the cluster indicate a poor fit with the isothermal ellipsoid model (Figure \ref{fig_cluster_identification}).

{\it LDN 1251B} is a SFR in the L 1251 cloud located at the eastern edge of the giant Cepheus Flare molecular complex \citep[][and references therein]{Kun2008}. MMA identifies a single compact cluster associated with a molecular clump (Figure \ref{fig_cluster_assignment_maps}). The data-model residual values are below $15$\% across the cluster (Figure \ref{fig_cluster_identification}). SFiNCs MST identifies the same single cluster, but adds to it another dozen of YSOs located outside the clump (Figure Set \ref{fig_mma_vs_mst}).

{\it Cep OB3b} is a rich SFR located at the interface between the Cepheus OB3 association and the giant Cepheus molecular cloud \citep[][and references therein]{Kun2008}. MMA identifies four clusters: the revealed A, B, and C clusters, and the cluster D embedded in the Cepheus B cloud (Figure Set \ref{fig_cluster_assignment_maps}). The data-model residual values vary from less than several percent at the cores to $<20$\% at the halos of the rich clusters A, B, and C (Figure \ref{fig_cluster_identification}). Parameters for the weak cluster D are unreliable. 

SFiNCs MST identifies the richer clusters A, B and C (Figure \ref{fig_mma_vs_mst}). MST breaks the cluster C into a main structure surrounded by a few sparse groups, each including several YSOs. One of the groups is associated with the molecular clump of the Cepheus F cloud. 

\citet{Allen2012} applied the N=11 nearest neighbor algorithm to a {\it Spitzer/Chandra} catalog of Cep OB3b. They identify two clusters that match the SFiNCs clusters A and C.

{\it Cep A} is an active SFR associated with one of the few dense and massive molecular clumps of the giant Cepheus molecular cloud \citep[][and references therein]{Kun2008}. MMA finds the single, rich cluster associated with a molecular clump (Figure \ref{fig_cluster_assignment_maps}). The data-model residuals are $<10$\% throughout the field (Figure \ref{fig_cluster_identification}).  The SFiNCs MST procedure also identifies the cluster A (Figure \ref{fig_mma_vs_mst}).

Due to the presence of high nebular MIR emission at the center of the region, the MIR YSO catalog of G09 misses numerous X-ray selected YSOs that lie projected against the center of the molecular clump (Figure 6 in \citet{Getman2017}). Their MST analysis fragments the main cluster into two subclusters; however this clustering result is likely affected by the aforementioned bias in their data.

{\it Cep C} is an active SFR associated with the most massive clump of the giant Cepheus molecular cloud \citep[][and references therein]{Kun2008}. MMA finds two clusters associated with molecular clumps and filaments (Figure \ref{fig_cluster_assignment_maps}). The data-model residual values are $<7$\% across the main cluster A (Figure \ref{fig_cluster_identification}). To the west of cluster A lies a high residual spot associated with a dozen YSOs that was not classified as a significant cluster by MMA. Cluster B has unreliable parameters due to the small sample ($N_{4,data}=4$).

SFiNCs MST breaks the main cluster A into two subclusters (Figure \ref{fig_mma_vs_mst}). The western component  corresponds to the aforementioned high residual spot.  MST also identifies the cluster B, but chains additional YSOs to it.  Some are projected against a molecular clump but some lie outside the clump.

Consistent with MMA, the MST analysis of G09 identifies the cores of the two SFiNCs clusters A and B. 



%%%%%%%%%%%%%%%%%%%%%%%%%%%%%%%%%%%%%%%%%%%%%%%%%%

% Don't change these lines
\bsp	% typesetting comment
\label{lastpage}
\end{document}